\def\IEEEsubmission{0}
\def\reviewColor{\color{black}}
\def\complexNumbers{\mathbb{C}}
\def\realNumbers{\mathbb{R}}
\def\figuresize{\textwidth/2-0.3in}
\def\constante{{\rm e}}

\def\expectationOperator[#1][#2]{{\mathbb{E}_{#2}}\left[#1\right]}
\def\indicatorFunction[#1]{\mathbb{I}\left[{#1}\right]}
\def\probability[#1]{\text{Pr}\left({#1}\right)}
\def\complexGaussian[#1][#2]{\mathcal{CN}({#1,#2})}
\def\gaussian[#1][#2]{\mathcal{N}({#1,#2})}
\def\meanG{\mu}
\def\sigmaG{\sigma}
\def\normalPDF[#1]{\phi\left(#1\right)}
\def\normalCDF[#1]{\Phi\left(#1\right)}

\def\gradientPowerCoefficient{\zeta}
\def\gradientPowerCoeffNegative{\zeta^{-}_{\indexGradient}}
\def\gradientPowerCoeffPositive{\zeta^{+}_{\indexGradient}}

\def\aVariable{x}

\def\indexSumKpKm{n}
\def\coeffErr[#1]{A(#1)}
\def\hyperGeometricFcn[#1][#2][#3][#4]{{_2F_1}\left(#1,#2;#3;#4\right)}
\def\numberOfEdgeDevices{K}
\def\timeDomainOFDM[#1]{s(#1)}

\def\numberOfActiveSubcarriers{M}
\def\idftSize{N}
\def\dataSymbols[#1]{d_{#1}}

\def\receivedSymbolAtSubcarrier[#1]{r_{#1}^{(\indexCommunicationRound)}}
\def\transmittedSymbolAtSubcarrier[#1]{t_{#1}^{(\indexCommunicationRound)}}
\def\randomSymbolAtSubcarrier[#1]{s_{#1}^{(\indexCommunicationRound)}}
\def\channelAtSubcarrier[#1]{h_{#1}^{(\indexCommunicationRound)}}
\def\noiseAtSubcarrier[#1]{n_{#1}^{(\indexCommunicationRound)}}
\def\numberOfOFDMSymbols{S}
\def\indexOFDMSymbol{m}
\def\asymbolFromED[#1]{d_{#1}}

\def\exponentialIntegral[#1]{{\rm Ei}(#1)}
\def\tciFactor[#1]{p_{#1}}
\def\mappingFunction{{f}}
\def\encoder[#1]{\psi_#1}
\def\symbolEnergy{E_{\rm s}}
\def\voteInTime[#1]{m^{#1}}
\def\voteInFrequency[#1]{l^{#1}}
\def\numberOFEDsForOptionOne{K^{+}_{\indexGradient}}
\def\numberOFEDsForOptionSecond{K^{-}_{\indexGradient}}
\def\noiseVariance{\sigma_{\rm n}^2}

\def\correctDecision[#1]{p_{#1}}
\def\zeroDecision[#1]{z_{#1}}
\def\zeroDecisionMax{z_{\rm max}}
\def\incorrectDecision[#1]{q_{#1}}
\def\aparameterForBer[#1]{\epsilon_{#1}}
\def\numberOfEDsWithCorrectChoice{X}
\def\numberOfEDsWithInCorrectChoice{Y}
\def\numberOfEDsWithZeroChoice{Z}
\def\probabilityIncorrect[#1]{P^{\rm err}_{#1}}
\def\meanOptionOne{\mu^{+}_{\indexGradient}}
\def\meanOptionTwo{\mu^{-}_{\indexGradient}}
\def\effectiveSNR{\xi}

\def\oneVector[#1]{\textbf{\textrm{1}}_{#1}}
\def\zeroVector[#1]{\textbf{\textrm{0}}_{#1}}
\def\identityMatrix[#1]{{\textbf{\textrm{I}}_{#1}}}

\def\dataset[#1]{\mathcal{D}_{#1}}

\def\datasetBatch[#1]{\mathcal{\tilde{D}}_{#1}}
\def\batchSize{n_{\rm b}}

\def\completeData{\mathcal{D}}
\def\numberOfModelParameters{Q}
\def\sampleData[#1]{{\textrm{\textbf{x}}}_{#1}}
\def\sampleLabel[#1]{{y}_{#1}}
\def\learningRate{\eta}

\def\maxGradient{g_{\text{max}}}

\def\deltaVectorAtIteration[#1][#2]{{\Delta}^{(#1)}_{#2}}
\def\deltaVectorAtIterationEle[#1]{{\bm \Delta}^{#1}}

\def\indexED{k}
\def\indexGradient{i}
\def\indexSampleData{{\ell}}
\def\indexCommunicationRound{t}

\def\modelParametersAtIteration[#1]{\textbf{w}^{(#1)}}
\def\modelParametersAtIterationEle[#1][#2]{w^{(#1)}_{#2}}

\def\modelParameters{\textbf{w}}
\def\modelParametersEle[#1]{{w}_{#1}}
\def\modelParametersOptimal{{\textbf{w}^{*}}}

\def\localGradientSign[#1][#2]{\bar{\textbf{g}}_{#1}^{(#2)}}
\def\localGradient[#1][#2]{\tilde{\textbf{g}}_{#1}^{(#2)}}
\def\localGradientNoIndex[#1]{\tilde{\textbf{g}}_{#1}}

\def\localGradientSignElement[#1][#2]{{\bar{g}}_{#1}^{(#2)}}
\def\localGradientElement[#1][#2]{{\tilde{g}}_{#1}^{(#2)}}
\def\localGradientNoIndexElement[#1]{{\tilde{g}}_{#1}}
\def\gradientWeight[#1]{\omega{\left(#1\right)}}

\def\encoderGradient[#1][#2][#3]{\psi_{#1}^{#2}(#3)}

\def\lossFunctionSample[#1]{f(#1)}
\def\lossFunctionLocal[#1][#2]{F_{#1}(#2)}
\def\lossFunctionGlobal[#1]{F(#1)}
\def\lossFunctionGlobalMinimum{F^*}

\def\majorityVoteEle[#1][#2]{{v}^{(#1)}_{#2}}
\def\majorityVote[#1]{\textbf{v}^{(#1)}}

\def\majorityVoteSchemeEle[#1][#2]{{\tilde{v}}^{(#1)}_{#2}}
\def\majorityVoteScheme[#1]{\tilde{\textbf{v}}^{(#1)}}

\def\globalGradient[#1]{{\textbf{{g}}}^{(#1)}}
\def\globalGradientElement[#1][#2]{{{g}}^{(#1)}_{#2}}
\def\globalGradientNoIndex{{\textbf{{g}}}}
\def\globalGradientElementNoIndex[#1]{{g_{#1}}}

\def\powerRef{P_{\text{ref}}}

\def\communicationRounds{T}

\def\metricForFirst[#1]{e_{#1}^{+}}
\def\metricForSecond[#1]{e_{#1}^{-}}

\def\nonnegativeConstants{\textbf{L}}
\def\nonnegativeConstantsEle[#1]{L_{#1}}
\def\varianceBound{{\bm \sigma}}
\def\varianceBoundEle[#1]{\sigma_{#1}}

\def\channelVector[#1]{\textbf{\textrm{h}}_{#1}^{(\indexCommunicationRound)}}

\def\receiveVector[#1]{\textbf{\textrm{r}}_{#1}^{(\indexCommunicationRound)}}
\def\receiveVectorEstimate[#1]{\tilde{\textbf{\textrm{{r}}}}_{#1}^{(\indexCommunicationRound)}}

\def\matrixForCut[#1]{\textbf{\textrm{C}}_{#1}}
\def\symbolVector[#1]{\textbf{\textrm{d}}_{#1}^{(\indexCommunicationRound)}}
\def\symbolVectorEstimate[#1]{\tilde{\textbf{\textrm{{d}}}}_{#1}^{(\indexCommunicationRound)}}
\def\receivedVector[#1]{\textbf{\textrm{r}}_{#1}^{(\indexCommunicationRound)}}
\def\noiseVector[#1]{\textbf{\textrm{n}}_{#1}^{(\indexCommunicationRound)}}
\def\noiseVectorOnSymbols[#1]{\tilde{\textbf{\textrm{n}}}_{#1}^{(\indexCommunicationRound)}}

\def\transmittedVector[#1]{\textbf{\textrm{t}}_{#1}^{(\indexCommunicationRound)}}
\def\channelVector[#1]{\textbf{\textrm{h}}_{#1}^{(\indexCommunicationRound)}}
\def\idftMatrix[#1]{\textbf{\textrm{F}}_{#1}^{\rm H}}
\def\dftMatrix[#1]{\textbf{\textrm{F}}_{#1}}
\def\transformPrecoder[#1]{\textbf{\textrm{T}}_{#1}}
\def\transformDecoder[#1]{\textbf{\textrm{T}}_{#1}^{\rm H}}
\def\dftPrecoder[#1]{\textbf{\textrm{D}}_{#1}}
\def\dftDecoder[#1]{\textbf{\textrm{D}}_{#1}^{\rm H}}

\def\frequencyMapping{\textbf{\textrm{M}}_{\textrm{f}}}

\def\channelImpulseResponse[#1]{\textbf{\textrm{h}}_{#1}^{(\indexCommunicationRound)}}
\def\channelMatrix[#1]{\textbf{\textrm{H}}_{#1}^{(\indexCommunicationRound)}}
\def\channelMatrixDiag[#1]{{\bf \Lambda}_{#1}^{(\indexCommunicationRound)}}

\def\samplePeriod{T_{\rm sample}}
\def\symbolSpacing{T_{\rm spacing}}

\def\pulseDuration{T_{\rm pulse}}
\def\syncError{T_{\rm sync}}
\def\frameSyncError{T_{\rm lock}}
\def\channelSpread{T_{\rm chn}}

\def\guardTimeUniform{T_{\rm g}}
\def\numberOfVotesPerDFTsOFDM{M_{\rm vote}}
\def\uniformGap{M_{\rm gap}}
\def\numberOfElementsInAPulse{M_{\rm seq}}
\def\symbolsActivatedBins{\textbf{p}}

\def\referenceDistance{R_{\rm ref}}
\def\minimumDistance{R_{\rm min}}
\def\cellRadius{R_{\rm max}}
\def\distanceED[#1]{r_{#1}}
\def\powerED[#1]{P_{#1}}
\def\pathlossExponent{\alpha}
\def\powerControl{\beta}
\def\effectivePathLossExponent{\alpha_{\rm eff}}
\def\largeScaleImpactOnLearning{\gamma}
\def\arandomvar{y}
\def\indexArea{u}
\def\Nerror{N_{\text{err}}}
\def\thresholdForZero{\mathscr{t}}
\def\steepnessFactor{\rho}
\def\allBatches{{[\datasetBatch[\indexED]]}}

\newcommand\mydots{\hbox to 1em{.\hss.\hss.}}

\if\IEEEsubmission1
\documentclass[journal,comsoc,12pt,onecolumn,draftclsnofoot,]{IEEEtran} \def\baselineSize{1.5}
\else
\documentclass[journal,comsoc]{IEEEtran}\def\baselineSize{1}
\fi

\usepackage{multicol}
\usepackage{lipsum}
\usepackage{mathtools}
\usepackage{amsthm}
\usepackage{acronym}
\usepackage[bookmarksopen=true]{hyperref}
\usepackage{stfloats}
\usepackage{amsfonts}
\usepackage{acronym}
\usepackage{cite}
\usepackage{multirow}
\usepackage{bm}
\usepackage{amsmath,amssymb}
\usepackage{graphicx}
\usepackage[table,xcdraw]{xcolor}
\usepackage[caption=false,font=footnotesize]{subfig}
\usepackage[geometry]{ifsym}
\usepackage{array}
\usepackage[utf8]{inputenc}
\usepackage[T1]{fontenc}

\let\norm\undefined 
\DeclarePairedDelimiter\norm{\lVert}{\rVert}

\usepackage{tikz}
\usepackage{lipsum}
\usetikzlibrary{calc,shadings,patterns}
\makeatletter
\tikzset{%
  remember picture with id/.style={%
    remember picture,
    overlay,
    save picture id=#1,
  },
  save picture id/.code={%
    \edef\pgf@temp{#1}%
    \immediate\write\pgfutil@auxout{%
      \noexpand\savepointas{\pgf@temp}{\pgfpictureid}}%
  },
  if picture id/.code args={#1#2#3}{%
    \@ifundefined{save@pt@#1}{%
      \pgfkeysalso{#3}%
    }{
      \pgfkeysalso{#2}%
    }
  }
}

\def\savepointas#1#2{%
  \expandafter\gdef\csname save@pt@#1\endcsname{#2}%
}

\def\tmk@labeldef#1,#2\@nil{%
  \def\tmk@label{#1}%
  \def\tmk@def{#2}%
}

\tikzdeclarecoordinatesystem{pic}{%
  \pgfutil@in@,{#1}%
  \ifpgfutil@in@%
    \tmk@labeldef#1\@nil
  \else
    \tmk@labeldef#1,(0pt,0pt)\@nil
  \fi
  \@ifundefined{save@pt@\tmk@label}{%
    \tikz@scan@one@point\pgfutil@firstofone\tmk@def
  }{%
  \pgfsys@getposition{\csname save@pt@\tmk@label\endcsname}\save@orig@pic%
  \pgfsys@getposition{\pgfpictureid}\save@this@pic%
  \pgf@process{\pgfpointorigin\save@this@pic}%
  \pgf@xa=\pgf@x
  \pgf@ya=\pgf@y
  \pgf@process{\pgfpointorigin\save@orig@pic}%
  \advance\pgf@x by -\pgf@xa
  \advance\pgf@y by -\pgf@ya
  }%
}

\makeatother

\newcounter{hatchNumber}
\DeclarePairedDelimiter\ceil{\lceil}{\rceil}
\DeclarePairedDelimiter\floor{\lfloor}{\rfloor}

\usepackage{cuted}
\makeatletter
\newif\ifAC@uppercase@first%
\def\Aclp#1{\AC@uppercase@firsttrue\aclp{#1}\AC@uppercase@firstfalse}%
\def\AC@aclp#1{%
	\ifcsname fn@#1@PL\endcsname%
	\ifAC@uppercase@first%
	\expandafter\expandafter\expandafter\MakeUppercase\csname fn@#1@PL\endcsname%
	\else%
	\csname fn@#1@PL\endcsname%
	\fi%
	\else%
	\AC@acl{#1}s%
	\fi%
}%
\def\Acp#1{\AC@uppercase@firsttrue\acp{#1}\AC@uppercase@firstfalse}%
\def\AC@acp#1{%
	\ifcsname fn@#1@PL\endcsname%
	\ifAC@uppercase@first%
	\expandafter\expandafter\expandafter\MakeUppercase\csname fn@#1@PL\endcsname%
	\else%
	\csname fn@#1@PL\endcsname%
	\fi%
	\else%
	\AC@ac{#1}s%
	\fi%
}%
\def\Acfp#1{\AC@uppercase@firsttrue\acfp{#1}\AC@uppercase@firstfalse}%
\def\AC@acfp#1{%
	\ifcsname fn@#1@PL\endcsname%
	\ifAC@uppercase@first%
	\expandafter\expandafter\expandafter\MakeUppercase\csname fn@#1@PL\endcsname%
	\else%
	\csname fn@#1@PL\endcsname%
	\fi%
	\else%
	\AC@acf{#1}s%
	\fi%
}%
\def\Acsp#1{\AC@uppercase@firsttrue\acsp{#1}\AC@uppercase@firstfalse}%
\def\AC@acsp#1{%
	\ifcsname fn@#1@PL\endcsname%
	\ifAC@uppercase@first%
	\expandafter\expandafter\expandafter\MakeUppercase\csname fn@#1@PL\endcsname%
	\else%
	\csname fn@#1@PL\endcsname%
	\fi%
	\else%
	\AC@acs{#1}s%
	\fi%
}%
\edef\AC@uppercase@write{\string\ifAC@uppercase@first\string\expandafter\string\MakeUppercase\string\fi\space}%
\def\AC@acrodef#1[#2]#3{%
	\@bsphack%
	\protected@write\@auxout{}{%
		\string\newacro{#1}[#2]{\AC@uppercase@write #3}%
	}\@esphack%
}%
\def\Acl#1{\AC@uppercase@firsttrue\acl{#1}\AC@uppercase@firstfalse}
\def\Acf#1{\AC@uppercase@firsttrue\acf{#1}\AC@uppercase@firstfalse}
\def\Ac#1{\AC@uppercase@firsttrue\ac{#1}\AC@uppercase@firstfalse}
\def\Acs#1{\AC@uppercase@firsttrue\acs{#1}\AC@uppercase@firstfalse}

\newtheorem{theorem}{Theorem}

\newtheorem{lemma}{Lemma}

\newtheorem{corollary}{Corollary}
\newtheorem{assumption}{Assumption}

\DeclareMathOperator{\sign}{sign}
\DeclareMathOperator{\diag}{diag}
\def\diagOperator[#1]{\diag\left(#1\right)}
\def\signNormal[#1]{\sign\left(#1\right)}
\def\signThreshold[#1][#2]{\sign_{#2}\left(#1\right)}

\acrodef{SNR}{signal-to-noise ratio}
\acrodef{RMSE}{root-mean-square error}
\acrodef{OFDM}{orthogonal frequency division multiplexing}
\acrodef{DFT}{discrete Fourier transform}
\acrodef{PSK}{phase-shift keying}
\acrodef{QAM}{quadrature amplitude modulation}
\acrodef{QPSK}{quadrature phase-shift keying}
\acrodef{PMEPR}{peak-to-mean envelope power ratio}
\acrodef{BER}{bit-error ratio}
\acrodef{SNR}{signal-to-noise ratio}
\acrodef{PSD}{power spectral density}
\acrodef{SE}{spectral efficiency}
\acrodef{CP}{cyclic prefix}
\acrodef{AWGN}{additive white Gaussian noise}
\acrodef{CFR}{channel frequency response}
\acrodef{CIR}{channel impulse response}
\acrodef{MMSE}{minimum mean square error}
\acrodef{LMMSE}{linear minimum mean square error}
\acrodef{BPSK}{binary phase shift keying}
\acrodef{BLER}{block-error rate}
\acrodef{ML}{maximum likelihood}
\acrodef{PHY}{physical layer}
\acrodef{PA}{power amplifier}
\acrodef{IDFT}{inverse DFT}
\acrodef{DoF}{degrees-of-freedom}
\acrodef{IoT}{Internet-of-Things}
\acrodef{FDE}{frequency-domain equalization}
\acrodef{FSK}{frequency-shift keying}
\acrodef{FSK-MV}{\ac{FSK}-based \ac{MV}}
\acrodef{RF}{radio-frequency}
\acrodef{IM}{index modulation}
\acrodef{BS}{base station}
\acrodef{MF}{matched filter}
\acrodef{PPM}{pulse-position modulation}

\acrodef{PPM-MV}[PPM-MV]{\ac{PPM}-based \ac{MV}}
\acrodef{DFT-s-OFDM}{\ac{DFT}-spread \ac{OFDM}}
\acrodef{SC}{single-carrier}

\acrodef{BAA}{broadband analog aggregation}
\acrodef{OBDA}{one-bit broadband digital aggregation}
\acrodef{FEEL}{federated edge learning}
\acrodef{FL}{federated learning}
\acrodef{ED}{edge device}
\acrodef{ES}{edge server}
\acrodef{UL}{uplink}
\acrodef{DL}{downlink}
\acrodef{OAC}{over-the-air computation}
\acrodef{TCI}{truncated-channel inversion}
\acrodef{MV}{majority vote}
\acrodef{CNN}{convolution neural network}
\acrodef{ReLU}{rectified-linear unit}
\acrodef{CSI}{channel state information}
\acrodef{PAPR}{peak-to-average power ratio}
\acrodef{iid}[IID]{independent and identically distributed}
\acrodef{5G}{Fifth Generation}
\acrodef{4G}{Fourth Generation}
\acrodef{NR}{New Radio}
\acrodef{LTE}{Long Term Evolution}
\acrodef{RACH}{random-access channel}
\acrodef{DNN}{deep nueral network}
\acrodef{SGD}{stochastic gradient descent}
\acrodef{signSGD}{sign stochastic gradient descent}

\acrodef{5G}{Fifth Generation}
\acrodef{4G}{Fourth Generation}
\acrodef{NR}{New Radio}
\acrodef{LTE}{Long Term Evolution}
\acrodef{PRACH}{physical random access channel}
\acrodef{PUCCH}{physical uplink control channel}
\acrodef{OFDMA}{orthogonal frequency division multiple access}

\acrodef{PDF}{probability density function}
\acrodef{CDF}{cummulative distribution function}

\acrodef{HP}{hard-coded participation}
\acrodef{HPA}{hard-coded participation with absentees}
\acrodef{SP}{soft-coded participation}
\def\BibTeX{{\rm B\kern-.05em{\sc i\kern-.025em b}\kern-.08em
    T\kern-.1667em\lower.7ex\hbox{E}\kern-.125emX}}
\begin{document}

\title{
	{ Distributed Learning over a Wireless Network with
		Non-coherent Majority Vote Computation
	}\\
	\thanks{Alphan~\c{S}ahin is with the Electrical  Engineering Department,
		University of South Carolina, Columbia, SC, USA. E-mail: asahin@mailbox.sc.edu}
	\thanks{This paper was presented in part at the IEEE International Conference on Advanced Communication Technologies and Networking 2021 \cite{sahinCommnet_2021} and IEEE Wireless Communications \& Networking Conference 2022 \cite{sahinWCNC_2022}.}
	\author{Alphan~\c{S}ahin,~\IEEEmembership{Member,~IEEE}} 
}
\maketitle

\begin{abstract}
In this study, we propose an \ac{OAC} scheme to calculate the \ac{MV} for \ac{FEEL}. 
With the proposed approach, \acp{ED} transmit the signs of local stochastic gradients, i.e., votes, by activating one of two orthogonal resources. The \acp{MV} at the \ac{ES} are obtained with non-coherent detectors by exploiting the accumulations on the resources. Hence, the proposed scheme eliminates the need for \ac{CSI} at the \acp{ED} and \ac{ES}. In this study, we analyze various gradient-encoding strategies through the weight functions and waveform configurations over  \ac{OFDM}. We show that specific weight functions that enable absentee \acp{ED} (i.e., \ac{HPA}) or weighted votes (i.e., \ac{SP})  can substantially reduce the probability of detecting the incorrect \ac{MV}. By taking path loss, power control, cell size, and fading channel  into account, we prove the convergence of the distributed learning for a non-convex function for \ac{HPA}.
Through simulations, we show that the proposed scheme  with \ac{HPA} and \ac{SP} can provide  high test accuracy even when the time-synchronization and the power control  are not ideal under heterogeneous data distribution scenarios.
\end{abstract}
\begin{IEEEkeywords}
Distributed learning, federated edge learning, FSK, OFDM, over-the-air computation, PMEPR, PPM.
\end{IEEEkeywords}
\section{Introduction}

\acresetall

\Ac{OAC} exploits the signal-superposition property of wireless multiple access channels to compute a mathematical function \cite{Gastpar_2003}. It was initially proposed for reliable communications in the interference channel \cite{Nazer_2007} and  wireless sensor networks
\cite{Goldenbaum_2013}. It has recently been applied to  wireless distributed learning \cite{Wanchun_2020} and wireless control systems \cite{Park_2021} to address the latency issues occurring when a larger number of \acp{ED} or \ac{IoT} devices access the limited wireless spectrum. In this study, we particularly consider  \ac{OAC} for \ac{FEEL} \cite{Guangxu_2020,Guangxu_2021}, i.e., one of the promising frameworks for wireless distributed learning.

\ac{FEEL} implements \ac{FL} \cite{pmlr-v54-mcmahan17a} in a wireless network to train a model such as a neural network \cite{Mingzhe_2021}. With \ac{FEEL}, to promote the data privacy, a large number of model parameters (or gradients) are communicated between many \acp{ED} and the \ac{ES} over a wireless channel for aggregation, instead of local data samples. 
However, typical user multiplexing methods such as \ac{OFDMA} can be inefficient in this scenario since the \ac{ES} is not interested in local information of the \acp{ED} but only in a function of them, that is often an arithmetic sum \cite{chen2021distributed}. 
Hence, \ac{OAC} is a prominent solution to address the per-round communication latency in \ac{FEEL} \cite{liu2021training} via snap-shot calculations. Nevertheless, ensuring the robustness of an \ac{OAC} scheme is a challenging task due to the wireless channel, imperfect power control, and time-synchronization errors in practice. Also, the state-of-the-art solutions often require \ac{CSI}  to be available at the \acp{ED} or the \ac{ES}. In this study, we propose an \ac{OAC} scheme to address these challenges by relying on distributed training by the \ac{MV} \cite{Bernstein_2018}.

\subsection{Related Work}
A reliable superposition in a wireless channel is one of the major challenges for \ac{OAC}. To address this issue, a majority of the solutions in the literature adopt pre-equalization techniques \cite{ Guangxu_2020, Guangxu_2021,sery2020overtheair, Amiri_2020, hellstrom2021overtheair,tang2021multislot,Liqun_2021,Zang_2020}. 
 For example, in \cite{Guangxu_2020}, \ac{BAA} over \ac{OFDM} is investigated. It is proposed to modulate  the \ac{OFDM} subcarriers with the model parameters at the \acp{ED}. To achieve a coherent superposition at the \ac{ES}, the symbols on the \ac{OFDM} subcarriers are multiplied with the inverse of the channel coefficients and  the subcarriers that fade are excluded from the transmissions, which is known as {\em \ac{TCI}} in the literature. 
In \cite{Amiri_2020}, the gradient estimates are sparsified  and the sparse vectors are projected into a low-dimensional space to reduce the bandwidth. The compressed data is transmitted with \ac{BAA}.
In \cite{hellstrom2021overtheair}, \ac{BAA}  is investigated  with power control and re-transmissions over static channels. In \cite{sery2020overtheair}, a time-varying precoder is used along with \ac{TCI}. 
In  \cite{tang2021multislot}, time diversity is exploited with a multi-slot \ac{OAC} framework to mitigate the impact of fading channel on \ac{OAC}. In \cite{Liqun_2021}, instead of \ac{TCI}, the parameters are multiplied with the conjugate of the channel coefficients to address the power instability due to the channel inversion.  In \cite{Zang_2020}, the channel inversion is optimized with a sum-power constraint to avoid potential interference issues. 
In \cite{Guangxu_2021}, \ac{OBDA} is proposed  to facilitate the implementation of \ac{FEEL}. In this method,  by considering distributed training by \ac{MV} with the  \ac{signSGD}~\cite{Bernstein_2018}, the \acp{ED} transmit \ac{QPSK} symbols  over \ac{OFDM} subcarriers along with \ac{TCI}, where the signs of the stochastic gradients, i.e., votes, are mapped to the real and imaginary parts of the \ac{QPSK} symbols. At the \ac{ES}, the signs of the real and imaginary parts of the superposed received symbols are calculated to obtain the \ac{MV}. However, \ac{OBDA} still relies on  \ac{TCI} as in \ac{BAA}.

A pre-equalizer can impose stringent requirements on underlying mechanisms to achieve sample-level time synchronization and accurate channel estimation, which can be hard to be met in practice \cite{Careem_2020,Haejoon_2021}. It can also cause power instabilities due to the inversion in frequency-selective channels. To eliminate the pre-equalization, one potential approach  is to employ a large number of antennas at the \ac{ES} and mitigate the impact of the wireless channel on \ac{OAC} with beamforming \cite{weiICC_2022,Amiria_2021}. Although \ac{CSI} is not used at the \acp{ED}, the sum of the superposed channel, needs to be available at the \ac{ES}. 
To the best of our knowledge, the state-of-the-art \ac{OAC} schemes for \ac{FEEL} do not address the case where \ac{CSI} is unavailable to both \acp{ED} and \ac{ES}.

\subsection{Contributions}

In this study, we introduce an \ac{OAC} scheme to calculate the \ac{MV} for \ac{FEEL}.  By extending our  work in \cite{sahinCommnet_2021} and \cite{sahinWCNC_2022}, our contributions  can be listed as follows:

{\bf Non-coherent MV computation}: 
Instead of forming \ac{QPSK} symbols based on the signs of the local stochastic gradients as in OBDA, 
we dedicate two sets of orthogonal resources to transmit the signs of local gradients. Thus, the votes from different \acp{ED} accumulate on the resources non-coherently in fading channel and the \ac{ES} obtains the \ac{MV} with an energy detector. Hence, \ac{CSI} is not needed at the \acp{ED} and \ac{ES} with the proposed method. Also, it eliminates the information loss and power instabilities due to the \ac{TCI} and channel estimation overhead.

{\bf Robustness against impairments}:
The proposed approach provides robustness against time-synchronization errors as it does not to encode the sign of local stochastic gradients into the phase of the transmitted symbols. Considering the randomness in fading channel, path loss, and imperfect power control in a cell, we prove the convergence of \ac{FEEL} in the presence of the proposed scheme for a non-convex loss function. We also show that it can be used with well-known \ac{PMEPR} reduction techniques. 

{\bf Gradient-encoding with various weight functions:}
We extend our initial work in \cite{sahinCommnet_2021} and \cite{sahinWCNC_2022} by generalizing the sign operation with weight functions, which leads to various gradient-encoding strategies, i.e., \ac{HP}, \ac{HPA}, and \ac{SP}. We show that both the probability of detecting the correct \ac{MV} and the convergence rate improve by reducing the impacts of \acp{ED} that have smaller absolute local stochastic gradients on the \ac{MV}. We demonstrate that this strategy can lead to high test accuracy under imperfect power control and heterogeneous data distribution. 

{\bf Compatible waveform configurations:}
We show that the proposed scheme can be configured as \ac{FSK} over \ac{OFDM} and \ac{PPM} over \ac{DFT-s-OFDM} used in \ac{4G} \ac{LTE} and \ac{5G} \ac{NR}. We evaluate their performances through comprehensive simulations.

The rest of the paper is organized as follows. In Section \ref{sec:system}, we provide our system model. In Section \ref{sec:fskMV}, we discuss the proposed scheme in detail. In Section \ref{sec:numerical}, we present numerical results and compare it with \ac{OBDA}. We conclude the paper in Section \ref{sec:conclusion}.

{\em Notation:} The complex  and real numbers are denoted by $\complexNumbers$ and  $\realNumbers$, respectively. 
$\expectationOperator[\cdot][x]$ is the expectation of its  argument over $x$. $\expectationOperator[\cdot][]$ denotes the expectation  over all random variables.
 The function $\signNormal[\cdot]$ results in $1$, $-1$, or $\pm1$ at random for a positive, a negative, or a zero-valued argument, respectively.  The $N$-dimensional all zero vector and the $N\times N$ identity matrix are  $\zeroVector[{N}]$ and $\identityMatrix[{N}]$, respectively. 
The notation  $(\textbf{a})_i^j$ denotes the vector $[{a}_i,{a}_{i+1},\mydots,{a}_j]^{\rm T}$. 
 The function $\indicatorFunction[\cdot]$ results in $1$ if  its argument holds, otherwise it is $0$.  $\probability[\cdot]$ is the probability of an event. 
 The zero-mean  multivariate  complex Gaussian distribution with the covariance matrix ${\textbf{\textrm{C}}_{\numberOfActiveSubcarriers}}$ of an $\numberOfActiveSubcarriers$-dimensional random vector $\textbf{\textrm{x}}\in\complexNumbers^{\numberOfActiveSubcarriers\times1}$ is denoted by
  $\textbf{\textrm{x}}\sim\complexGaussian[\zeroVector[\numberOfActiveSubcarriers]][{\textbf{\textrm{C}}_{\numberOfActiveSubcarriers}}]$. $\gaussian[\meanG][\sigmaG^2]$ is the normal distribution with the mean $\meanG$ and the variance $\sigmaG^2$. The distribution function of the standard normal distribution is $\normalCDF[x]$. Ordinary hypergeometric function is $\hyperGeometricFcn[a][b][c][z]$.

\section{System Model}
\label{sec:system}
\subsection{Deployment and Power Control}
Consider a wireless network with $\numberOfEdgeDevices$ \acp{ED} that are connected to an \ac{ES}, where each \ac{ED} and the \ac{ES}  are equipped with a single antenna. To model power control in the network, let the \ac{SNR} of an \ac{ED} at the \ac{ES} be {\reviewColor $\powerRef/\noiseVariance$} when the link distance between an \ac{ED} and the \ac{ES} is equal to the reference distance $\referenceDistance$. We set the received signal power of the $\indexED$th \ac{ED} at the \ac{ES} as
\begin{align}
\powerED[\indexED]=\left(\frac{{\distanceED[\indexED]}}{\referenceDistance}\right)^{-(\pathlossExponent-\powerControl)}{\reviewColor \powerRef}~,
\label{eq:pathloss}
\end{align}
where {\reviewColor $\powerRef$ is the received signal power for the link distance $\referenceDistance$}, $\distanceED[\indexED]$ is the link distance between the $\indexED$th \ac{ED} and the \ac{ES}, $\pathlossExponent$ is the path loss exponent, and $\powerControl\in [0,\pathlossExponent]$ is a coefficient that determines the amount of the path loss compensated. While $\powerControl=0$ means that there is no power control in the network, $\powerControl=\pathlossExponent$ leads to a system with perfect power control. We define the effective path loss exponent $\effectivePathLossExponent$  as $\effectivePathLossExponent\triangleq\pathlossExponent-\powerControl$. {\reviewColor Without loss of generality, we assume $\powerRef=1$ Watt.}

We assume that the \acp{ED} are deployed uniformly in a circular cell, where the cell radius is $\cellRadius$~meters and the minimum distance between the \ac{ES} and the \acp{ED} is $\minimumDistance$~meters for $\minimumDistance\ge\referenceDistance$. We do not consider the impact of multiple cells (e.g., inter-cell interference) or a more complicated large-scale channel model (e.g., shadowing) on learning in this work to provide insights into the proposed scheme with a tractable analysis. We refer the reader to \cite{mohammadICC_2022}  for our preliminary results on multi-cell non-coherent \ac{MV} computation.

\subsection{Signal Model}
For \ac{OAC}, the \acp{ED} access the wireless channel  on the same time-frequency resources {\em simultaneously} with $\numberOfOFDMSymbols$ \ac{OFDM}-based symbols consisting of $\numberOfActiveSubcarriers$ active subcarriers.  
We express
the $\indexOFDMSymbol$th transmitted baseband precoded \ac{OFDM} symbol for the $\indexED$th \ac{ED} as 
\begin{align}
	\transmittedVector[\indexED,\indexOFDMSymbol] = \idftMatrix[\idftSize]\frequencyMapping \transformPrecoder[\numberOfActiveSubcarriers]\symbolVector[\indexED,\indexOFDMSymbol]~,
	\label{eq:transmitSymbol}
\end{align}
where $\idftMatrix[\idftSize]\in\complexNumbers^{\idftSize\times\idftSize}$ is the normalized $\idftSize$-point \ac{IDFT} matrix (i.e., $\idftMatrix[\idftSize]\dftMatrix[\idftSize]=\identityMatrix[\idftSize]$),  $\transformPrecoder[\numberOfActiveSubcarriers]\in\complexNumbers^{\numberOfActiveSubcarriers\times\numberOfActiveSubcarriers}$ is an orthonormal linear precoder, $\frequencyMapping\in\realNumbers^{\idftSize\times\numberOfActiveSubcarriers}$ is the mapping matrix that maps the precoder output to a set of contiguous subcarriers,  and $\symbolVector[\indexED,\indexOFDMSymbol]\in\complexNumbers^{\numberOfActiveSubcarriers}$ contains the symbols on $\numberOfActiveSubcarriers$ bins  for  $\indexOFDMSymbol\in\{0,1,\mydots,\numberOfOFDMSymbols-1\}$.  For $\transformPrecoder[\numberOfActiveSubcarriers]=\identityMatrix[\numberOfActiveSubcarriers]$, the vector $\transmittedVector[\indexED,\indexOFDMSymbol]$ is an \ac{OFDM} symbol. If the precoder is the normalized {\reviewColor$\numberOfActiveSubcarriers$-point} \ac{DFT} matrix, i.e., $\transformPrecoder[\numberOfActiveSubcarriers]=\dftPrecoder[\numberOfActiveSubcarriers]$,  the vector $\transmittedVector[\indexED,\indexOFDMSymbol]$ becomes a  \ac{DFT-s-OFDM} symbol.  Note that  \ac{DFT-s-OFDM} is a special \ac{SC} waveform using circular convolution \cite{Sahin_2016}, where the symbol spacing in time is $\symbolSpacing=\idftSize\samplePeriod/\numberOfActiveSubcarriers$~seconds, the pulse shape is Dirichlet sinc \cite{Kakkavas_2017},  and $\samplePeriod$ is the sample period.

In this study, we assume that the \ac{CP} duration is larger than the maximum-excess delay denoted by $\channelSpread$~seconds and the multipath channels between the \ac{ES} and the \acp{ED} are independent from each other. Assuming that the transmissions from the \acp{ED} arrive at the \ac{ES} within the \ac{CP} duration, the  $\indexOFDMSymbol$th received baseband signal in discrete-time can be written as
\begin{align}
	\receivedVector[\indexOFDMSymbol] =\sum_{\indexED=1}^{\numberOfEdgeDevices}\sqrt{{\powerED[\indexED]}}\channelMatrix[\indexED]\transmittedVector[\indexED,\indexOFDMSymbol]+\noiseVector[\indexOFDMSymbol]~,
	\label{eq:superposition}
\end{align}
where $\channelMatrix[\indexED]\in\complexNumbers^{\idftSize\times\idftSize}$ is a circular-convolution matrix based on the \ac{CIR}  between the $\indexED$th \ac{ED} and the \ac{ES}, i.e., $\channelImpulseResponse[\indexED]$, and $\noiseVector[\indexOFDMSymbol]\sim\mathcal{CN}(\zeroVector[{\idftSize}],\noiseVariance\identityMatrix[{\idftSize}])$ is the \ac{AWGN}. At the \ac{ES}, we calculate the aggregated symbols on the bins for the  $\indexOFDMSymbol$th  precoded \ac{OFDM} symbol  as
\begin{align}
	\symbolVectorEstimate[\indexOFDMSymbol] =\transformDecoder[\numberOfActiveSubcarriers]\frequencyMapping^{\rm H}\dftMatrix[\idftSize]\receivedVector[\indexOFDMSymbol]=\sum_{\indexED=1}^{\numberOfEdgeDevices}\sqrt{{\powerED[\indexED]}}\transformPrecoder[\numberOfActiveSubcarriers]^{\rm H}\channelMatrixDiag[\indexED]\transformPrecoder[\numberOfActiveSubcarriers]\symbolVector[\indexED,\indexOFDMSymbol]+\noiseVectorOnSymbols[\indexOFDMSymbol]~,
	\label{eq:rx}
\end{align}
where $\channelMatrixDiag[\indexED]
=\diagOperator[{\sqrt{\idftSize}\frequencyMapping^{\rm H}\dftMatrix[\idftSize]\channelImpulseResponse[\indexED]}]
\in \complexNumbers^{\numberOfActiveSubcarriers\times \numberOfActiveSubcarriers}$ is a diagonal matrix based on \ac{CFR}{\reviewColor\cite[Chapter 12.4]{goldsmith_2005}} and $\noiseVectorOnSymbols[\indexOFDMSymbol]\sim\mathcal{CN}(\zeroVector[{\numberOfActiveSubcarriers}],\noiseVariance\identityMatrix[{\numberOfActiveSubcarriers}])$. Note that we do not use \ac{FDE} in \eqref{eq:rx} as we use \ac{OFDM} framework for the \ac{OAC} in this study.

\subsubsection{Time-Synchronization Errors}
In this study,  we consider two different time synchronization errors described as follows:

\paragraph{Time of arrivals}  In practice, the time of transmissions of an \acp{ED} may not be precise, which can cause random time of arrivals at the \ac{ES}. To model this impairment, the time of arrivals of the \acp{ED}' signals at the \ac{ES} location is a random variable with uniform distribution between $0$ and $\syncError$~seconds, where $\syncError$ is the maximum difference among time of arrivals  and it is equal to the reciprocal to the signal bandwidth. 
\paragraph{Frame synchronization}
The time synchronization at the \ac{ES} may also not be precise in practice. To model this, we assume that the point where the \ac{DFT} starts is backed off by $\Nerror$ samples within the \ac{CP} window, i.e., $\frameSyncError=\Nerror\samplePeriod$. Note that  the uncertainty of the synchronization point within the \ac{CP} window is often not an issue for traditional communications due to the channel estimation. However, it can cause a non-negligible impact on \ac{OAC} since equalization is often not used at the receiver for an \ac{OAC} scheme.

We embed aforementioned time-synchronization errors into the \ac{CFR}, i.e., the diagonal elements of $\channelMatrixDiag[\indexED]$, with additional phase rotations since a translation in the time domain results in phase rotations in the frequency domain.

\subsubsection{Frequency-Synchronization Errors}
We assume that the frequency synchronization in the network is handled with a control mechanism as done in 3GPP \ac{4G} \ac{LTE} and/or \ac{5G} \ac{NR} with \ac{RACH} and/or \ac{PUCCH}  \cite{10.5555/3294673}.

%

\subsection{Learning Model}
 Let $\dataset[\indexED]$ denote the local data containing labeled data samples at the $\indexED$th \ac{ED} as $\{(\sampleData[\indexSampleData], \sampleLabel[\indexSampleData] )\}\in\dataset[\indexED]$, $\forall\indexED$, where $\sampleData[\indexSampleData]$ and $\sampleLabel[\indexSampleData]$ are $\indexSampleData$th data sample and its associated label, respectively. The centralized learning problem can  be expressed as 
\begin{align}
	\modelParametersOptimal=\arg\min_{\modelParameters} \lossFunctionGlobal[\modelParameters]=\arg\min_{\modelParameters} \frac{1}{|\completeData|}\sum_{\forall(\sampleData[], \sampleLabel[] )\in\completeData} \lossFunctionSample[{\modelParameters,\sampleData[],\sampleLabel[]}]~,
	\label{eq:clp}
\end{align}
where $\completeData=\dataset[1]\cup\dataset[2]\cup\cdots\cup\dataset[K]$ and  $\lossFunctionSample[{\modelParameters,\sampleData[],\sampleLabel[]}]$ is the sample loss function that measures the labeling error for $(\sampleData[], \sampleLabel[])$ for the parameters $\modelParameters\triangleq[\modelParametersEle[1],\mydots,\modelParametersEle[\numberOfModelParameters]]^{\rm T}\in\realNumbers^{\numberOfModelParameters}$, and $\numberOfModelParameters$ is the number of parameters. With full-batch gradient descent, a local optimum point can be  obtained as
\begin{align}
\modelParametersAtIteration[\indexCommunicationRound+1] = \modelParametersAtIteration[\indexCommunicationRound] - \learningRate  \globalGradient[\indexCommunicationRound]~,
\end{align}
where $\learningRate$ is the learning rate and
\begin{align}
	\globalGradient[\indexCommunicationRound] =  \nabla \lossFunctionGlobal[{\modelParametersAtIteration[\indexCommunicationRound]}]
	= \frac{1}{|\completeData|}\sum_{\forall(\sampleData[], \sampleLabel[] )\in\completeData} \nabla 
	\lossFunctionSample[{\modelParametersAtIteration[\indexCommunicationRound],\sampleData[],\sampleLabel[]}]
	~,
	\label{eq:GlobalGradient}
\end{align}
where the $\indexGradient$th element of  $\globalGradient[\indexCommunicationRound]\triangleq[\globalGradientElement[\indexCommunicationRound][1],\mydots,\globalGradientElement[\indexCommunicationRound][\numberOfModelParameters]]^{\rm T}$ is the gradient of $\lossFunctionGlobal[{\modelParametersAtIteration[\indexCommunicationRound]}]$ with respect to $\modelParametersAtIterationEle[\indexCommunicationRound][\indexGradient]$.

In  \cite{Bernstein_2018}, in the context of parallel processing, distributed training by \ac{MV} with \ac{signSGD} is investigated to solve \eqref{eq:clp}. In this method, for the $\indexCommunicationRound$th communication round, the $\indexED$th \ac{ED} first calculates the local stochastic gradient  $\localGradient[\indexED][\indexCommunicationRound]\triangleq[\localGradientElement[\indexED,1][\indexCommunicationRound],\mydots,\localGradientElement[\indexED,\numberOfModelParameters][\indexCommunicationRound]]^{\rm T}$ as 
\begin{align}
	\localGradient[\indexED][\indexCommunicationRound] =  \nabla  \lossFunctionLocal[\indexED][{\modelParametersAtIteration[\indexCommunicationRound]}] 
	= \frac{1}{\batchSize} \sum_{\forall(\sampleData[], \sampleLabel[] )\in\datasetBatch[\indexED]} \nabla 
	\lossFunctionSample[{\modelParametersAtIteration[\indexCommunicationRound],\sampleData[],\sampleLabel[]}]
	~,
	\label{eq:LocalGradientEstimate}
\end{align}
where $\datasetBatch[\indexED]\subset\dataset[\indexED]$ is the selected data batch from the local data set and $\batchSize=|\datasetBatch[\indexED]|$ as the batch size. Afterwards, instead of the actual values of local stochastic gradients, the $\indexED$th \ac{ED} sends their signs, i.e., $\localGradientSign[\indexED][\indexCommunicationRound]\triangleq[\localGradientSignElement[\indexED,1][\indexCommunicationRound],\mydots,\localGradientSignElement[\indexED,\numberOfModelParameters][\indexCommunicationRound]]^{\rm T}$, $\forall\indexED$, to the \ac{ES}, where the $\indexGradient$th element of the vector $\localGradientSign[\indexED][\indexCommunicationRound]$ is
$\localGradientSignElement[\indexED,\indexGradient][\indexCommunicationRound]=\sign(\localGradientElement[\indexED,\indexGradient][\indexCommunicationRound])$. 
The \ac{ES}  obtains  the \ac{MV} for the $\indexGradient$th gradient as
 \begin{align}
 	\majorityVoteEle[\indexCommunicationRound][\indexGradient]\triangleq\sign\left(\sum_{\indexED=1}^{\numberOfEdgeDevices} \localGradientSignElement[\indexED,\indexGradient][\indexCommunicationRound]\right)~.
 	\label{eq:majorityVote}
 \end{align} 
Subsequently, the \ac{ES} sends $\majorityVote[\indexCommunicationRound]=[\majorityVoteEle[\indexCommunicationRound][1],\mydots,\majorityVoteEle[\indexCommunicationRound][\numberOfModelParameters]]^{\rm T}$ back to the \acp{ED}  and the models at the \acp{ED} are updated as
$
\modelParametersAtIteration[\indexCommunicationRound+1] = \modelParametersAtIteration[\indexCommunicationRound] - \learningRate  \majorityVote[\indexCommunicationRound]$.
This procedure is repeated consecutively until a predetermined convergence criterion is achieved.  

 For \ac{FEEL}, the optimization problem can also be expressed as \eqref{eq:clp} for a scenario where the local data samples and their labels are not available at the \ac{ES} and the link between an \ac{ED} and the \ac{ES} experiences a  wireless channel. To solve \eqref{eq:clp} under these constraints,  we adopt the same procedure summarized for the distributed training by the \ac{MV}  with two major differences: 
  	1)~We calculate the \ac{MV} for each gradient with the proposed \ac{OAC} scheme, leading a different expression from \eqref{eq:majorityVote}. 
 	2)~We investigate various gradient-encoding operations that are different from the sign operation, which improves the learning performance, as discussed in Section~\ref{sec:fskMV}.

\section{Majority Vote with Non-coherent Detection}
\label{sec:fskMV}

\subsection{Edge Device - Transmitter}
With the proposed \ac{OAC} scheme, the \acp{ED} perform a low-complexity operation to transmit the signs of the  gradients given in \eqref{eq:LocalGradientEstimate}:
Let $\mappingFunction$ be a mapping function that maps $\indexGradient\in\{1,2,\mydots,\numberOfModelParameters\}$ to  a distinct pair of $(\voteInTime[+],\voteInFrequency[+])$ and $(\voteInTime[-],\voteInFrequency[-])$ that indicate the resources for $\voteInTime[+],\voteInTime[-]\in\{0,1,\mydots,\numberOfOFDMSymbols-1\}$, $\voteInFrequency[+],\voteInFrequency[-]\in\{0,1,\mydots,2\numberOfVotesPerDFTsOFDM-1\}$, and $\numberOfModelParameters=\numberOfOFDMSymbols\numberOfVotesPerDFTsOFDM$. For all $\indexGradient$, we  determine the elements of the symbol vector based on $\localGradientElement[\indexED,\indexGradient][\indexCommunicationRound]$ as
\if\IEEEsubmission0
\begin{align}
	(\symbolVector[\indexED,{\voteInTime[+]}])_{\voteInFrequency[+](\numberOfElementsInAPulse+\uniformGap)+1}^{\voteInFrequency[+](\numberOfElementsInAPulse+\uniformGap)+\numberOfElementsInAPulse} \nonumber&\\&\hspace{-17mm}
	=  \sqrt{\symbolEnergy}\symbolsActivatedBins \randomSymbolAtSubcarrier[\indexED,\indexGradient]
	\gradientWeight[{\localGradientElement[\indexED,\indexGradient][\indexCommunicationRound]}]\indicatorFunction[{{\signNormal[{\localGradientElement[\indexED,\indexGradient][\indexCommunicationRound]}]}=1}]~,
	\label{eq:symbolOne}
\end{align}
and
\begin{align}
	(\symbolVector[\indexED,{\voteInTime[-]}])_{\voteInFrequency[-](\numberOfElementsInAPulse+\uniformGap)+1}^{\voteInFrequency[-](\numberOfElementsInAPulse+\uniformGap)+\numberOfElementsInAPulse} \nonumber&\\&\hspace{-17mm} =\sqrt{\symbolEnergy}\symbolsActivatedBins \randomSymbolAtSubcarrier[\indexED,\indexGradient]
	\gradientWeight[{\localGradientElement[\indexED,\indexGradient][\indexCommunicationRound]}]\indicatorFunction[{{\signNormal[{\localGradientElement[\indexED,\indexGradient][\indexCommunicationRound]}]}=-1}]~,
	\label{eq:symbolTwo}
\end{align}
\else
\begin{align}
	(\symbolVector[\indexED,{\voteInTime[+]}])_{\voteInFrequency[+](\numberOfElementsInAPulse+\uniformGap)+1}^{\voteInFrequency[+](\numberOfElementsInAPulse+\uniformGap)+\numberOfElementsInAPulse} 
	=  \sqrt{\symbolEnergy}\symbolsActivatedBins \randomSymbolAtSubcarrier[\indexED,\indexGradient]
	\gradientWeight[{\localGradientElement[\indexED,\indexGradient][\indexCommunicationRound]}]\indicatorFunction[{{\signNormal[{\localGradientElement[\indexED,\indexGradient][\indexCommunicationRound]}]}=1}]~,
	\label{eq:symbolOne}
\end{align}
and
\begin{align}
	(\symbolVector[\indexED,{\voteInTime[-]}])_{\voteInFrequency[-](\numberOfElementsInAPulse+\uniformGap)+1}^{\voteInFrequency[-](\numberOfElementsInAPulse+\uniformGap)+\numberOfElementsInAPulse}  =\sqrt{\symbolEnergy}\symbolsActivatedBins \randomSymbolAtSubcarrier[\indexED,\indexGradient]
	\gradientWeight[{\localGradientElement[\indexED,\indexGradient][\indexCommunicationRound]}]\indicatorFunction[{{\signNormal[{\localGradientElement[\indexED,\indexGradient][\indexCommunicationRound]}]}=-1}]~,
	\label{eq:symbolTwo}
\end{align}
\fi
where $\gradientWeight[\cdot]$ is an even-symmetric weight function that ranges from $0$ to $1$, $\symbolsActivatedBins\in\complexNumbers^{\numberOfElementsInAPulse\times 1}$ is a sequence with $\norm{\symbolsActivatedBins}_2^2={\numberOfElementsInAPulse}$,  $\randomSymbolAtSubcarrier[\indexED,\indexGradient]$ is a randomization symbol on the unit-circle, $\symbolEnergy=2(\numberOfElementsInAPulse+\uniformGap)/\numberOfElementsInAPulse$ is an energy normalization factor, and $\numberOfElementsInAPulse\ge1$ and $\uniformGap\ge0$ are the parameters that determine the sequence length for each gradient and the gap between the sequences, respectively. 

{\reviewColor With \eqref{eq:symbolOne} and \eqref{eq:symbolTwo}, two sets of orthogonal resources are allocated for the $\indexGradient$th gradient at the $\indexED$th ED and either of two resources is activated, i.e., {\em orthogonal signaling}, based on the sign of the gradient, where the activation is expressed via the function $\indicatorFunction[\cdot]$. Since $2(\numberOfElementsInAPulse+\uniformGap)$ resources are allocated for each gradient,  the maximum number of gradients that can be carried for each precoded-\ac{OFDM} symbol can be calculated as}
\begin{align}
	\numberOfVotesPerDFTsOFDM = \floor*{\frac{\numberOfActiveSubcarriers}{2(\numberOfElementsInAPulse+\uniformGap)}}~. 
	\label{eq:numberOfVotes}
\end{align}

The proposed scheme can used with various weight functions and  waveform configurations. 
{\reviewColor While the weight function in  \eqref{eq:symbolOne} and \eqref{eq:symbolTwo} primarily addresses heterogeneous data distribution scenarios, the randomization symbols and the sequence $\symbolsActivatedBins$  along with the precoder $\transformPrecoder[\numberOfActiveSubcarriers]$ determine the waveform characteristics as discussed  in Section~\ref{subsec:encoding} and Section~\ref{subsec:waveform} in detail.}

\subsubsection{Gradient-encoding strategies}
\label{subsec:encoding}
We analyze two different weight functions to map the local gradients to the dedicated resources, i.e., \ac{HP} and \ac{HPA}. We discuss a generalization of the weight function in Section~\ref{subsec:extensions}.
\paragraph{\ac{HP}}
With this strategy, the $\indexED$th \ac{ED} always activates one of two dedicated resources for the $\indexGradient$th gradient and the weight is constant, i.e.,
\begin{align}
\gradientWeight[{\localGradientElement[\indexED,\indexGradient][\indexCommunicationRound]}] = 1~.
	\nonumber
\end{align}
Thus, the \acp{ED} always participate in the \ac{MV} calculation even when the local gradient is close to zero or equal to zero. 

\paragraph{\ac{HPA}}
This strategy allows an \ac{ED}  to be absent in the \ac{MV} calculation if the absolute of the $\indexGradient$th local gradient is less than or equal to a pre-determined threshold $\thresholdForZero$ as
\begin{align}
	\gradientWeight[{\localGradientElement[\indexED,\indexGradient][\indexCommunicationRound]}] = \begin{cases}
1,&|\localGradientElement[\indexED,\indexGradient][\indexCommunicationRound]|>\thresholdForZero\\
0,&|\localGradientElement[\indexED,\indexGradient][\indexCommunicationRound]|\le\thresholdForZero
		\end{cases}~,
	\nonumber
\end{align}
where $\thresholdForZero$ is a non-negative constant. Therefore, if the magnitude of the local gradient is less than or equal to $\thresholdForZero$, neither of the two resources are activated with \ac{HPA}. If the threshold is set to $0$ and $\probability[{\localGradientElement[\indexED,\indexGradient][\indexCommunicationRound]}=0]=0$ hold, \ac{HPA} corresponds to \ac{HP}. \ac{HPA} allows the \ac{MV} computation to consider only the \acp{ED} with large gradient magnitudes  by eliminating the \acp{ED} that have weaker positions (i.e., converging \acp{ED}) on the correct gradient direction (i.e., $\signNormal[{\globalGradientElement[\indexCommunicationRound][\indexGradient]}]$). Hence, \ac{HPA} can improve the probability of detecting the correct  gradient direction as elaborated theoretically in Section~\ref{subsec:ErrProT}.

\subsubsection{Waveform configurations}
\label{subsec:waveform}
To transmit the encoded gradients, we configure the precoder $\transformPrecoder[\numberOfActiveSubcarriers]$ to obtain two modulation schemes that also lead to fundamentally different time and frequency characteristics discussed as follows:
\paragraph{FSK configuration}
For the \ac{FSK} configuration, the precoder $\transformPrecoder[\numberOfActiveSubcarriers]$ is set to $\identityMatrix[\numberOfActiveSubcarriers]$, i.e., the transmitted signals from \acp{ED}  consist of \ac{OFDM} symbols and the resources for the encoded gradients are the subcarriers. This configuration provides robustness against time-synchronization errors since the time-synchronization errors  within the \ac{CP} duration cause merely phase rotations in the frequency domain and the proposed \ac{OAC} scheme does not carry the information on the phase. To maximize $\numberOfVotesPerDFTsOFDM$, we set $\symbolsActivatedBins=[1]$, $\numberOfElementsInAPulse=1$, and $\uniformGap=0$.  Under this specific configuration, two subcarriers are dedicated with \eqref{eq:symbolOne} and \eqref{eq:symbolTwo}. As a special case of the mapping function  $\mappingFunction$, if  $\voteInTime[-]=\voteInTime[+]$ and $\voteInFrequency[-]=\voteInFrequency[+]+1$ hold for all $\indexGradient$, the adjacent subcarriers of $\voteInTime[+]$th \ac{OFDM} symbol are used for the $\indexGradient$th gradient, which corresponds to \ac{FSK} over \ac{OFDM} subcarriers.  
We refer to the proposed scheme under this specific mapping as {\ac{FSK-MV}}.  

{\reviewColor It is well-known that the entries
	of a stochastic gradient vector can be highly correlated due to the over-parameterization of neural networks \cite{Xue_2022}. Hence, without any precaution, the entries of $\symbolVector[\indexED,\indexOFDMSymbol]$ can be correlated and the resulting signal can have large power fluctuations. To decrease the correlation in the frequency domain, we use random \ac{QPSK} symbols \cite{Jawhar_2019} for $\randomSymbolAtSubcarrier[\indexED,\indexGradient]$.}

\paragraph{PPM configuration}
\label{par:PPMconfTX}
For the \ac{PPM} configuration, the precoder $\transformPrecoder[\numberOfActiveSubcarriers]$ is set to $\dftDecoder[\numberOfActiveSubcarriers]$. Hence, the transmitted signals from \acp{ED} are based on \ac{DFT-s-OFDM} symbols and the resources utilized for the encoded gradients are wide-band pulses.  We synthesize the pulse in a \ac{PPM} symbol by activating consecutive $\numberOfElementsInAPulse$ bins of \ac{DFT-s-OFDM}, which effectively corresponds to a pulse with the duration of $\pulseDuration\approx\numberOfElementsInAPulse\symbolSpacing$~seconds by combining $\numberOfElementsInAPulse$ shifted versions of the Dirichlet sinc functions in time.  In this study, we choose $\symbolsActivatedBins$ as $[1,-1,1,-1,\cdots]^{\rm T}$ since this sequence yields a rectangular-like pulse in the time domain for \ac{DFT-s-OFDM}, as illustrated in Section~\ref{sec:numerical}.  It is worth noting that the proposed framework allows one to design $\symbolsActivatedBins$ for various pulse shapes, which can be considered for further optimization of the proposed scheme.   If $\voteInTime[+]=\voteInTime[-]$ and $\voteInFrequency[+]=\voteInFrequency[-]+1$ for all $\indexGradient$, the adjacent resources of the $\voteInTime[+]$th precoded \ac{OFDM} symbol are used for voting and we refer to this configuration as \ac{PPM-MV}.

As compared to the \ac{FSK-MV}, the \ac{PPM-MV} configuration requires   a guard period  between the adjacent pulses to accommodate the time-synchronization errors  and the delay spread. To address this issue, we deactivate the following $\uniformGap$ bins after $\numberOfElementsInAPulse$ active bins, which results in a guard period with the duration of $\guardTimeUniform\approx\uniformGap\symbolSpacing$~seconds, where the condition given by
\begin{align}
\uniformGap\ge \ceil*{\frac{\channelSpread+\syncError+\frameSyncError}{\symbolSpacing}}~,
\label{eq:condition}
\end{align}
must hold. 
Under the condition \eqref{eq:condition},  $\numberOfVotesPerDFTsOFDM$ with the \ac{PPM-MV} is smaller than the one with the \ac{FSK-MV}, as can be deduced from \eqref{eq:numberOfVotes}. Nevertheless, the \ac{PPM-MV} brings two distinct features: 1) It leads to a trade-off between the \ac{PMEPR} and the resource utilization. For a given $\uniformGap$,  the pulse energy distributes more evenly in time with increasing $\numberOfElementsInAPulse$. Hence, the amplitude of the baseband signal decreases as less votes are carried. This results in a decreasing \ac{PMEPR}, but at the expense of more resource consumption. 2) 
The multi-path channel affects all of the encoded local gradients of an \ac{ED} {\em similarly} with the \ac{PPM-MV}, whereas it amplifies or attenuates them in the frequency domain  with the \ac{FSK-MV} due to the frequency selectivity.
On the other hand, the orthogonality of the \ac{PPM} symbols within a \ac{DFT-s-OFDM} symbol is lost in a frequency-selective channel, while the \ac{FSK-MV} ensures the orthogonality of \ac{FSK} symbols.  Nevertheless, the interference among \ac{PPM} symbols due to the multi-path channel can be maintained negligibly low under the condition \eqref{eq:condition}.

\subsection{Edge Server - Receiver}
\begin{figure*}[t]
	\centering
	{\includegraphics[width =6.4in]{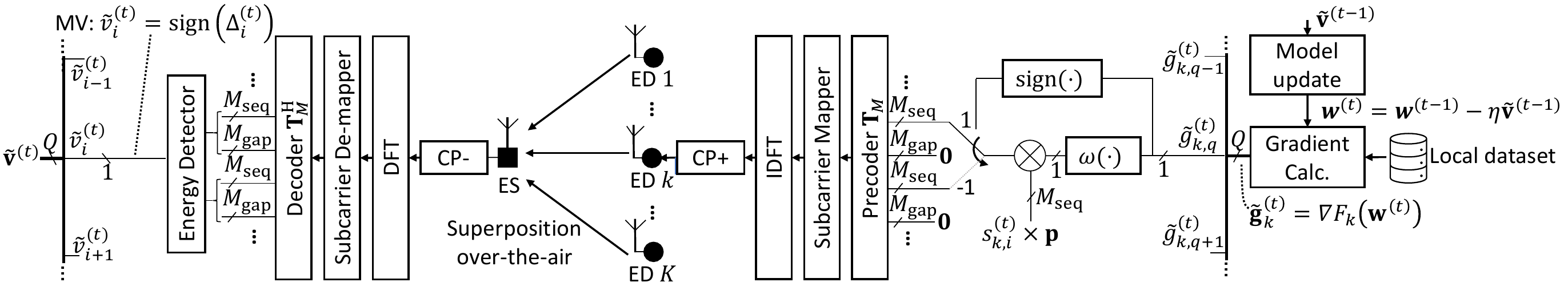}
	} 
	\caption{Transmitter and receiver diagrams  for the proposed \ac{OAC} scheme.}
	\label{fig:feelBlockDiagram}
\end{figure*}

At the ES, we first identify the pairs $(\voteInTime[+],\voteInFrequency[+])$ and $(\voteInTime[-],\voteInFrequency[-])$  based on $\mappingFunction$ for a given $\indexGradient$. We then obtain the \ac{MV}  for the $\indexGradient$th gradient with an energy detector as
\begin{align}
	\majorityVoteSchemeEle[\indexCommunicationRound][\indexGradient] = \signNormal[{\deltaVectorAtIteration[\indexCommunicationRound][\indexGradient]}]~,
	\label{eq:detector}
\end{align}
where $\deltaVectorAtIteration[\indexCommunicationRound][\indexGradient]\triangleq{\metricForFirst[\indexGradient]-\metricForSecond[\indexGradient]}$ for
$\metricForFirst[\indexGradient]\triangleq  \norm{\matrixForCut[{\voteInFrequency[+]}]\symbolVectorEstimate[{\voteInTime[+]}]}_2^2$ and $\metricForSecond[\indexGradient] \triangleq  \norm{\matrixForCut[{\voteInFrequency[-]}]\symbolVectorEstimate[{\voteInTime[-]}]}_2^2$, where $\matrixForCut[{\voteInFrequency[]}]\in\realNumbers^{\numberOfElementsInAPulse+\uniformGap\times\numberOfActiveSubcarriers}$ is a de-mapping matrix that takes the corresponding symbols from the received precoded OFDM symbol for a given $\voteInFrequency[]\in\{\voteInFrequency[-],\voteInFrequency[+]\}$. After the detection, the \ac{ES} broadcasts  $\majorityVoteScheme[\indexCommunicationRound]=[\majorityVoteSchemeEle[\indexCommunicationRound][1],\mydots,\majorityVoteSchemeEle[\indexCommunicationRound][\numberOfModelParameters]]^{\rm T}$ and the models at the \acp{ED} are updated as
\begin{align}
	\modelParametersAtIteration[\indexCommunicationRound+1] = \modelParametersAtIteration[\indexCommunicationRound] - \learningRate  \majorityVoteScheme[\indexCommunicationRound]~.
	\label{eq:MVsignSGDsche}
\end{align}

The non-coherent \ac{MV} detection in \eqref{eq:detector} is valid for all gradient encoders and waveform configurations,  discussed in Section~\ref{subsec:encoding} and Section~\ref{subsec:waveform}, respectively. For \ac{FSK-MV}, $\metricForFirst[\indexGradient]$ and $\metricForSecond[\indexGradient]$
are the energies of the superposed symbols on adjacent subcarriers as $\numberOfElementsInAPulse=1$ and $\uniformGap=0$. For \ac{PPM-MV}, since the multipath channel disperses the pulses in the time domain and the synchronization error changes the position of the pulse in time, the calculations of $\metricForFirst[\indexGradient]$ and $\metricForSecond[\indexGradient]$ consider $\numberOfElementsInAPulse+\uniformGap$ bins through $\matrixForCut[{\voteInFrequency[]}]$ after the \ac{DFT-s-OFDM} receive processing is completed.

The transmitter and receiver block diagrams based on the aforementioned discussions are provided in \figurename~\ref{fig:feelBlockDiagram}.

\section{Why Does It Work without CSI?}
\label{ssec:why}
The proposed scheme leads to a fundamentally different training strategy since the \ac{MV} is determined in a probabilistic manner by comparing ${\metricForFirst[\indexGradient]}$ and ${\metricForSecond[\indexGradient]}$  in \eqref{eq:detector}. To elaborate this difference, we  analyze the proposed scheme from three different perspectives: average received signal power, error probability, and convergence rate.

\subsection{Average Received Signal Power}
Let $\numberOFEDsForOptionOne$ and $\numberOFEDsForOptionSecond$ be the number of \acp{ED} with $\localGradientElement[\indexED,\indexGradient][\indexCommunicationRound]>0$ and $\localGradientElement[\indexED,\indexGradient][\indexCommunicationRound]<0$, respectively, such that $\gradientWeight[{\localGradientElement[\indexED,\indexGradient][\indexCommunicationRound]}]\neq0$ holds. We  define the average received signal power as
$
	\meanOptionOne\triangleq
	\expectationOperator[{\metricForFirst[\indexGradient]}][{\distanceED[\indexED],\channelMatrix[\indexED],\noiseVector[\indexOFDMSymbol],\datasetBatch[\indexED]}]
$
and 
$
	\meanOptionTwo\triangleq
	\expectationOperator[{\metricForSecond[\indexGradient]}][{\distanceED[\indexED],\channelMatrix[\indexED],\noiseVector[\indexOFDMSymbol],\datasetBatch[\indexED]}]
$.
We obtain the expressions of $\meanOptionOne$ and $\meanOptionTwo$ with the following lemma:

\begin{lemma}[Average received signal power] \rm  For given $\numberOFEDsForOptionOne$ and $\numberOFEDsForOptionSecond$, $\meanOptionOne$ and $\meanOptionTwo$ are
	\begin{align}
		\meanOptionOne\approx\numberOfElementsInAPulse\symbolEnergy\numberOFEDsForOptionOne\largeScaleImpactOnLearning\gradientPowerCoeffPositive+(\numberOfElementsInAPulse + \uniformGap)\noiseVariance~,
		\label{eq:energyFirst}
	\end{align}
	and 
	\begin{align}
		\meanOptionTwo\approx\numberOfElementsInAPulse\symbolEnergy\numberOFEDsForOptionSecond\largeScaleImpactOnLearning\gradientPowerCoeffNegative+(\numberOfElementsInAPulse + \uniformGap)\noiseVariance~,		
		\label{eq:energySecond}
	\end{align}
	respectively, where $\gradientPowerCoeffPositive,\gradientPowerCoeffNegative\in[0,1]$ and $\largeScaleImpactOnLearning\in[0,1]$ are given by
	\begin{align}
		\gradientPowerCoeffPositive\triangleq	\expectationOperator[|\gradientWeight[{\localGradientElement[\indexED,\indexGradient][\indexCommunicationRound]}]|^2|\gradientWeight[{\localGradientElement[\indexED,\indexGradient][\indexCommunicationRound]}]\neq0,{\localGradientElement[\indexED,\indexGradient][\indexCommunicationRound]}>0 ][{\datasetBatch[\indexED]}]~,
		\label{eq:weightimpactPositive}
	\end{align}	 
	\begin{align}
	\gradientPowerCoeffNegative\triangleq	\expectationOperator[|\gradientWeight[{\localGradientElement[\indexED,\indexGradient][\indexCommunicationRound]}]|^2|\gradientWeight[{\localGradientElement[\indexED,\indexGradient][\indexCommunicationRound]}]\neq0,{\localGradientElement[\indexED,\indexGradient][\indexCommunicationRound]}<0][{\datasetBatch[\indexED]}]~,
	\label{eq:weightimpactNegative}
\end{align}	 
and
	\begin{align}
		\largeScaleImpactOnLearning \triangleq \begin{cases}
			\frac{2\referenceDistance^{\effectivePathLossExponent}{\reviewColor \powerRef}}{\cellRadius^2-\minimumDistance^2} \frac{\minimumDistance^{2-\effectivePathLossExponent}-\cellRadius^{2-\effectivePathLossExponent}}{\effectivePathLossExponent-2}~, & \effectivePathLossExponent\neq2\\
			\frac{2\referenceDistance^{\effectivePathLossExponent}{\reviewColor \powerRef}}{\cellRadius^2-\minimumDistance^2}\ln{\frac{\cellRadius}{\minimumDistance}}~, & \effectivePathLossExponent=2\\
		\end{cases}~.
	\label{eq:pathlossimpact}
	\end{align}
	\label{lemma:exp}
\end{lemma}
The proof is given in Appendix~\ref{app:lemma:exp}. 

Note that while  $\gradientPowerCoeffNegative$ and $\gradientPowerCoeffPositive$ are less than or equal to 1 for a general weight function, they are equal to $1$ for \ac{HP} and \ac{HPA}. Also, under perfect power control, the coefficient $\largeScaleImpactOnLearning$ is equal to 1.

Based on Lemma~\ref{lemma:exp}, \eqref{eq:detector} is likely to obtain the \ac{MV} because $\meanOptionOne$ and $\meanOptionTwo$ are linear functions of $\numberOFEDsForOptionOne$ and $\numberOFEDsForOptionSecond$, respectively. However, the detection performance depends on the parameter $\largeScaleImpactOnLearning\in[0,1]$ that captures the impacts of power control, path loss, and cell size on ${\metricForFirst[\indexGradient]}$ and ${\metricForSecond[\indexGradient]}$. 

\begin{figure}[t]
	\centering
	{\includegraphics[width =3.3in]{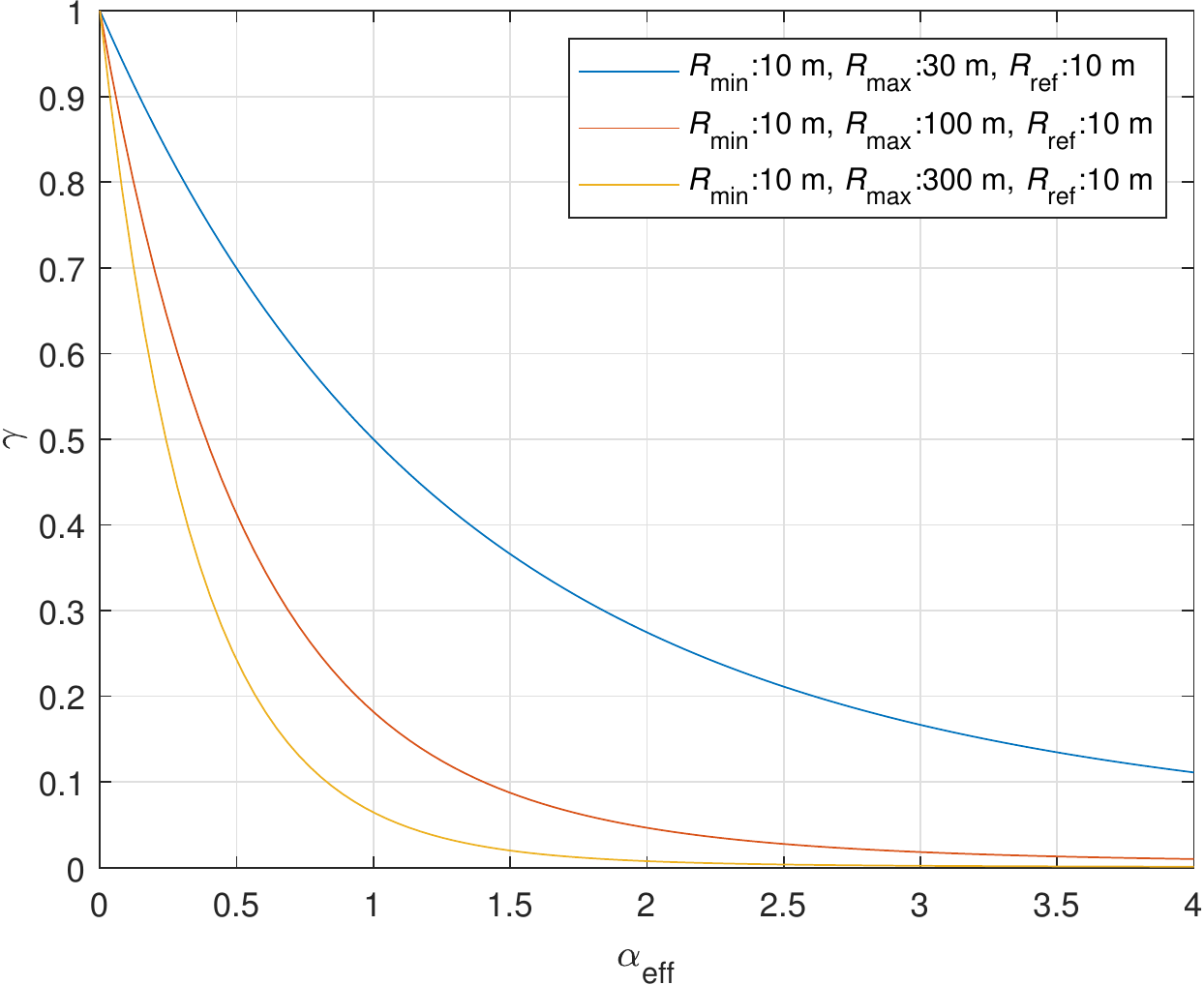}
	} 
	\caption{ The impact of cell size and effective path loss exponent on the factor $\largeScaleImpactOnLearning$.}
	\label{fig:lambda}
\end{figure}

In \figurename~\ref{fig:lambda}, we plot $\largeScaleImpactOnLearning$ for different cell sizes for a given $\effectivePathLossExponent$. As can be seen from \figurename~\ref{fig:lambda}, for a better power control or a smaller cell size, $\largeScaleImpactOnLearning$ increases to $1$. Heuristically, both of these cases imply a better detection performance under noise, which leads to a lower error probability and a better convergence rate as discussed in Section~\ref{subsec:ErrProT} and Section~\ref{subsec:Convergence}, respectively.

\subsection{Error Probability}
\label{subsec:ErrProT}
We define the error probability $\probabilityIncorrect[\indexGradient]$ as the probability of misidentifying the correct gradient sign for the $\indexGradient$th gradient, i.e.,
\begin{align}
	\probabilityIncorrect[\indexGradient]\triangleq\probability[{\signNormal[{\deltaVectorAtIteration[\indexCommunicationRound][\indexGradient]}]\neq \signNormal[{\globalGradientElement[\indexCommunicationRound][\indexGradient]}]}]~.
\end{align}
To obtain an analytical expression for the error probability, we make the following assumptions:
\begin{assumption}[Exponentially-distributed received signal power]
	\rm 
	For given $\numberOFEDsForOptionOne$ and $\numberOFEDsForOptionSecond$,		${\metricForFirst[\indexGradient]}$ and ${\metricForSecond[\indexGradient]}$ are exponential random variables, where their means are ${\meanOptionOne}$ and ${\meanOptionTwo}$, respectively.
	\label{assump:exp}
\end{assumption}
\begin{assumption}[Independent, identical, and unbiased gradients \cite{Bernstein_2018}]
	\rm The local stochastic  gradient estimates  are independent and  unbiased, i.e.,
		$
		\expectationOperator[{\localGradientElement[\indexED,\indexGradient][\indexCommunicationRound]}][{\datasetBatch[\indexED]}]=\globalGradientElement[\indexCommunicationRound][\indexGradient],~\forall\indexED,\indexGradient.
		$	
	\label{assump:iid}
\end{assumption}

Assumption~\ref{assump:exp} holds true when the power control is ideal under Rayleigh fading. It is a weak assumption under imperfect power control due to the central limit theorem. 

\def\numberOfCorrectEDs{k_{\rm c}}
\def\numberOfIncorrectEDs{k_{\rm i}}
Let $\correctDecision[\indexGradient]$, $\zeroDecision[\indexGradient]$, and $\incorrectDecision[\indexGradient]$ be the probabilities defined by
\begin{align}
	\correctDecision[\indexGradient]&\triangleq\probability[{\signNormal[{\localGradientElement[\indexED,\indexGradient][\indexCommunicationRound]}]=\signNormal[{\globalGradientElement[\indexCommunicationRound][\indexGradient]}]|\gradientWeight[{\localGradientElement[\indexED,\indexGradient][\indexCommunicationRound]}]\neq0}],~\forall\indexED,\label{eq:prP}\\
	\zeroDecision[\indexGradient]&\triangleq\probability[{\gradientWeight[{\localGradientElement[\indexED,\indexGradient][\indexCommunicationRound]}]=0}],~\forall\indexED,\label{eq:prZ}\\
	\incorrectDecision[\indexGradient]&\triangleq\probability[{\signNormal[{\localGradientElement[\indexED,\indexGradient][\indexCommunicationRound]}]\neq\signNormal[{\globalGradientElement[\indexCommunicationRound][\indexGradient]}]|\gradientWeight[{\localGradientElement[\indexED,\indexGradient][\indexCommunicationRound]}]\neq0}],~\forall\indexED,\label{eq:prQ}
\end{align}
where $\correctDecision[\indexGradient]+\zeroDecision[\indexGradient]+\incorrectDecision[\indexGradient]=1$. 
$\probabilityIncorrect[\indexGradient]$ for \ac{HPA} can be obtained as follows:
\begin{lemma}[Error probability for HPA]
\rm  Suppose that the gradient encoder is \ac{HPA} and Assumption~\ref{assump:exp} and Assumption~\ref{assump:iid} hold. The error probability in Rayleigh fading channel  is
\if\IEEEsubmission 0
\begin{align}
\probabilityIncorrect[\indexGradient]&=
{\reviewColor\sum_{\indexSumKpKm=0}^{\numberOfEdgeDevices}\frac{\indexSumKpKm\frac{\incorrectDecision[\indexGradient]}{1-\zeroDecision[\indexGradient]}+\frac{\noiseVariance}{2\largeScaleImpactOnLearning}}{\indexSumKpKm+\frac{\noiseVariance}{\largeScaleImpactOnLearning}}\binom{\numberOfEdgeDevices}{\indexSumKpKm}{\zeroDecision[\indexGradient]}^{\numberOfEdgeDevices-\indexSumKpKm}	(1-\zeroDecision[\indexGradient])^{\indexSumKpKm}}\nonumber\\
&=
\frac{\incorrectDecision[\indexGradient]}{1-\zeroDecision[\indexGradient]}(1-\coeffErr[{\zeroDecision[\indexGradient]}])+\frac{1}{2}\coeffErr[{\zeroDecision[\indexGradient]}]~,
	\label{eq:errorPrHPA}
\end{align}
\else
\begin{align}
	\probabilityIncorrect[\indexGradient]&=
	{\reviewColor\sum_{\indexSumKpKm=0}^{\numberOfEdgeDevices}\frac{\indexSumKpKm\frac{\incorrectDecision[\indexGradient]}{1-\zeroDecision[\indexGradient]}+\frac{\noiseVariance}{2\largeScaleImpactOnLearning}}{\indexSumKpKm+\frac{\noiseVariance}{\largeScaleImpactOnLearning}}\binom{\numberOfEdgeDevices}{\indexSumKpKm}{\zeroDecision[\indexGradient]}^{\numberOfEdgeDevices-\indexSumKpKm}	(1-\zeroDecision[\indexGradient])^{\indexSumKpKm}}=
	\frac{\incorrectDecision[\indexGradient]}{1-\zeroDecision[\indexGradient]}(1-\coeffErr[{\zeroDecision[\indexGradient]}])+\frac{1}{2}\coeffErr[{\zeroDecision[\indexGradient]}]~,
	\label{eq:errorPrHPA}
\end{align}
\fi
where  $\coeffErr[{\zeroDecision[]}]\in[0,1]$ is defined by
\if\IEEEsubmission 0
\begin{align}
	\coeffErr[{\zeroDecision[]}]
&\triangleq
{\reviewColor\sum_{\indexSumKpKm=0}^{\numberOfEdgeDevices}\frac{\frac{\noiseVariance}{\largeScaleImpactOnLearning}}{\indexSumKpKm+\frac{\noiseVariance}{\largeScaleImpactOnLearning}}\binom{\numberOfEdgeDevices}{\indexSumKpKm}{\zeroDecision[]}^{\numberOfEdgeDevices-\indexSumKpKm}	(1-\zeroDecision[])^{\indexSumKpKm}}\nonumber\\
&	
	=
\begin{cases}
	{\zeroDecision[]}^\numberOfEdgeDevices \hyperGeometricFcn[\frac{\noiseVariance}{\largeScaleImpactOnLearning}][{-\numberOfEdgeDevices}][{\frac{\noiseVariance}{\largeScaleImpactOnLearning}+1}][{1-\frac{1}{\zeroDecision[]}}]~, & \zeroDecision[]\neq 0 \\
	\frac{\noiseVariance}{\noiseVariance+\largeScaleImpactOnLearning\numberOfEdgeDevices}~, & \zeroDecision[]= 0 
\end{cases}~.
\label{eq:coeffErr}
\end{align}
\else
\begin{align}
	\coeffErr[{\zeroDecision[]}]
	&\triangleq
	{\reviewColor\sum_{\indexSumKpKm=0}^{\numberOfEdgeDevices}\frac{\frac{\noiseVariance}{\largeScaleImpactOnLearning}}{\indexSumKpKm+\frac{\noiseVariance}{\largeScaleImpactOnLearning}}\binom{\numberOfEdgeDevices}{\indexSumKpKm}{\zeroDecision[]}^{\numberOfEdgeDevices-\indexSumKpKm}	(1-\zeroDecision[])^{\indexSumKpKm}}=
	\begin{cases}
		{\zeroDecision[]}^\numberOfEdgeDevices \hyperGeometricFcn[\frac{\noiseVariance}{\largeScaleImpactOnLearning}][{-\numberOfEdgeDevices}][{\frac{\noiseVariance}{\largeScaleImpactOnLearning}+1}][{1-\frac{1}{\zeroDecision[]}}]~, & \zeroDecision[]\neq 0 \\
		\frac{\noiseVariance}{\noiseVariance+\largeScaleImpactOnLearning\numberOfEdgeDevices}~, & \zeroDecision[]= 0 
	\end{cases}~.
	\label{eq:coeffErr}
\end{align}
\fi
\label{lemma:errorPrHPA}
\end{lemma}
The proof is given in Appendix~\ref{app:lemma:errorPrHPA}. 

{\reviewColor
Lemma~\ref{lemma:errorPrHPA} implies the following results:
\begin{corollary}[Legitimate EDs]\rm For  $\incorrectDecision[\indexGradient]<\correctDecision[\indexGradient]$, ${\incorrectDecision[\indexGradient]}/{(1-\zeroDecision[\indexGradient])}={\incorrectDecision[\indexGradient]}/(\incorrectDecision[\indexGradient]+\correctDecision[\indexGradient])\le\probabilityIncorrect[\indexGradient]<1/2$ holds.
	\label{corr:fairEDs}
\end{corollary}
}

\ac{HPA} with $\thresholdForZero=0$ and $\zeroDecision[\indexGradient]=0$ corresponds to \ac{HP}. Hence, Lemma~\ref{lemma:errorPrHPA} can also be used for obtaining the error probability for \ac{HP} as follows:
\begin{corollary}[Error probability for HP]
	\rm
Suppose that the gradient encoder is \ac{HP}. Under Assumption~\ref{assump:exp} and Assumption~\ref{assump:iid}, the error probability is given by
\begin{align}
	\probabilityIncorrect[\indexGradient]={\incorrectDecision[\indexGradient]}\frac{\largeScaleImpactOnLearning\numberOfEdgeDevices}{\noiseVariance+\largeScaleImpactOnLearning\numberOfEdgeDevices}+\frac{1}{2}\frac{\noiseVariance}{\noiseVariance+\largeScaleImpactOnLearning\numberOfEdgeDevices}~,
	\label{eq:errorPrHP}
\end{align}
where $\largeScaleImpactOnLearning\in[0,1]$ is given in \eqref{eq:pathlossimpact}.
\end{corollary}
{\reviewColor
	
\begin{corollary}[Large zero-weight probability]
	\rm
	For $\zeroDecision[]=1-\epsilon$ for $\epsilon\ll1$, $\coeffErr[{\zeroDecision[]}]\approx\zeroDecision[]^{\numberOfEdgeDevices}$, implying
	$
\probabilityIncorrect[\indexGradient]\approx\incorrectDecision[\indexGradient]({1-\zeroDecision[\indexGradient]^{\numberOfEdgeDevices}})/({1-\zeroDecision[\indexGradient]})+\zeroDecision[\indexGradient]^{\numberOfEdgeDevices}/2	$.
\end{corollary}	
	
\subsubsection{The impacts of the number of EDs and power control on the error probability}
\def\ruleK{R_1}
\def\ruleGamma{R_2}
\def\ratioK{\rho_1}
\def\scaleGamma{a}
\def\ratioGamma{\rho_2}
To understand how the error probability $\probabilityIncorrect[\indexGradient]$ changes with $\numberOfEdgeDevices$ and $\largeScaleImpactOnLearning$, let $\ruleK$ and $\ruleGamma$ be the scaling rules defined by $\ruleK\triangleq\probabilityIncorrect[\indexGradient](\numberOfEdgeDevices+1,\largeScaleImpactOnLearning)/\probabilityIncorrect[\indexGradient](\numberOfEdgeDevices,\largeScaleImpactOnLearning)$ and $\ruleGamma\triangleq\probabilityIncorrect[\indexGradient](\numberOfEdgeDevices,\largeScaleImpactOnLearning\scaleGamma)/\probabilityIncorrect[\indexGradient](\numberOfEdgeDevices,\largeScaleImpactOnLearning)$, respectively. By using Lemma~\ref{lemma:errorPrHPA} and the recurrence relation of binomial coefficients, we can then express $\ruleK$ and $\ruleGamma$ as
\begin{align}
\ruleK =& \zeroDecision[\indexGradient] + (1-\zeroDecision[\indexGradient])\frac{\sum_{\indexSumKpKm=0}^{\numberOfEdgeDevices}\frac{(\indexSumKpKm+1)\frac{\incorrectDecision[\indexGradient]}{1-\zeroDecision[\indexGradient]}+\frac{\noiseVariance}{2\largeScaleImpactOnLearning}}{\indexSumKpKm+1+\frac{\noiseVariance}{\largeScaleImpactOnLearning}}\binom{\numberOfEdgeDevices}{\indexSumKpKm}{\zeroDecision[\indexGradient]}^{\numberOfEdgeDevices-\indexSumKpKm}	(1-\zeroDecision[\indexGradient])^{\indexSumKpKm}}{\sum_{\indexSumKpKm=0}^{\numberOfEdgeDevices}\frac{\indexSumKpKm\frac{\incorrectDecision[\indexGradient]}{1-\zeroDecision[\indexGradient]}+\frac{\noiseVariance}{2\largeScaleImpactOnLearning}}{\indexSumKpKm+\frac{\noiseVariance}{\largeScaleImpactOnLearning}}\binom{\numberOfEdgeDevices}{\indexSumKpKm}{\zeroDecision[\indexGradient]}^{\numberOfEdgeDevices-\indexSumKpKm}	(1-\zeroDecision[\indexGradient])^{\indexSumKpKm}}~, \label{eq:ratioK}\\
\ruleGamma = & \frac{\sum_{\indexSumKpKm=0}^{\numberOfEdgeDevices}\frac{\indexSumKpKm\frac{\incorrectDecision[\indexGradient]}{1-\zeroDecision[\indexGradient]}+\frac{\noiseVariance}{2\largeScaleImpactOnLearning\scaleGamma}}{\indexSumKpKm+\frac{\noiseVariance}{\largeScaleImpactOnLearning\scaleGamma}}\binom{\numberOfEdgeDevices}{\indexSumKpKm}{\zeroDecision[\indexGradient]}^{\numberOfEdgeDevices-\indexSumKpKm}	(1-\zeroDecision[\indexGradient])^{\indexSumKpKm}}{\sum_{\indexSumKpKm=0}^{\numberOfEdgeDevices}\frac{\indexSumKpKm\frac{\incorrectDecision[\indexGradient]}{1-\zeroDecision[\indexGradient]}+\frac{\noiseVariance}{2\largeScaleImpactOnLearning}}{\indexSumKpKm+\frac{\noiseVariance}{\largeScaleImpactOnLearning}}\binom{\numberOfEdgeDevices}{\indexSumKpKm}{\zeroDecision[\indexGradient]}^{\numberOfEdgeDevices-\indexSumKpKm}	(1-\zeroDecision[\indexGradient])^{\indexSumKpKm}}~. \label{eq:ratioGamma}
\end{align}
If a larger $\numberOfEdgeDevices$ (i.e., more participants) or a larger  $\largeScaleImpactOnLearning$ (i.e., better power control) decrease the error probability, 
we need to show $\ruleK<1$ and $\ruleGamma<1$, respectively.
By comparing the weights of the summands on the numerators and denominators in \eqref{eq:ratioK} and \eqref{eq:ratioGamma}, the required conditions are obtained as $\incorrectDecision[\indexGradient]<\correctDecision[\indexGradient]$ (i.e., Corollary~\ref{corr:fairEDs}) and $\scaleGamma>1$ (i.e., better power control), respectively

\subsubsection{Threshold optimization for HPA}
\def\arbitraryPDF[#1][#2]{f_{#1}(#2)}
\def\arbitraryCDF[#1][#2]{F(#2)}
\def\thresholdForZeroOpt{\hat{\thresholdForZero}}

Suppose that the global gradient $\globalGradientElement[\indexCommunicationRound][\indexGradient]$ is a positive value. Let $\arbitraryCDF[x][x]$ denote the \ac{CDF} of $\localGradientElement[\indexED,\indexGradient][\indexCommunicationRound]$. For HPA, we can then express $\correctDecision[\indexGradient]$, $\zeroDecision[\indexGradient]$, and $\incorrectDecision[\indexGradient]$ as functions of $\thresholdForZero$ as
$
	\correctDecision[\indexGradient](\thresholdForZero)=\probability[{
		\localGradientElement[\indexED,\indexGradient][\indexCommunicationRound]>\thresholdForZero
	}]=1-\arbitraryCDF[x][\thresholdForZero]$, $
	\zeroDecision[\indexGradient](\thresholdForZero)=\probability[{
		|\localGradientElement[\indexED,\indexGradient][\indexCommunicationRound]|\le\thresholdForZero
	}]=\arbitraryCDF[x][\thresholdForZero]-\arbitraryCDF[x][-\thresholdForZero]$,
	$\incorrectDecision[\indexGradient](\thresholdForZero)=\probability[{
		\localGradientElement[\indexED,\indexGradient][\indexCommunicationRound]<-\thresholdForZero
	}]=\arbitraryCDF[x][-\thresholdForZero]$, $\forall\indexED$,
respectively (see \figurename~\ref{fig:errorPrHPA}). Hence, the optimal threshold can be obtained as 
\begin{align}
	\thresholdForZeroOpt = \arg\min_{\thresholdForZero} \frac{\incorrectDecision[\indexGradient](\thresholdForZero)}{1-\zeroDecision[\indexGradient](\thresholdForZero)}(1-\coeffErr[{\zeroDecision[\indexGradient](\thresholdForZero)}])+\frac{1}{2}\coeffErr[{\zeroDecision[\indexGradient](\thresholdForZero)}].
	\label{eq:optTheProblem}
\end{align} 
Due to the hypergeometric function in $\coeffErr[{\zeroDecision[\indexGradient](\thresholdForZero)}]$, in general, it is not tractable to obtain $\thresholdForZeroOpt$ in closed form. Nevertheless, $\thresholdForZeroOpt$ can be obtained numerically by simply sweeping $\thresholdForZero$. We discuss a numerical example in Section~\ref{subsec:ErrPro}, which demonstrates that  $\probabilityIncorrect[\indexGradient]$ is not a monotonically increasing function over the range of threshold $\thresholdForZero$ (i.e., a non-zero $\thresholdForZeroOpt$ exists) and  the error probability for \ac{HPA} can be much lower than that of \ac{HP}.

\begin{figure}[t]
	\centering
	{\includegraphics[width =1.7in]{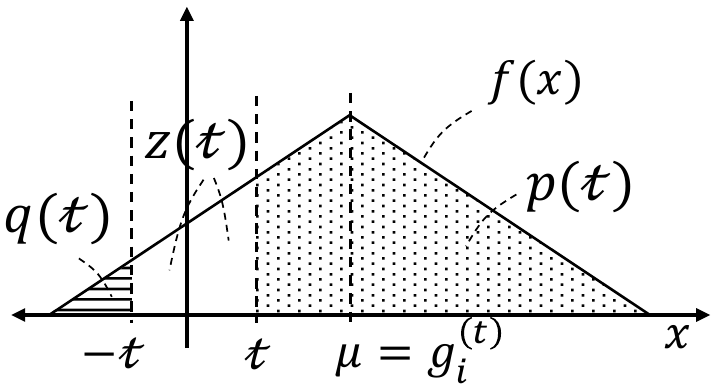}
	} 
\reviewColor
	\caption{The definitions of $\correctDecision[\indexGradient](\thresholdForZero)$, $\zeroDecision[\indexGradient](\thresholdForZero)$, and $\incorrectDecision[\indexGradient](\thresholdForZero)$ for HPA when $\globalGradientElement[\indexCommunicationRound][\indexGradient]>0$. $f(x)$ denotes the \ac{PDF} of $\localGradientElement[\indexED,\indexGradient][\indexCommunicationRound]$}
	\label{fig:errorPrHPA}
\end{figure}

It is worth noting that distribution of $\localGradientElement[\indexED,\indexGradient][\indexCommunicationRound]$, i.e., $\arbitraryCDF[x][x]$, is unknown and changes over the communication rounds in practice \cite{Zhang_2021}. Hence, solving \eqref{eq:optTheProblem} for HPA or obtaining the optimum weight function for the general case are currently difficult problems without making strong assumptions. 
}





\subsection{Convergence Rate}
\label{subsec:Convergence}
 For the convergence analysis, we make several assumptions given as follows:
\begin{assumption}[Bounded loss function \cite{Bernstein_2018}]
	\rm There exists a constant $\lossFunctionGlobalMinimum$ such that
	$\lossFunctionGlobal[\modelParameters]\ge \lossFunctionGlobalMinimum,\forall\modelParameters$.
	\label{assump:boundedLoss}
\end{assumption}
\begin{assumption}[Smoothness {\cite{Bernstein_2018},\cite[Lemma 1.2.3]{nesterov2004introductory}}] 
	\rm 
	Let $\globalGradientNoIndex$ be the gradient of $\lossFunctionGlobal[\modelParameters]$ evaluated at $\modelParameters$. For all $\modelParameters$ and $\modelParameters'$, the expression given by
	\begin{align}
		\left| \lossFunctionGlobal[\modelParameters'] - (\lossFunctionGlobal[\modelParameters]+\globalGradientNoIndex^{\rm T}(\modelParameters'-\modelParameters)) \right| \le \frac{1}{2}\sum_{\indexGradient=1}^{\numberOfModelParameters} \nonnegativeConstantsEle[\indexGradient](\modelParametersEle[\indexGradient]'-\modelParametersEle[\indexGradient])^2~,
		\nonumber
	\end{align}	
	holds for a non-negative constant vector
	$\nonnegativeConstants=[\nonnegativeConstantsEle[1],\mydots,\nonnegativeConstantsEle[\numberOfModelParameters]]^{\rm T}$.
	\label{assump:smoothness}
\end{assumption}
\begin{assumption}[Variance bound \cite{Bernstein_2018}]
	\rm A local stochastic  gradient estimate has a coordinate bounded variance, i.e.,
	$
	\expectationOperator[{(\localGradientElement[\indexED,\indexGradient][\indexCommunicationRound]-\globalGradientElement[\indexCommunicationRound][\indexGradient])^2}][{\datasetBatch[\indexED]}]\le\varianceBoundEle[\indexGradient]^2/\batchSize,~\forall\indexED,\indexGradient,		
	$
	where  $\varianceBound = [\varianceBoundEle[1],\mydots,\varianceBoundEle[\numberOfModelParameters]]^{\rm T}$ is a non-negative constant  vector.
	\label{assump:varBound}
\end{assumption}
\begin{assumption}[Unimodal, symmetric gradient noise \cite{Bernstein_2018}]
	\rm
	For any given $\modelParametersAtIteration[\indexCommunicationRound]$,  ${\localGradientElement[\indexED,\indexGradient][\indexCommunicationRound]}$, $\forall\indexED,\indexGradient$, has a unimodal distribution that is also symmetric around its mean.
	\label{assump:mmodal}
\end{assumption}
\begin{assumption}[Bounded zero-weight probability]
	\rm There exists a constant $\zeroDecisionMax\in[0,1]$ such that
	$		\probability[{\gradientWeight[{\localGradientElement[\indexED,\indexGradient][\indexCommunicationRound]}]=0}]\le\zeroDecisionMax,~\forall\indexED,\indexGradient$.
	\label{assump:zeroBound}
\end{assumption}
\begin{assumption}[Maximum absolute]
	\rm The magnitude of the local stochastic  gradient estimates  are less than or equal to the non-negative constant $\maxGradient$, i.e.,
$
		|{\localGradientElement[\indexED,\indexGradient][\indexCommunicationRound]}|\le\maxGradient,~\forall\indexED,\indexGradient
$~.
	\label{assump:maxBound}
\end{assumption}

With the considerations of path loss, power control, and cell size, the convergence rate in the presence of  the proposed scheme with \ac{HPA} in Rayleigh fading channel can obtained as follows:
\begin{theorem}[Convergence rate for  HPA]
	\rm Suppose Assumptions~\ref{assump:exp}-\ref{assump:maxBound} hold true and the gradient encoder is based on \ac{HPA}. The convergence rate of the distributed training by the \ac{MV} based on the proposed scheme in Rayleigh fading channel is
	\if\IEEEsubmission0
	\begin{align}
	\expectationOperator[\frac{1}{\communicationRounds}\sum_{\indexCommunicationRound=0}^{\communicationRounds-1} \norm{\globalGradient[\indexCommunicationRound]}_1][]\nonumber\le&\frac{	\lossFunctionGlobal[{\modelParametersAtIteration[0]}]- \lossFunctionGlobalMinimum}{{\learningRate\communicationRounds(1-\coeffErr[{\zeroDecisionMax}])}}+\frac{\learningRate\norm{\nonnegativeConstants}_1}{2(1-\coeffErr[{\zeroDecisionMax}])} \nonumber\\& +\frac{\sqrt{3}}{\sqrt{\batchSize}}\frac{1}{1-\zeroDecisionMax}\frac{\maxGradient}{ \thresholdForZero+\maxGradient }\norm{\varianceBound}_1~.
		\label{eq:convergence}
	\end{align}
	\else
	\begin{align}
	\expectationOperator[\frac{1}{\communicationRounds}\sum_{\indexCommunicationRound=0}^{\communicationRounds-1} \norm{\globalGradient[\indexCommunicationRound]}_1][]\le\frac{	\lossFunctionGlobal[{\modelParametersAtIteration[0]}]- \lossFunctionGlobalMinimum}{{\learningRate\communicationRounds(1-\coeffErr[{\zeroDecisionMax}])}}+\frac{\learningRate\norm{\nonnegativeConstants}_1}{2(1-\coeffErr[{\zeroDecisionMax}])}  +\frac{\sqrt{3}}{\sqrt{\batchSize}}\frac{1}{1-\zeroDecisionMax}\frac{\maxGradient}{ \thresholdForZero+\maxGradient }\norm{\varianceBound}_1~.
	\label{eq:convergence}
	\end{align}
	\fi
where  $\coeffErr[{\zeroDecision[]}]\in[0,1]$ is given in \eqref{eq:coeffErr}.
	\label{th:convergence}
\end{theorem}
The proof is given in Appendix~\ref{app:th:conv}.

Since \ac{HPA} with $\thresholdForZero=0$ under the condition of $\zeroDecisionMax=0$ corresponds to \ac{HP}, Theorem~\ref{th:convergence} also be used for obtaining the convergence rate for \ac{HP}:
\begin{corollary}[Convergence rate for HP]
	\rm
	Under Assumptions~Assumptions~\ref{assump:exp}-\ref{assump:maxBound}, the convergence rate of the distributed training by the \ac{MV} based on the proposed scheme with \ac{HP} in Rayleigh fading channel is equal to \eqref{eq:convergence} for $\thresholdForZero=0$ and $\zeroDecisionMax=0$, i.e.,
	\begin{align}
		\expectationOperator[\frac{1}{\communicationRounds}\sum_{\indexCommunicationRound=0}^{\communicationRounds-1} \norm{\globalGradient[\indexCommunicationRound]}_1][]\le\frac{	\lossFunctionGlobal[{\modelParametersAtIteration[0]}]- \lossFunctionGlobalMinimum}{{\learningRate\communicationRounds\frac{\numberOfEdgeDevices\largeScaleImpactOnLearning}{\numberOfEdgeDevices\largeScaleImpactOnLearning+\noiseVariance}}}+\frac{\learningRate\norm{\nonnegativeConstants}_1}{2
			\frac{\numberOfEdgeDevices\largeScaleImpactOnLearning}{\numberOfEdgeDevices\largeScaleImpactOnLearning+\noiseVariance}}  +\frac{\sqrt{3}\norm{\varianceBound}_1}{\sqrt{\batchSize}}~.
		\label{eq:convergenceHP}
	\end{align}	
\end{corollary}

Based on Theorem~\ref{th:convergence}, we can infer the followings: 
\begin{itemize}
\item For a larger \ac{SNR} (i.e., a larger $\powerRef/\noiseVariance$)  and a large number of \acp{ED} (i.e., a larger $\numberOfEdgeDevices$), the convergence rate improves. 

\item The power control improves the convergence rate since $\largeScaleImpactOnLearning$ increases with a lower $\effectivePathLossExponent$. 
 Another way of improving the convergence rate is to reduce the cell size, yielding a larger $\largeScaleImpactOnLearning$ as illustrated in \figurename~\ref{fig:lambda}. This leads a practical trade-off:  While the number of \acp{ED} may be larger for a larger cell, the power control becomes a harder task. 

\item 
Both $\coeffErr[{\zeroDecisionMax}]$ and $\zeroDecisionMax$ tend to increase with  $\thresholdForZero$, as exemplified in Section~\ref{subsec:ErrPro} for \ac{HPA}. Therefore, for a larger  $\thresholdForZero$, the first two terms of the right-hand side of \eqref{eq:convergence} become larger. On the other hand, the last term of  the right-hand side of \eqref{eq:convergence} is not a monotonic function over the range of threshold $\thresholdForZero$, which is similar to the error probability in \eqref{eq:errorPrHPA}. This implies that the threshold  $\thresholdForZero$ can be non-zero to minimize the right-hand side of \eqref{eq:convergence}. Hence,  \ac{HPA} with the optimum threshold can improve the convergence rate. 

\item If $\learningRate=1/\sqrt{\communicationRounds}$ and $\batchSize=\communicationRounds$, the convergence rate is similar to the one with \ac{signSGD} in an ideal channel \cite[Theorem~1]{Bernstein_2018} where the training requires $\mathcal{O}(\sqrt{\communicationRounds})$ iterations.

\end{itemize}

\subsection{Extensions}
\label{subsec:extensions}
As a generalization, the weight function $\gradientWeight[\aVariable]$ can be chosen as a continuous function to achieve a soft participation. For example, consider the weight function given by
\begin{align}
	\gradientWeight[{\localGradientElement[\indexED,\indexGradient][\indexCommunicationRound]}] = \begin{cases}
		1,&|\localGradientElement[\indexED,\indexGradient][\indexCommunicationRound]|>\thresholdForZero(1+\steepnessFactor)\\		
		0,&|\localGradientElement[\indexED,\indexGradient][\indexCommunicationRound]|\le\thresholdForZero(1-\steepnessFactor)\\
		\frac{1}{2}+\frac{1}{2}\cos\left(\frac{\pi(|\localGradientElement[\indexED,\indexGradient][\indexCommunicationRound]|-\thresholdForZero(1+\steepnessFactor))}{2\steepnessFactor\thresholdForZero}\right),&\text{otherwise}
	\end{cases}~,
	\label{eq:SPweight}
\end{align}
where $\steepnessFactor\in[0,1]$ is a factor that determines the steepness of weight function. With this weight function, all \acp{ED} can participate in the \ac{MV} calculation (e.g., for $\steepnessFactor=1$), but their impacts on the \ac{MV} is proportional to the magnitude of the local gradients, i.e., weighted votes. 
Note that the \ac{SP} with \eqref{eq:SPweight} for $\steepnessFactor=0$ is identical to \ac{HPA}.  As demonstrated in Section~\ref{subsec:ErrPro}, a smoother weight function can lower the error probability further as compared to the weight function for \ac{HPA}.

Another way extending the proposed scheme is to change the weight function over the communication rounds. For example, for \ac{HPA}, the threshold $\thresholdForZero$ can be reduced as function of $\indexCommunicationRound$ or chosen as a function of the local stochastic gradients by exploiting fact that the gradients tend to decrease over the communication rounds. {\reviewColor In certain cases, it may be desirable to consider the cardinality of the datasets in the \ac{MV} calculation in \eqref{eq:majorityVote} (e.g., scaling with the sign information with $|\dataset[\indexED]|$). To address this case,  designing the weight function  for each ED as a function of the cardinality of the local dataset is a potential extension.}  While these adaptations potentially improve the performance, we leave the analyses of such cases as future study.

\subsection{Comparisons}
\label{subsec:comp}
\subsubsection{Robustness against time-varying fading channel} As opposed to the approaches in \cite{Guangxu_2020} and \cite{Guangxu_2021}, the proposed scheme does not utilize the \ac{CSI} for \ac{TCI} at the \acp{ED}. Hence, it is compatible with time-varying channels (e.g., as in mobile networks \cite{Zeng_2020}) and does not lose the gradient information due to the \ac{TCI}. As a trade-off, it increases the number of time-frequency resources for \ac{OAC} as compared to  \ac{OBDA}. As compared to the approaches in \cite{Yang_2020} and \cite{Amiria_2021}, the proposed scheme also does not require \ac{CSI} at the \ac{ES} or multiple antennas.

\subsubsection{Robustness against synchronization errors} 
\ac{FSK-MV} and \ac{PPM-MV} are more robust against time-synchronization errors as compared to \ac{OBDA}
because the time misalignment among the \acp{ED} or the uncertainty on the receiver synchronization within the \ac{CP} window cause phase rotations in the frequency domain and \ac{FSK-MV} or \ac{PPM-MV} do not encode gradient information on the amplitude or phase. Also, the proposed scheme does not use any channel-related information at the \acp{ED} and the \ac{ES}. 


\subsubsection{Robustness against imperfect power control and data heterogeneity}
\label{subsec:heterobust}
Our results in Section~\ref{sec:numerical} shows that the proposed scheme with \ac{HPA} and \ac{SP} can improve the convergence rate in the case of imperfect power control and/or heterogeneous data distribution. This is because the  \ac{HPA} and \ac{SP} gradually reduce the impact of converging \acp{ED} on the \ac{MV} calculation. For example, consider a scenario where there is no power control. Without any weight function (i.e., \ac{HP}), the \ac{MV} is highly biased towards the decisions of the nearby \acp{ED} as their received signal powers are much larger than the ones far from the \ac{ES}. Hence, under a heterogeneous data distribution scenario, the model is more likely to learn to classify the labels of the nearby \acp{ED}. On the other hand, with \ac{HPA} and \ac{SP}, the model initially learns the nearby \acp{ED}' labels. However, since the absolute value of the gradients  of the converging nearby \acp{ED}  tend to be smaller in the later stages of the communication rounds, the impacts of the nearby \acp{ED} on the \ac{MV} are reduced with \ac{HPA} and \ac{SP}, which allows the model to learn the labels at the far \acp{ED}. {\reviewColor In another scenario, some of the labels may be available in a small number of EDs. In this case, these labels may not be learned well as the other EDs aggressively vote even if the magnitudes of their gradients are so small. With HPA, the EDs that have the common labels do not cast votes in the later stages of the learning process, which allows the neural network to learn the labels that are available at few EDs.}

\subsubsection{Robustness against power-amplifier non-linearity} 
{\reviewColor As shown in Section~\ref{sec:numerical}, \ac{OBDA} can suffer from high \ac{PMEPR} due to the correlated gradients. The proposed scheme addresses this issue with simple techniques regardless of the correlation of gradients.} For \ac{FSK-MV}, we use random \ac{QPSK} symbols for $\randomSymbolAtSubcarrier[\indexED,\indexGradient]$ to improve the waveform characteristics. The \ac{PMEPR} for \ac{PPM-MV} is lowered by increasing the pulse duration.


\section{Numerical Results}
\label{sec:numerical}
In this section, we first analyze  error probability with different weight functions numerically. Afterwards, we  evaluate the performance of \ac{FEEL} with the proposed scheme with the consideration of path loss, imperfect power control, and time-synchronization errors for both homogeneous and heterogeneous data distribution scenarios.

\subsection{Error Probability for HP, HPA, and SP}
\label{subsec:ErrPro}
\begin{figure}[t]
\centering
{\includegraphics[width =\figuresize]{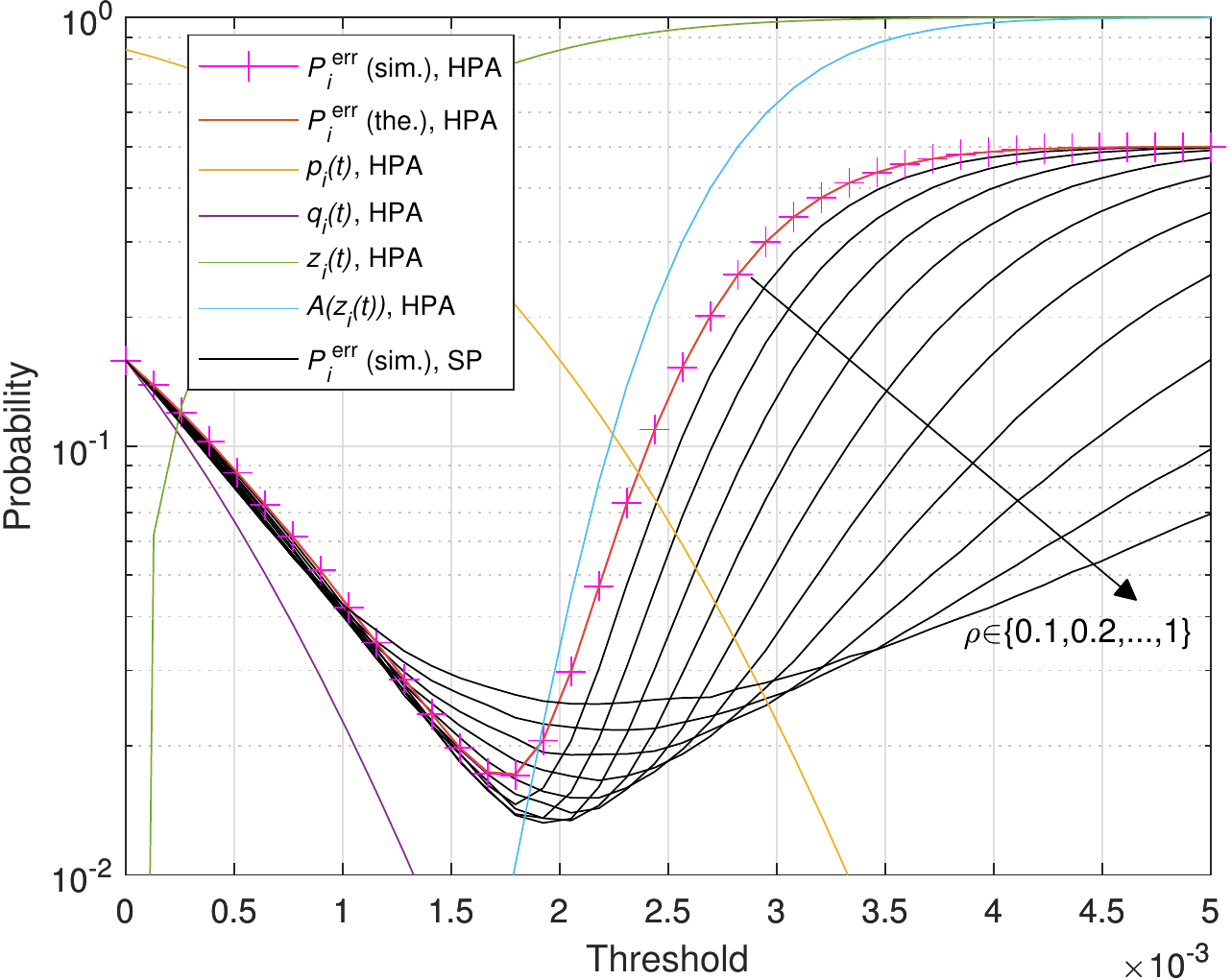}
} 
\caption{
	The error probability can decrease considerably with \ac{HPA} and \ac{SP} as compared to \ac{HP} (\ac{HPA} with $\thresholdForZero=0$ is \ac{HP}). }
\label{fig:errorPr}
\end{figure}

\rm 
To demonstrate the impact of the weight function on the error probability, we assume that ${\localGradientElement[\indexED,\indexGradient][\indexCommunicationRound]}\sim\gaussian[\meanG][\sigmaG^2]$, $\forall\indexED$ (i.e., $\globalGradientElement[\indexCommunicationRound][\indexGradient]=\meanG$), and compare  \ac{HP}, \ac{HPA}, and \ac{SP} with the weight function in \eqref{eq:SPweight} in Rayleigh  channel for $\numberOfEdgeDevices=20$ \acp{ED}, $\noiseVariance=0.01$, $\sigmaG=0.001$, $\meanG=0.001$. First, consider \ac{HPA}.
The  probabilities  $\incorrectDecision[\indexGradient]$ and $\zeroDecision[\indexGradient]$  a function of threshold $\thresholdForZero$ can be expressed as
$\incorrectDecision[\indexGradient](\thresholdForZero)=\normalCDF[({-\thresholdForZero-|\meanG|})/{\sigmaG}]$ and
$\zeroDecision[\indexGradient](\thresholdForZero)=\normalCDF[{(\thresholdForZero-|\meanG|})/{\sigmaG}]-\normalCDF[({-\thresholdForZero-|\meanG|})/{\sigmaG}]$, 
respectively.  By numerically evaluating expressions of $\incorrectDecision[\indexGradient](t)$ and $\zeroDecision[\indexGradient](t)$, we plot the error probability for \ac{HPA} given in \eqref{eq:errorPrHPA} as a function of $\thresholdForZero$  in \figurename~\ref{fig:errorPr}. While \ac{HP} (i.e., \ac{HPA} with $\thresholdForZero=0$) gives $\probabilityIncorrect[\indexGradient]=0.1588$, \ac{HPA} with the threshold $\thresholdForZero=0.0017$ remarkably decreases the error to $\probabilityIncorrect[\indexGradient]=0.017$. The results with Monte Carlo simulation  also align well with the theoretical values. 

For the \ac{SP}, we numerically obtain the error probability by sweeping the factor $\steepnessFactor$ from $0.1$ to $1$ with the step size $0.1$. As can been seen from \figurename~\ref{fig:errorPr}, the error probability decreases further to $0.013$ for $\steepnessFactor=0.3$ and $\thresholdForZero=0.019$. This result indicates there exist weight functions that can achieve a lower error probability than the one with \ac{HPA}.

\subsection{Federated Edge Learning}

We consider the learning task of handwritten-digit recognition with $\numberOfEdgeDevices=50$ \acp{ED} for $\minimumDistance=10$~meters and $\cellRadius=100$~meters. To demonstrate the impact of the imperfect power control on distributed learning, we choose $\pathlossExponent=4$ and $\powerControl\in\{2,4\}$ and set the \ac{SNR}, i.e., $\powerRef/\noiseVariance$, to be $20$~dB at $\referenceDistance=10$~meters. 
The link distance between the $\indexED$th \ac{ED} and the \ac{ES} is determined by $\distanceED[\indexED]=\sqrt{\minimumDistance^2 + (\indexED-1)(\cellRadius^2-\minimumDistance^2)/(\numberOfEdgeDevices-1)}$ to represent a uniform deployment in a circular area.
For the channel, we consider ITU Extended Pedestrian A (EPA) with no mobility. We regenerate the channels between the \ac{ES} and the \acp{ED} independently for each communication round to capture the long-term variations.
The subcarrier spacing,  the sample rate, and the \ac{IDFT} size are  $15$~kHz, $30.72$~Msps, and $\idftSize=2048$, respectively. We use  $\numberOfActiveSubcarriers=1200$ subcarriers (i.e., the signal bandwidth is $18$~MHz). For the synchronization errors, we assume that the maximum time difference between the arriving \ac{ED} signals is  $\syncError=55.6$~ns and the  synchronization uncertainty at the \ac{ES} is $\Nerror=3$ samples, i.e., $\frameSyncError=97.6$~ns. Otherwise, these parameters are set to $0$.

\begin{figure}[t]
	\centering
	\subfloat[\reviewColor Homogeneous data distribution in the cell. All \acp{ED} have data samples for 10 different digits.]{\includegraphics[width =\figuresize]{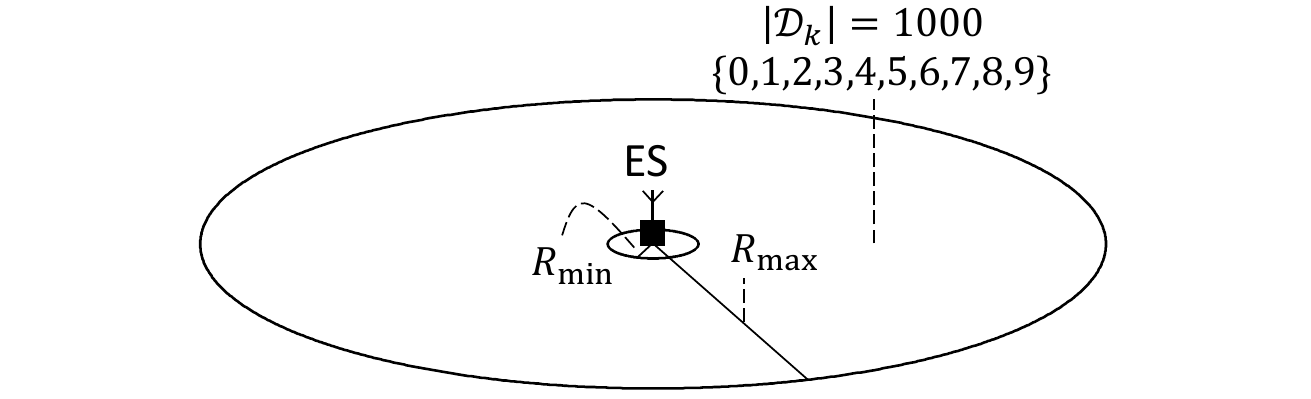}}
	\if\IEEEsubmission0\\
	\else
		~
	\fi
	\subfloat[\reviewColor Heterogeneous data distribution in the cell. The available digits at the \acp{ED} changes based on their locations. ]{\includegraphics[width =\figuresize]{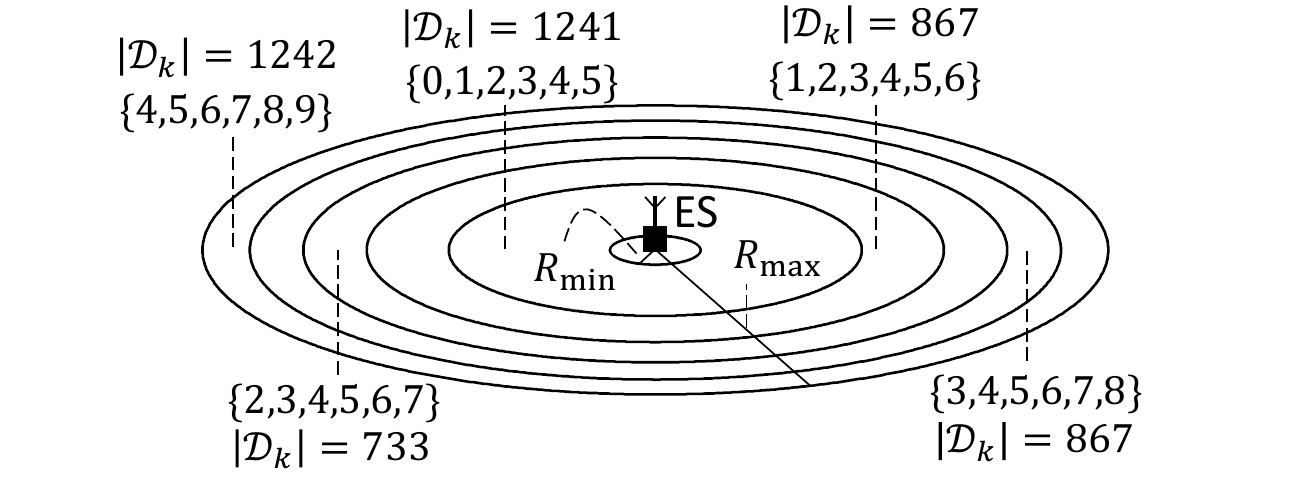}}
	\caption{\reviewColor Homogeneous versus heterogeneous data distributions considered for the numerical analyses. The available digits and dataset size in an area are shown in the figure. The radius of the concentric circles are $\{10,45.6,63.7,77.7,89.6,100\}$ meters.}
	\label{fig:niid}
\end{figure}
For the local data at the \acp{ED}, we use the MNIST database that contains labeled handwritten-digit images size of $28\times28$ from digit 0 to digit 9. We consider both homogeneous and heterogeneous data distributions in the cell. To prepare the data, we first choose $|\completeData|=50000$ training images from the database, where each digit has distinct $5000$ images.  
For homogeneous data distribution, we assume that each \ac{ED} has $100$ distinct images for each digit. For heterogeneous data distribution,   we assume that the distribution of the images depends on the  \ac{ED} locations to test the scheme in a more challenging scenario. To this end, we divide the cell into 5 areas with concentric circles and the \acp{ED} located in $\indexArea$th area have the data samples with the labels $\{\indexArea-1,\indexArea,1+\indexArea,2+\indexArea,3+\indexArea,4+\indexArea\}$ for $\indexArea\in\{1,\mydots,5\}$.  Hence, the availability of the labels gradually changes based on the link distance.  The areas between two adjacent concentric circles are identical and the number of \acp{ED} in each area is $10$.  Note that the labels $0$, $1$, $8$, and $9$ are available at less number of \acp{ED} as compared to other labels. {\reviewColor Also, since we distribute 5000 images/digit equally based on the number of EDs, the local dataset size changes as $1241$, $867$, $783$, $867$, and $1242$, respectively, when $\indexArea$ increases from 1 to 5. For instance, for $\indexArea=1$, the dataset has $500$, $250$, $166$, $125$, $100$ and $100$ images for the digits 0, 1, 2, 3, 4, and 5, respectively.} The homogeneous and heterogeneous data distributions in the area are illustrated in \figurename~\ref{fig:niid}. 

\if\IEEEsubmission0
\renewcommand{\baselinestretch}{1}
\else
\renewcommand{\baselinestretch}{0.8}
\fi
\begin{table}[t]
	\centering
	\caption{Neural network at the EDs.}
	\begin{tabular}{l|l|l}
		Layer               				& Learnables & Activations \\ \hline\hline
		\begin{tabular}[c]{@{}l@{}}Input \\- $28\times28\times1$ images \end{tabular}& N/A & $28\times28\times1$\\\hline
		\begin{tabular}[c]{@{}l@{}}Convolution 2D \\- $5\times5\times1$, $20$ filters\\ - Stride: [1 1]\\ - Padding: [0 0 0 0]    \end{tabular}	& \begin{tabular}[c]{@{}l@{}}Weights: $5\times5\times1\times20$\\ Bias: $1\times1\times20$ \end{tabular} & $24\times24\times20$ \\\hline
		Batchnorm & \begin{tabular}[c]{@{}l@{}}Offset: $1\times1\times20$\\ Scale: $1\times1\times20$ \end{tabular} & $24\times24\times20$ \\ \hline
		ReLU &  N/A& $24\times24\times20$\\\hline
		\begin{tabular}[c]{@{}l@{}}Convolution 2D \\- $3\times3\times1$, $20$ filters\\ - Stride: [1 1]\\ - Padding: [1 1 1 1]    \end{tabular}  	& \begin{tabular}[c]{@{}l@{}}Weights: $3\times3\times20\times20$\\ Bias: $1\times1\times20$ \end{tabular} & $24\times24\times20$ \\\hline
		Batchnorm & \begin{tabular}[c]{@{}l@{}}Offset: $1\times1\times20$\\ Scale: $1\times1\times20$ \end{tabular} & $24\times24\times20$\\ \hline
		ReLU &  N/A& $24\times24\times20$\\		\hline
		\begin{tabular}[c]{@{}l@{}}Convolution 2D \\- $3\times3\times1$, $20$ filters\\ - Stride: [1 1]\\ - Padding: [1 1 1 1]    \end{tabular}  	& \begin{tabular}[c]{@{}l@{}}Weights: $3\times3\times20\times20$\\ Bias: $1\times1\times20$ \end{tabular} & $24\times24\times20$\\\hline
		Batchnorm & \begin{tabular}[c]{@{}l@{}}Offset: $1\times1\times20$\\ Scale: $1\times1\times20$ \end{tabular} & $24\times24\times20$\\ \hline
		ReLU &  N/A& $24\times24\times20$\\		\hline
		\begin{tabular}[c]{@{}l@{}} Fully-connected layer \\ - $10$  outputs     \end{tabular} 					&    \begin{tabular}[c]{@{}l@{}}Weights: $10\times11520$\\ Bias: $10\times1$ \end{tabular}       & $1\times1\times10$ \\\hline
		Softmax &  N/A& $1\times1\times10$ \\\hline 
	\end{tabular}
	\label{table:layout}
\end{table}
\renewcommand{\baselinestretch}{\baselineSize}
For the model, we consider a \ac{CNN}  given in \tablename~\ref{table:layout}. Our model has $\numberOfModelParameters=123090$ learnable parameters that result in $\numberOfOFDMSymbols=206$ and $\numberOfOFDMSymbols=52$ OFDM symbols for the \ac{FSK-MV} and \ac{OBDA}, respectively. 
The  maximum-excess delay of the EPA channel is $410$~ns. Hence,  for \ac{PPM-MV}, we set $\uniformGap$ to $11$ to ensure the condition in \eqref{eq:condition} for $\symbolSpacing=55.6$~ns. The number of \ac{DFT-s-OFDM} symbols for $\numberOfElementsInAPulse=1$, $\numberOfElementsInAPulse=4$, and $\numberOfElementsInAPulse=9$ can then be  calculated as $2462$, $3078$, and $4103$, respectively.  For \ac{OBDA}, the \ac{TCI} threshold is $0.2$. The learning rate is $0.001$. The batch size $\batchSize$ is set to $64$. For the test accuracy calculations, we use $10000$ test samples available in the MNIST database. The simulations are performed in MATLAB.

\begin{figure*}
	\centering
	\subfloat[Homogeneous data, ideal power control ($\effectivePathLossExponent=0$).]{\includegraphics[width =\figuresize]{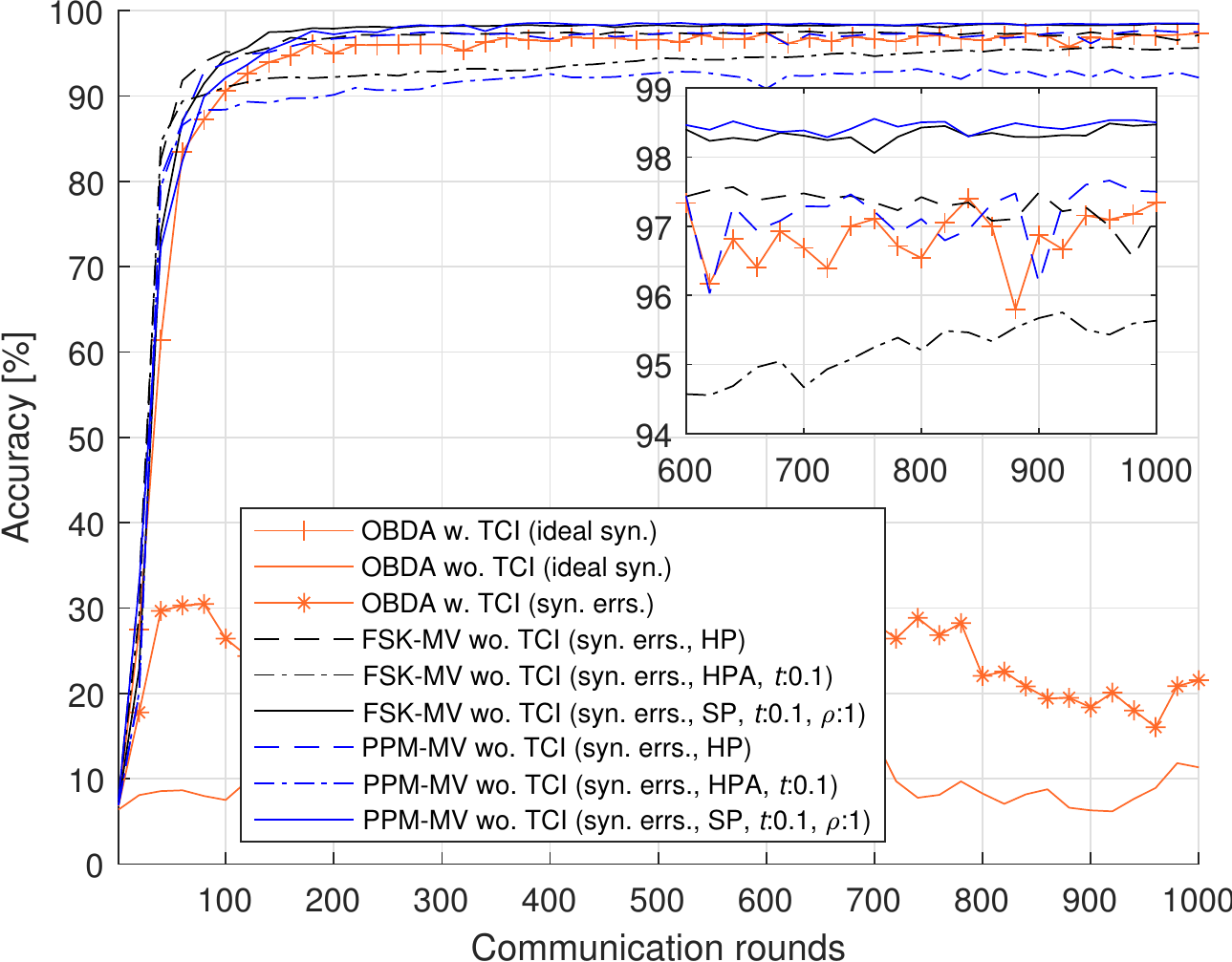}
		\label{subfig:acc_iid_aeff_zero}}~
	\subfloat[Homogeneous data, imperfect power control  ($\effectivePathLossExponent=2$).]{\includegraphics[width =\figuresize]{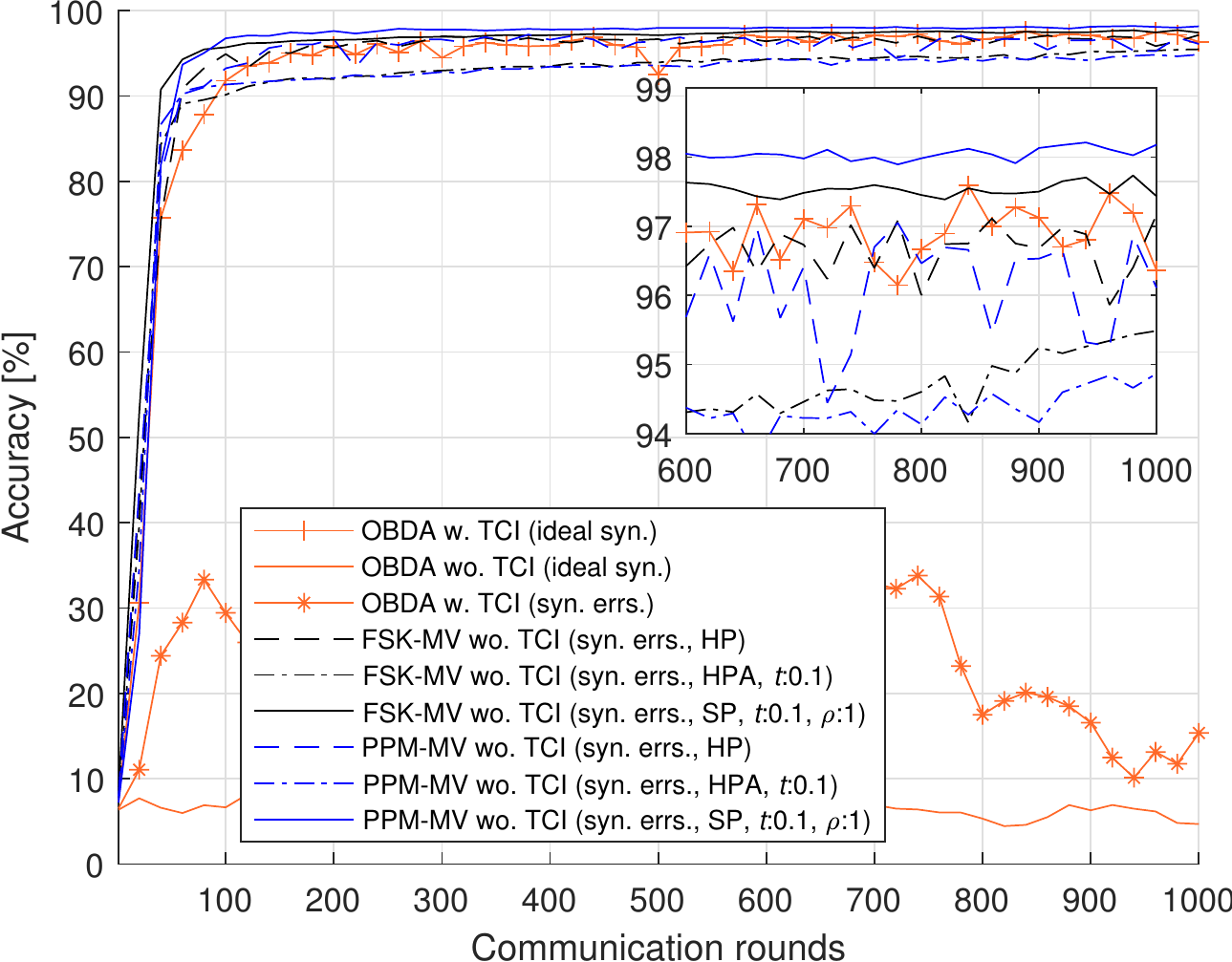}
		\label{subfig:acc_iid_aeff_two}}\\
	\subfloat[Heterogeneous data, ideal power control  ($\effectivePathLossExponent=0$).]{\includegraphics[width =\figuresize]{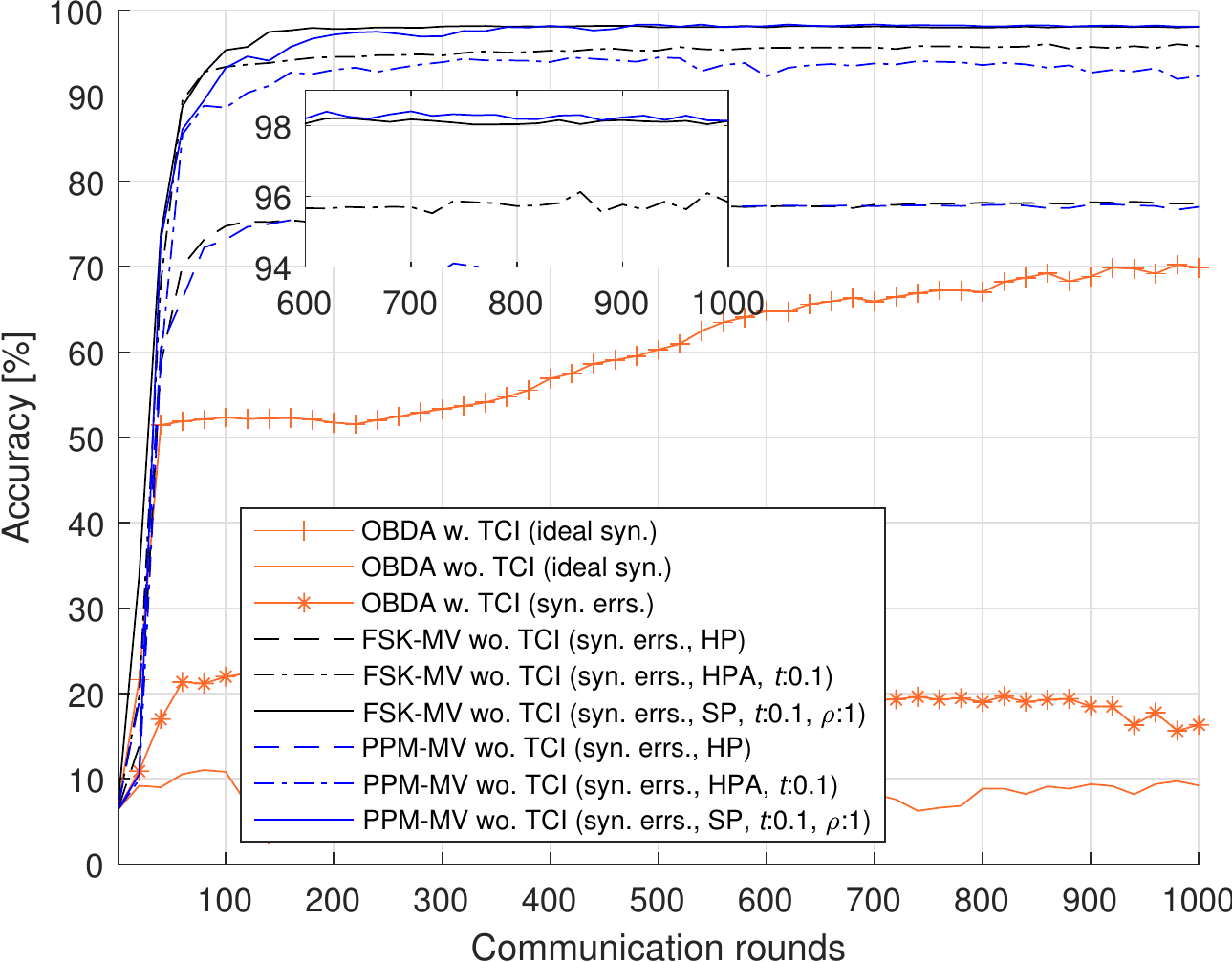}
		\label{subfig:acc_niid_aeff_zero}}~	
	\subfloat[Heterogeneous data, imperfect power control  ($\effectivePathLossExponent=2$).]{\includegraphics[width =\figuresize]{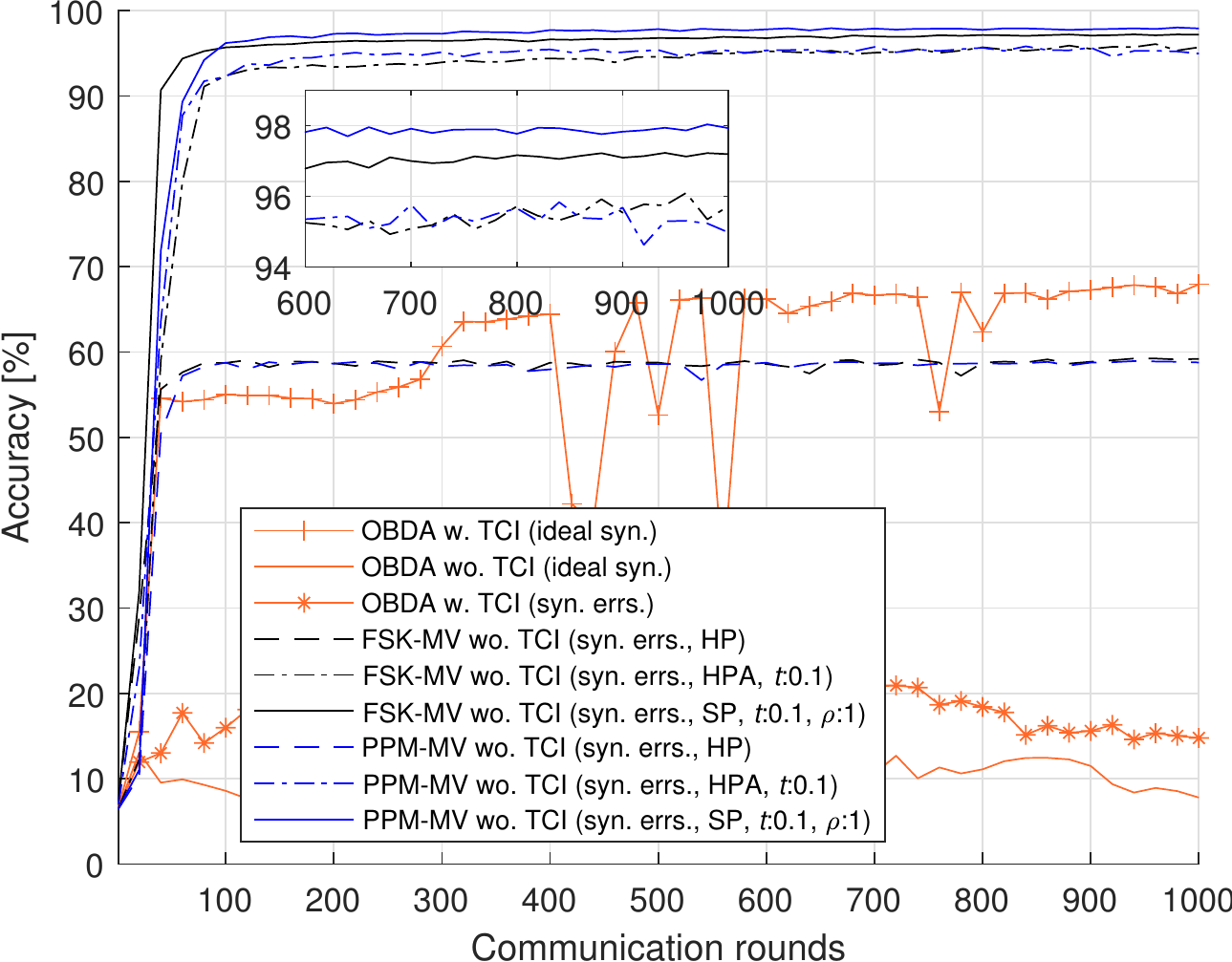}
		\label{subfig:acc_niid_aeff_two}}		
	\caption{Test accuracy versus communication rounds. The proposed scheme with HPA and SP  provides robustness against time-synchronization errors, heterogeneous data distribution even when the power control is imperfect.}
	\label{fig:testAcc}
\end{figure*}
\begin{figure*}
	\centering
	\subfloat[Homogeneous data, ideal power control ($\effectivePathLossExponent=0$).]{\includegraphics[width =\figuresize]{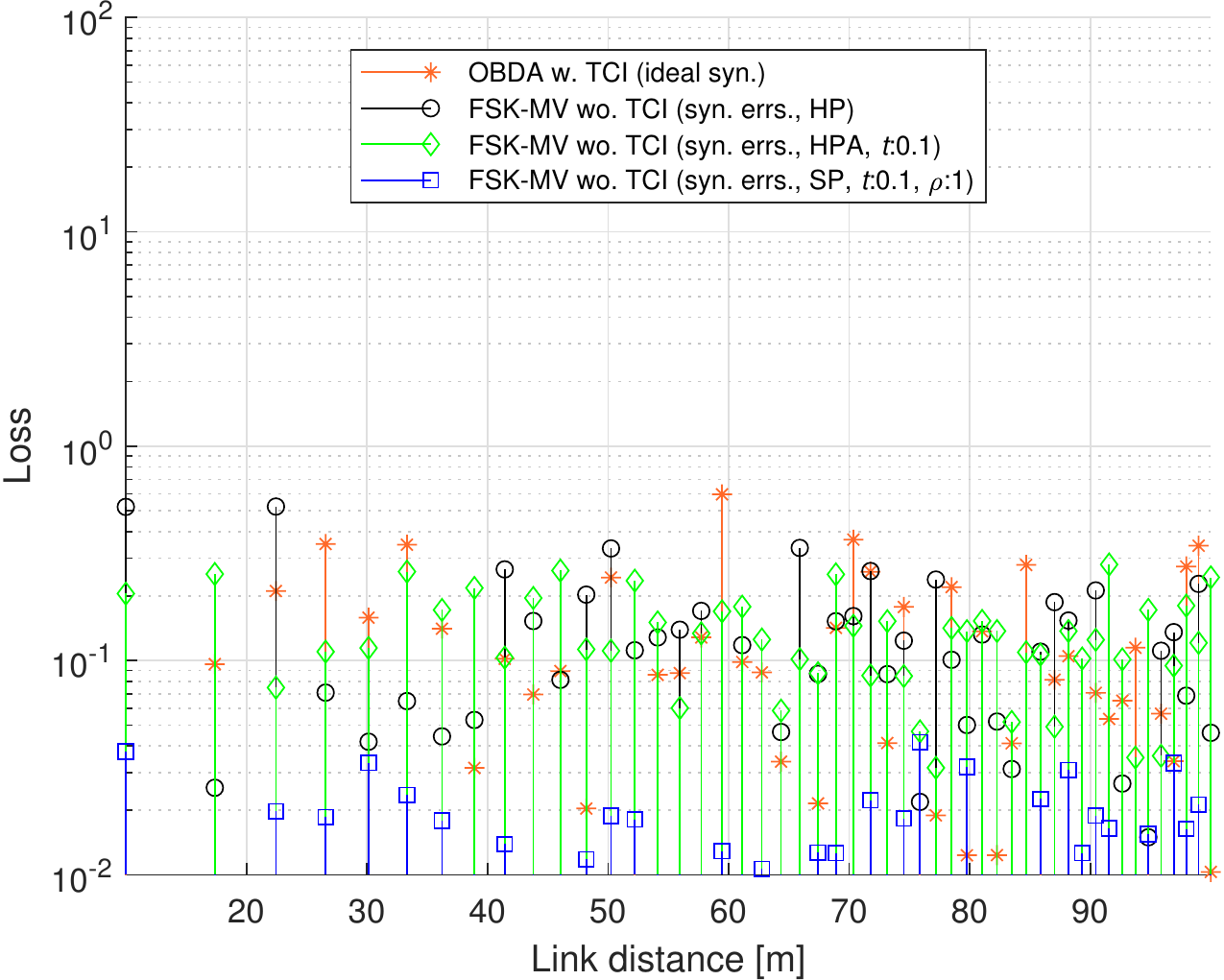}
		\label{subfig:ld_iid_aeff_zero}}~
	\subfloat[Homogeneous data, imperfect power control ($\effectivePathLossExponent=2$).]{\includegraphics[width =\figuresize]{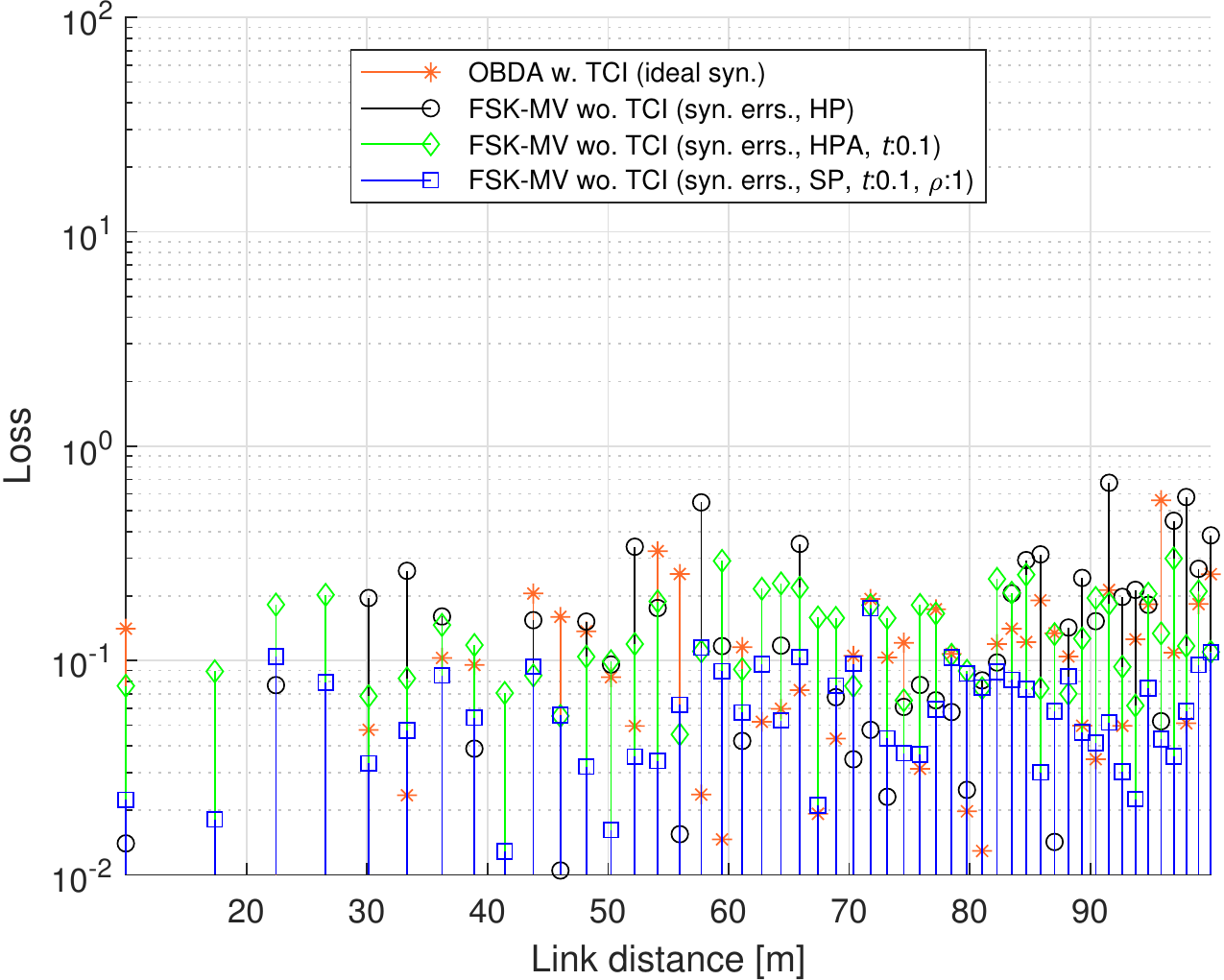}
		\label{subfig:ld_iid_aeff_two}}\\
	\subfloat[Heterogeneous data, ideal power control ($\effectivePathLossExponent=0$).]{\includegraphics[width =\figuresize]{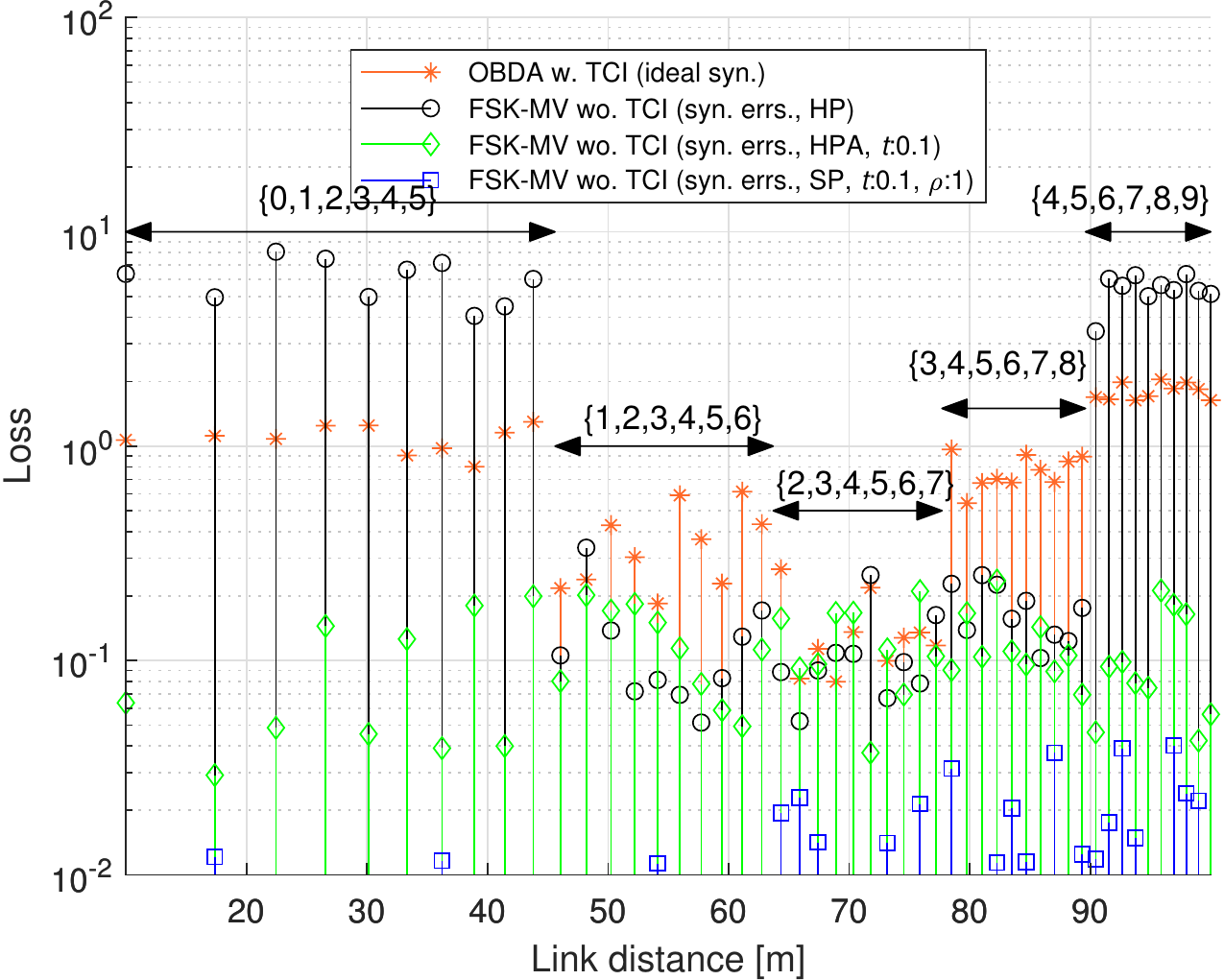}
		\label{subfig:ld_niid_aeff_zero}}~	
	\subfloat[Heterogeneous data, imperfect power control ($\effectivePathLossExponent=2$).]{\includegraphics[width =\figuresize]{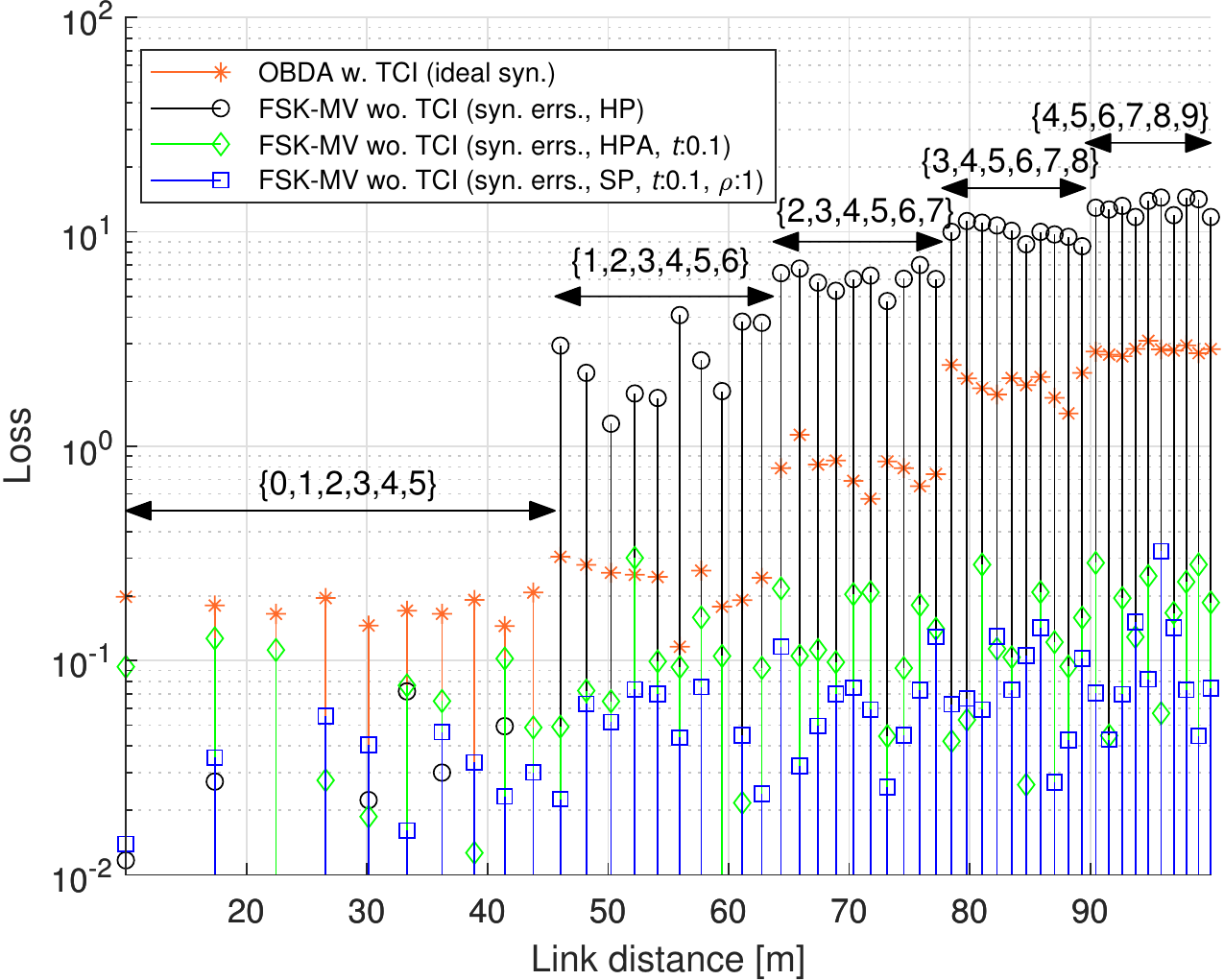}
		\label{subfig:ld_niid_aeff_two}}		
	\caption{Local loss versus link distance. The available labels are indicated as $\{\cdots\}$.}
	\label{fig:lossVsDistance}
\end{figure*}

\subsubsection{Test accuracy and loss under imperfections}

In \figurename~\ref{fig:testAcc}, we provide the test accuracy results  by taking time-synchronization errors and imperfect power control into account. 
 While we consider the homogeneous data distribution in \figurename~\ref{fig:testAcc}\subref{subfig:acc_iid_aeff_zero} and \figurename~\ref{fig:testAcc}\subref{subfig:acc_iid_aeff_two}, we evaluate the scenarios with the heterogeneous data distribution in \figurename~\ref{fig:testAcc}\subref{subfig:acc_niid_aeff_zero} and \figurename~\ref{fig:testAcc}\subref{subfig:acc_niid_aeff_two}. 
 For the same configurations, we provide the local losses at the \acp{ED} as a function of link distance in \figurename~\ref{fig:lossVsDistance} after $\communicationRounds=1000$  rounds. We only provide the local loss for \ac{FSK-MV} as the results with \ac{PPM-MV} are similar to that of \ac{FSK-MV}. The results in \figurename~\ref{fig:testAcc} and \figurename~\ref{fig:lossVsDistance} can be interpreted jointly as follows:
 
  \paragraph{Homogeneous data, perfect power control}
 In \figurename~\ref{fig:testAcc}\subref{subfig:acc_iid_aeff_zero}, the power control is assumed to be perfect (i.e., $\effectivePathLossExponent=0$). Under this configuration, \ac{OBDA} results in high test accuracy when the time synchronization is ideal and the \ac{CSI} is available at the \acp{ED} for \ac{TCI}. However, OBDA without \ac{TCI} or its utilization under imperfect time synchronization cause drastic reductions in the performance. On the other hand, both  \ac{FSK-MV} and \ac{PPM-MV} provide robustness against the time-synchronization errors and result in a high test accuracy without using \ac{CSI} at the \acp{ED}. For this case, \ac{HP} and \ac{SP} are better than \ac{HPA}. This is expected as both \ac{HP} and \ac{SP} enable full participation for the \ac{MV} calculation. The corresponding local losses at the \acp{ED} are given in \figurename~\ref{fig:lossVsDistance}\subref{subfig:ld_iid_aeff_zero}. While \ac{SP} is superior to \ac{HP} and \ac{HPA}, all options provide small loss values. 
 
  \paragraph{Homogeneous data, imperfect power control}
 In \figurename~\ref{fig:testAcc}\subref{subfig:acc_iid_aeff_two}, the received signal power at the \ac{ES} is not ideal (i.e., $\effectivePathLossExponent=2$). Although the test accuracy with \ac{OBDA}  (with \ac{TCI} and ideal synchronization) or \ac{FSK-MV}/\ac{PPM-MV} (without \ac{TCI} and ideal synchronization) reach to 97\%, \figurename~\ref{fig:lossVsDistance}\subref{subfig:ld_iid_aeff_two} indicates that the local losses slightly increase as compared to the ones in \figurename~\ref{fig:lossVsDistance}\subref{subfig:ld_iid_aeff_zero}. In this configuration, the \ac{FEEL} exploits the homogeneous data distribution in the cell, which also benefits to the far \acp{ED} that have the similar data distributions to the ones at the nearby \acp{ED}. 
 
\paragraph{Heterogeneous data, perfect power control}
In \figurename~\ref{fig:testAcc}\subref{subfig:acc_niid_aeff_zero}, we observe a non-negligible impact of the heterogeneous data distribution on the test accuracy. Although the power control is ideal in this case, the maximum test accuracy reduces to $77\%$ from $97\%$ for the proposed scheme with \ac{HP} and \ac{OBDA}. On the other hand, the proposed scheme with \ac{HPA} and \ac{SP} still provide a remarkably high test accuracy. As discussed in Section~\ref{subsec:heterobust}, this is because both \ac{HPA} and \ac{SP} reduce the impact of converging \acp{ED} on the \ac{MV} calculation.
 From \figurename~\ref{fig:lossVsDistance}\subref{subfig:ld_niid_aeff_zero}, we can identify the digits that are not learned well. We observe that the digit 0 and the digit 9 are not learned well since these digits are available in less number of \acp{ED} as compared to other digits. Similar issue occurs for the digit 1 and digit 8. Hence, the \ac{MV} is highly biased towards the labels that are available at large in the cell under \ac{HP} and \ac{OBDA}. 

\paragraph{Heterogeneous data, imperfect power control}
In \figurename~\ref{fig:testAcc}\subref{subfig:acc_niid_aeff_two},  in addition to the data heterogeneity, the power control is not ideal and we observe severe degradation in accuracy.
The maximum test accuracy reduces less than $70\%$ for the proposed scheme with \ac{HP} and \ac{OBDA}, while the test accuracy is still high with \ac{HPA} and \ac{SP}. Under this configuration, the local loss  tend to increase with the distance, i.e., the cell-edge \acp{ED}' labels are harder to learn, as shown in \figurename~\ref{fig:lossVsDistance}\subref{subfig:ld_niid_aeff_two}. As the cell-edge \acp{ED}' received signal powers are weak as compared the ones for the nearby \acp{ED}, the \ac{MV} is biased toward the nearby \acp{ED}' local data. Therefore, the digits available at the cell-edge \acp{ED}, e.g.,  digits 6, 7, 8, and 9, are not learned well  for \ac{HP} and \ac{OBDA}. On the other hand, the loss is reasonably small for both \ac{HPA} and \ac{SP} as these approaches tend to address the bias by reducing the impact of converging \ac{ED} on the \ac{MV}.

\begin{figure}
	\centering
	\subfloat[Varying $\thresholdForZero$ ($\steepnessFactor=1$).]{\includegraphics[width =\figuresize]{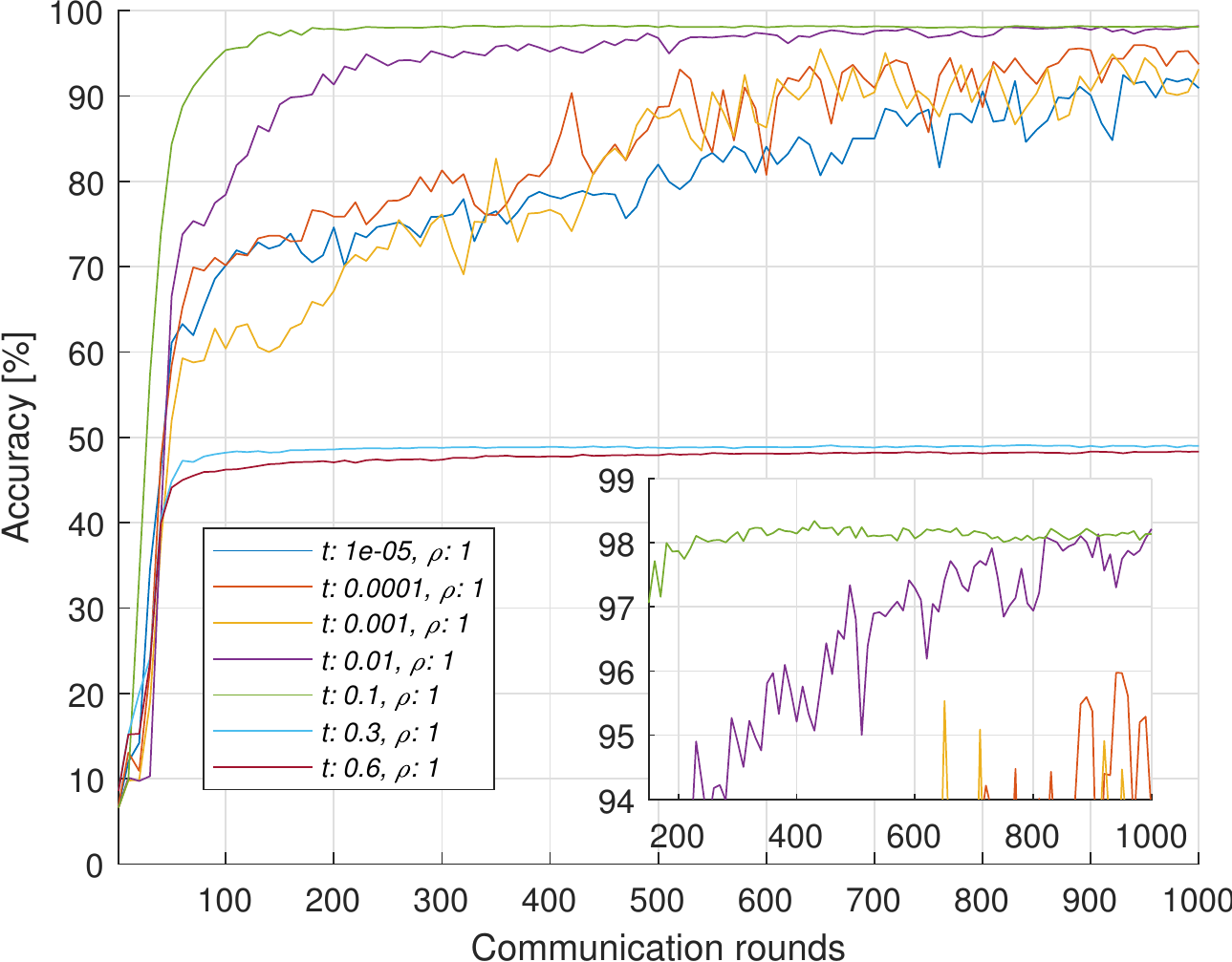}
		\label{subfig:sweept}}
	\if\IEEEsubmission0
		\\
	\fi
	\subfloat[Varying $\steepnessFactor$ ($\thresholdForZero=0.1$).]{\includegraphics[width =\figuresize]{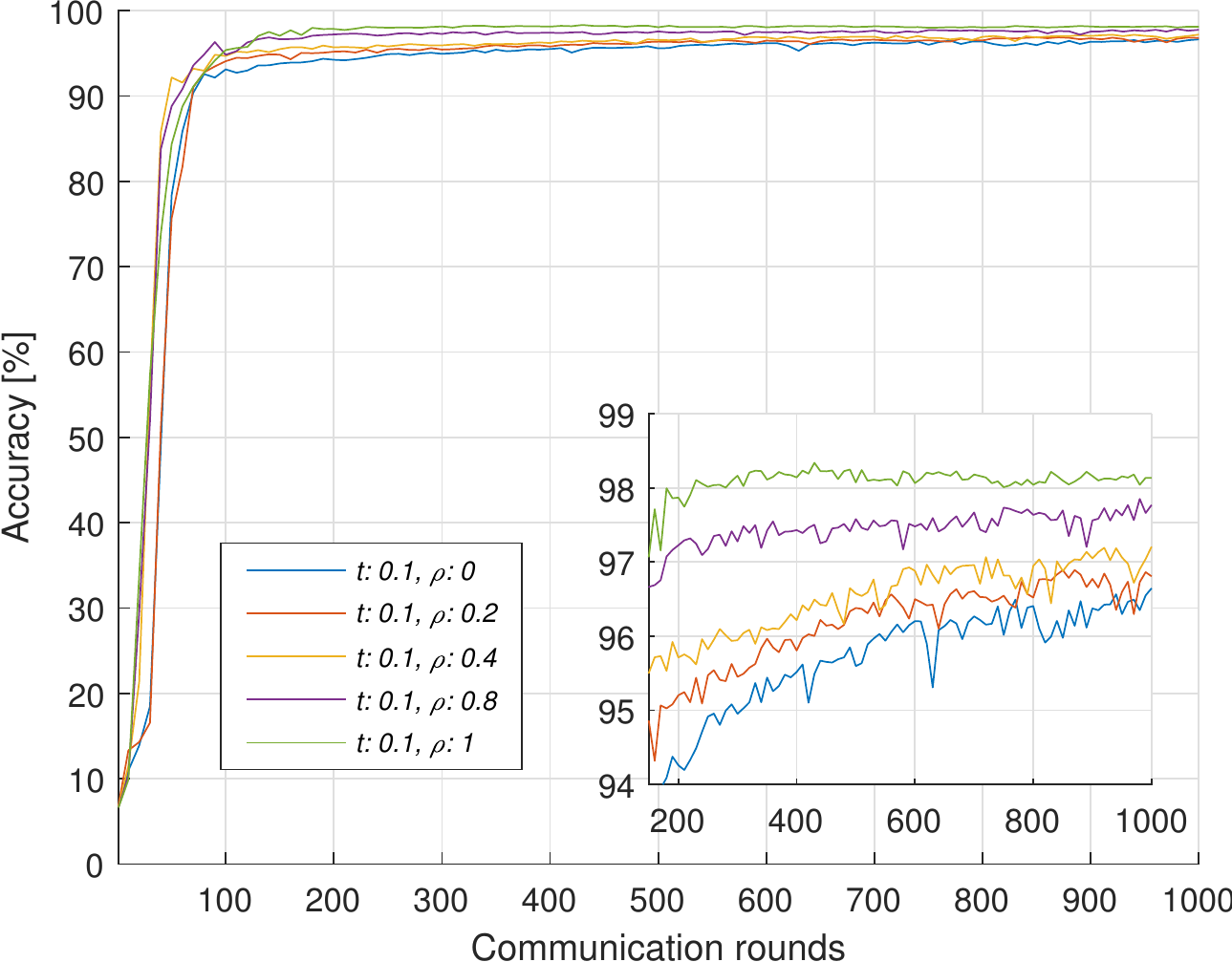}
		\label{subfig:sweeprho}}	
	\caption{Test accuracy for different $\thresholdForZero$ and $\steepnessFactor$ for \ac{SP} under imperfect power control  ($\effectivePathLossExponent=2$) and heterogeneous data distribution.}
	\label{fig:sweep}
\end{figure}
\subsubsection{Test accuracy for different SP parameters}
In \figurename~\ref{fig:sweep}, we analyze how the test accuracy over the communication rounds changes for different steepness and thresholds for \ac{SP}. For this analysis, we assume that the data distribution is heterogeneous and the power control is not ideal, i.e., ($\effectivePathLossExponent=2$).  In \figurename~\ref{fig:sweep}\subref{subfig:sweept}, we set $\steepnessFactor$ to $1$ and gradually  increase the threshold $\thresholdForZero$ from $1e-6$ to $0.6$. As can be seen from \figurename~\ref{fig:sweep}\subref{subfig:sweept}, the test accuracy performance improves for increasing threshold and reaches $98$\% for  {\reviewColor$\thresholdForZero=0.01$ and $\thresholdForZero=0.1$}.  However, it sharply reduces to $50$\% for $\thresholdForZero>0.1$. In \figurename~\ref{fig:sweep}\subref{subfig:sweeprho}, for $\thresholdForZero=0.1$, we alter $\steepnessFactor$ under the same configuration. The result implies that a larger  $\steepnessFactor$ yields a better performance, which implies that \ac{SP} can improve accuracy further by employing a smoother weight function. This result is also in line with the one in \figurename~\ref{fig:errorPr}.

\subsubsection{Waveform characteristics}
\begin{figure}
	\centering
	\subfloat[The waveforms in the time domain. The mean sample power is set to 1 for comparison.]{\includegraphics[width =\figuresize+0.2in]{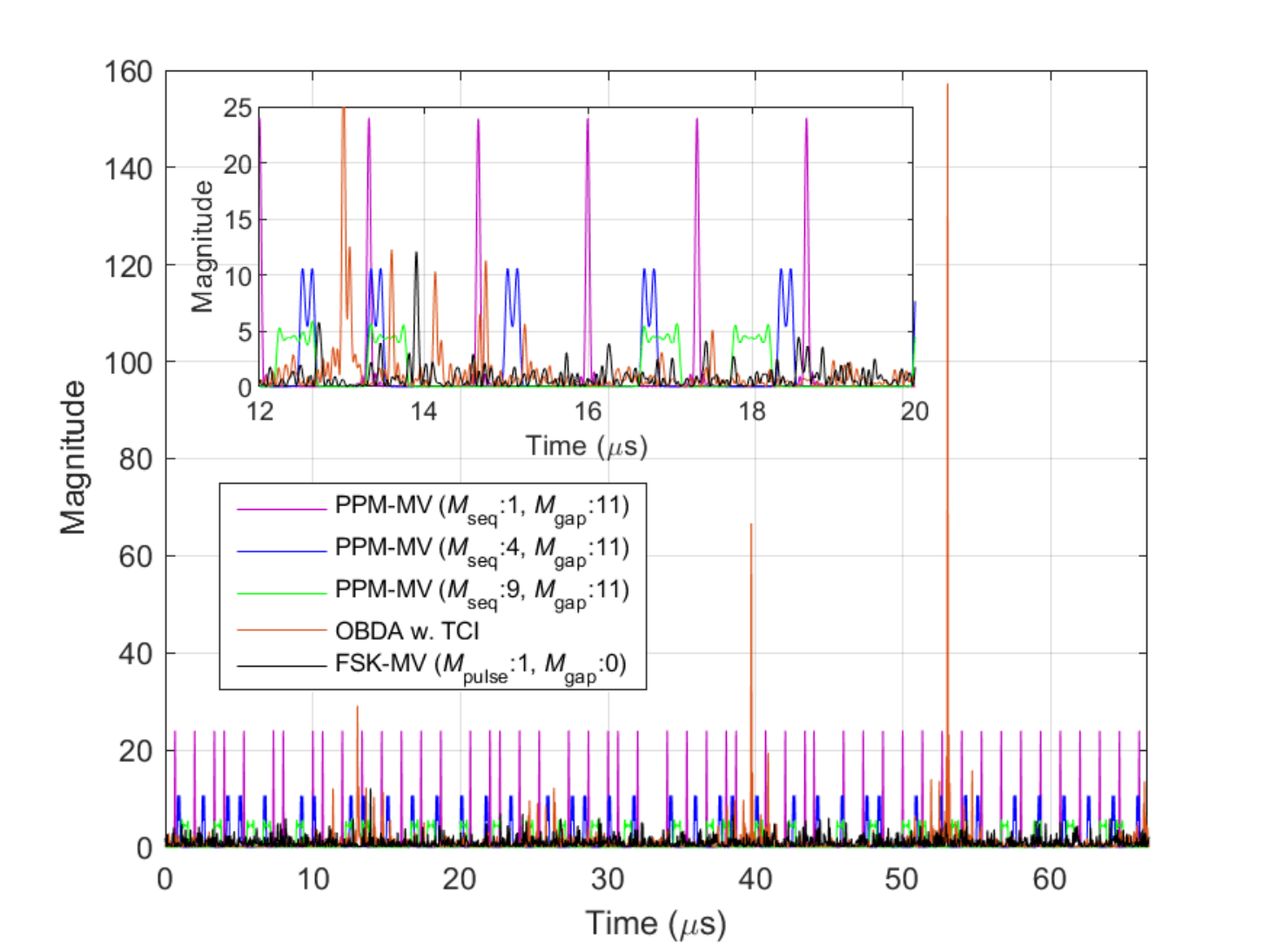}
		\label{subfig:temp}}
	\if\IEEEsubmission0
	\\
	\fi
	\subfloat[PMEPR distributions.]{\includegraphics[width =\figuresize+0.2in]{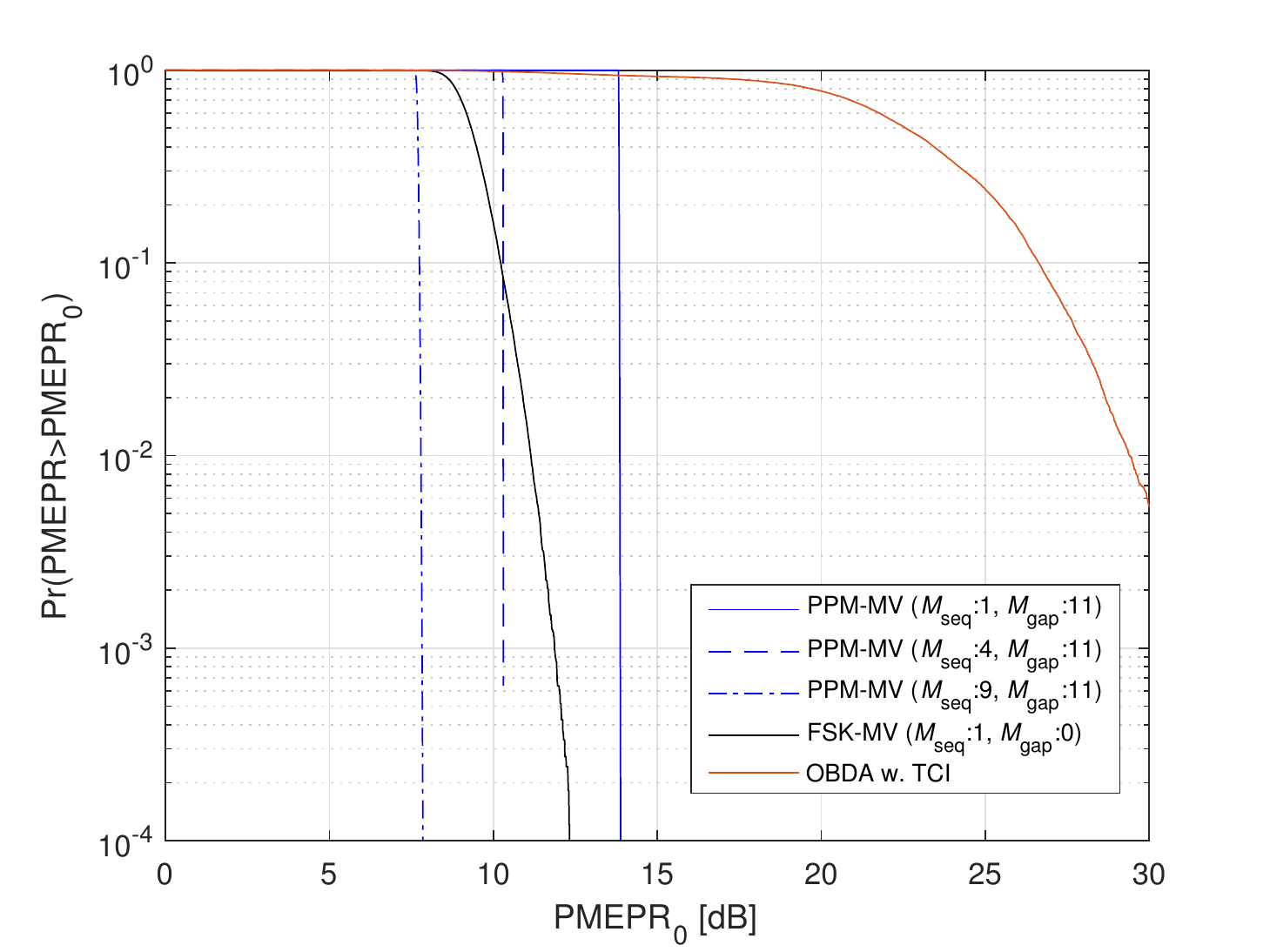}
		\label{subfig:pmepr}}	
	\caption{Temporal waveform characteristics. While the randomization symbols in \ac{FSK-MV} lowers the \ac{PMEPR}, the \ac{PMEPR} depends on the pulse width for \ac{PPM-MV}. For \ac{OBDA}, the \ac{PMEPR} can be very high since the local  gradients are correlated.}
	\label{fig:tempWave}
\end{figure}


\figurename~\ref{fig:tempWave}\subref{subfig:temp} details the temporal characteristics of \ac{OBDA}, \ac{FSK-MV}, and \ac{PPM-MV}. We see that the signal can be very peaky  with \ac{OBDA} when all the \ac{QPSK} symbols can be similar to each other. For \ac{PPM-MV}, this is not an issue as the votes are represented as separated pulses in time. For a large $\numberOfElementsInAPulse$, the shape of the magnitude of the pulse is similar to a rectangular window. For \ac{FSK-MV}, the temporal behavior is noise-like as it is based on \ac{OFDM}.

Finally, we compare the \ac{PMEPR} distributions in \figurename~\ref{fig:tempWave}\subref{subfig:pmepr} for \ac{OBDA}, \ac{FSK-MV}, and \ac{PPM-MV}, which is an important for factor for radios equipped with non-linear power amplifiers. 
Since the \ac{FSK-MV} introduces randomness in the frequency domain with the randomization symbols, it exhibits a similar behavior to a typical \ac{OFDM} transmission in terms of \ac{PMEPR}. On the other hand, the \ac{OBDA} can cause substantially high \ac{PMEPR} as the gradients can be correlated. The \ac{PMEPR} for 
\ac{PPM-MV} depends on $\numberOfElementsInAPulse$ and diminishes at the expense of more resources in time.

\section{Concluding Remarks}
\label{sec:conclusion}
In this study, we propose an \ac{OAC} scheme to compute the \ac{MV} for \ac{FEEL}. The proposed approach uses orthogonal time-frequency resources, i.e., subcarriers for \ac{FSK-MV} and wide-band pulses for \ac{PPM-MV}, to indicate the sign of the local stochastic gradients. Thus, it allows the \ac{ES} to detect the \ac{MV} with a non-coherent detector and eliminates the need for \ac{CSI} at the \acp{ED}  at the expense of a larger number of time and frequency resources.
We investigate various gradient-encoding strategies with weight functions to reduce the impact of an \ac{ED} that has a smaller absolute local stochastic gradient on the corresponding \ac{MV}. We theoretically show that enabling absentees via the weight function, e.g., \ac{HPA}, can considerably  improve the probability of detecting the correct \ac{MV}.  
We also prove the convergence of the \ac{FEEL} by taking path loss, power control, and cell size into account under \ac{HP} and \ac{HPA}. Through simulations, we demonstrate that the proposed method can provide a high test accuracy in fading channel even when the power control and the time synchronization are imperfect while resulting in an acceptable \ac{PMEPR} distribution. We also provide insights into the scenarios where local data distribution depends on the  \acp{ED} locations. Our results indicate that the proposed \ac{OAC} scheme with \ac{SP} and \ac{HPA} can address the bias in the \ac{MV} under the heterogeneous data distribution and/or imperfect power control.

{\reviewColor The proposed scheme opens up several research directions. We show the existence of non-zero threshold for \ac{HPA} to reduce the error probability in Lemma~\ref{lemma:errorPrHPA}. 
 However, deriving an optimal (or sufficiently good) threshold is currently an open problem due to the unknown gradient distributions. Secondly, 
 more theoretical results  are needed to understand the performance of the proposed scheme in heterogeneous  data distribution scenarios. Also, 
 the weight function along with the rule in \eqref{eq:majorityVote} can be optimized further to address unbalanced scenarios where \acp{ED} have non-identical dataset sizes or non-stationary cases where \acp{ED} leave or re-join the network.}


\appendices
\section{Proof of Lemma~\ref{lemma:exp}}
\label{app:lemma:exp}
\begin{IEEEproof}
	Due to the independent random variables, by using \eqref{eq:rx}, $\meanOptionOne$ can be calculated as
	\if\IEEEsubmission0
		\begin{align}
		&\expectationOperator[{\metricForFirst[\indexGradient]}][{\distanceED[\indexED],\channelMatrix[\indexED],\noiseVector[\indexOFDMSymbol],\datasetBatch[\indexED]}]
		=\nonumber\\&\hspace{5mm}\sum_{\substack{\indexED=1\\\localGradientElement[\indexED,\indexGradient][\indexCommunicationRound]>0}}^{\numberOfEdgeDevices}\underbrace{\expectationOperator[{{\powerED[\indexED]}}][{\distanceED[\indexED]}]}_{\triangleq\largeScaleImpactOnLearning}\underbrace{\expectationOperator[\norm{\matrixForCut[{\voteInFrequency[+]}]\transformPrecoder[\numberOfActiveSubcarriers]^{\rm H}\channelMatrixDiag[\indexED]\transformPrecoder[\numberOfActiveSubcarriers]\symbolVector[\indexED,\indexOFDMSymbol]}^2_2][{\channelMatrix[\indexED],\datasetBatch[\indexED]}]}_{\approx\numberOfElementsInAPulse\symbolEnergy\gradientPowerCoefficient}\nonumber\\&\hspace{50mm}+\underbrace{\expectationOperator[\norm{\matrixForCut[{\voteInFrequency[+]}]\noiseVectorOnSymbols[\indexOFDMSymbol]}^2_2][{\noiseVector[\indexOFDMSymbol]}]}_{=(\numberOfElementsInAPulse + \uniformGap)\noiseVariance}\nonumber\\
		&\approx{\numberOfElementsInAPulse\symbolEnergy}\numberOFEDsForOptionOne\largeScaleImpactOnLearning\gradientPowerCoefficient+{(\numberOfElementsInAPulse + \uniformGap)\noiseVariance}~.\nonumber
	\end{align}
\else
		\begin{align}
	\expectationOperator[{\metricForFirst[\indexGradient]}][{\distanceED[\indexED],\channelMatrix[\indexED],\noiseVector[\indexOFDMSymbol],\datasetBatch[\indexED]}]
	&=\sum_{\substack{\indexED=1\\\localGradientElement[\indexED,\indexGradient][\indexCommunicationRound]>0}}^{\numberOfEdgeDevices}\underbrace{\expectationOperator[{{\powerED[\indexED]}}][{\distanceED[\indexED]}]}_{\triangleq\largeScaleImpactOnLearning}\underbrace{\expectationOperator[\norm{\matrixForCut[{\voteInFrequency[+]}]\transformPrecoder[\numberOfActiveSubcarriers]^{\rm H}\channelMatrixDiag[\indexED]\transformPrecoder[\numberOfActiveSubcarriers]\symbolVector[\indexED,\indexOFDMSymbol]}^2_2][{\channelMatrix[\indexED],\datasetBatch[\indexED]}]}_{\approx\numberOfElementsInAPulse\symbolEnergy\gradientPowerCoefficient}+\underbrace{\expectationOperator[\norm{\matrixForCut[{\voteInFrequency[+]}]\noiseVectorOnSymbols[\indexOFDMSymbol]}^2_2][{\noiseVector[\indexOFDMSymbol]}]}_{=(\numberOfElementsInAPulse + \uniformGap)\noiseVariance}\nonumber\\
	&\approx{\numberOfElementsInAPulse\symbolEnergy}\numberOFEDsForOptionOne\largeScaleImpactOnLearning\gradientPowerCoefficient+{(\numberOfElementsInAPulse + \uniformGap)\noiseVariance}~.\nonumber
\end{align}
\fi
	The approximation given for the second expectation is exact for the \ac{FSK-MV} since $\transformPrecoder[\numberOfActiveSubcarriers]$ is an identity matrix, but it approximately holds for the \ac{PPM-MV} as the interference between the \ac{PPM} symbols is negligibly low under the condition given in \eqref{eq:condition}. 
	Based on \eqref{eq:pathloss}, $\largeScaleImpactOnLearning$ can be expressed as
	$	\largeScaleImpactOnLearning=\expectationOperator[{\left({{\distanceED[\indexED]}}/{\referenceDistance}\right)^{-\effectivePathLossExponent}{\reviewColor \powerRef}}][{\distanceED[\indexED]}]$.
	Due to the uniform deployment within the cell, the link distance distribution is
		$f(\distanceED[])={2\distanceED[]}/({\cellRadius^2-\minimumDistance^2})$.
	Hence, the distribution of $\arandomvar\triangleq\distanceED[]^{-\effectivePathLossExponent}$ can obtained as
	\begin{align}
		f(\arandomvar)=\frac{	f(\distanceED[])}{|\frac{d\arandomvar}{d\distanceED[]}|}\bigg|_{\distanceED[]=\arandomvar^{-\frac{1}{\effectivePathLossExponent}}}=\frac{2\arandomvar^{-\frac{\effectivePathLossExponent+2}{\effectivePathLossExponent}}}{(\cellRadius^2-\minimumDistance^2)\effectivePathLossExponent}.
		\label{eq:fdist}
	\end{align}
	From \eqref{eq:fdist}, $\largeScaleImpactOnLearning$ is equal to \eqref{eq:pathlossimpact}. Also,  $\gradientPowerCoefficient\le1$ as $\gradientWeight[x]\le1$. The same analysis can be done for $\meanOptionTwo$.
\end{IEEEproof}

\section{Proof of Lemma~\ref{lemma:errorPrHPA}}
\label{app:lemma:errorPrHPA}

Let $\numberOfEDsWithCorrectChoice$, $\numberOfEDsWithZeroChoice$, and $\numberOfEDsWithInCorrectChoice$ be random variables for counting the number of EDs that vote for the correct direction (i.e., the sign of $\globalGradientElement[\indexCommunicationRound][\indexGradient]$), do not cast a vote, and vote for the incorrect direction (i.e., the sign of $-\globalGradientElement[\indexCommunicationRound][\indexGradient]$), respectively. We can express the error probability $	\probabilityIncorrect[\indexGradient]$ as
\if \IEEEsubmission 0
\begin{align}
	\probabilityIncorrect[\indexGradient]=
	\sum_{\numberOFEDsForOptionOne=0}^{\numberOfEdgeDevices}
	\sum_{\numberOFEDsForOptionSecond=0}^{\numberOfEdgeDevices-\numberOFEDsForOptionOne}&
	\probability[{\signNormal[{\deltaVectorAtIteration[\indexCommunicationRound][\indexGradient]}]\neq \signNormal[{\globalGradientElement[\indexCommunicationRound][\indexGradient]}]}|\numberOfEDsWithCorrectChoice=\numberOFEDsForOptionOne,\numberOfEDsWithInCorrectChoice=\numberOFEDsForOptionSecond]\nonumber\\&\times\probability[\numberOfEDsWithCorrectChoice=\numberOFEDsForOptionOne,\numberOfEDsWithInCorrectChoice=\numberOFEDsForOptionSecond]~.
	\label{eq:Perrq}
\end{align}
\else
\begin{align}
	\probabilityIncorrect[\indexGradient]=
	\sum_{\numberOFEDsForOptionOne=0}^{\numberOfEdgeDevices}
	\sum_{\numberOFEDsForOptionSecond=0}^{\numberOfEdgeDevices-\numberOFEDsForOptionOne}&
	\probability[{\signNormal[{\deltaVectorAtIteration[\indexCommunicationRound][\indexGradient]}]\neq \signNormal[{\globalGradientElement[\indexCommunicationRound][\indexGradient]}]}|\numberOfEDsWithCorrectChoice=\numberOFEDsForOptionOne,\numberOfEDsWithInCorrectChoice=\numberOFEDsForOptionSecond]\times\probability[\numberOfEDsWithCorrectChoice=\numberOFEDsForOptionOne,\numberOfEDsWithInCorrectChoice=\numberOFEDsForOptionSecond]~.
	\label{eq:Perrq}
\end{align}
\fi
Based on Assumption~\ref{assump:iid}, we can express the independent probability in \eqref{eq:Perrq} with a binomial distribution as
\if\IEEEsubmission0
\begin{align}
	&\probability[\numberOfEDsWithCorrectChoice=\numberOFEDsForOptionOne,\numberOfEDsWithInCorrectChoice=\numberOFEDsForOptionSecond] \nonumber\\&~~~~~~~~~~~~~~~= \binom{\numberOfEdgeDevices}{\numberOfEdgeDevices+\numberOFEDsForOptionSecond}\binom{\numberOfEdgeDevices+\numberOFEDsForOptionSecond}{\numberOFEDsForOptionSecond}\correctDecision[\indexGradient]^{\numberOFEDsForOptionOne}\incorrectDecision[\indexGradient]^{\numberOFEDsForOptionSecond}{\zeroDecision[\indexGradient]}^{\numberOfEdgeDevices-\numberOFEDsForOptionOne-\numberOFEDsForOptionSecond}.
	\label{eq:binProb}
\end{align}
\else
\begin{align}
	&\probability[\numberOfEDsWithCorrectChoice=\numberOFEDsForOptionOne,\numberOfEDsWithInCorrectChoice=\numberOFEDsForOptionSecond] = \binom{\numberOfEdgeDevices}{\numberOfEdgeDevices+\numberOFEDsForOptionSecond}\binom{\numberOfEdgeDevices+\numberOFEDsForOptionSecond}{\numberOFEDsForOptionSecond}\correctDecision[\indexGradient]^{\numberOFEDsForOptionOne}\incorrectDecision[\indexGradient]^{\numberOFEDsForOptionSecond}{\zeroDecision[\indexGradient]}^{\numberOfEdgeDevices-\numberOFEDsForOptionOne-\numberOFEDsForOptionSecond}.
	\label{eq:binProb}
\end{align}
\fi

To calculate the posterior  in \eqref{eq:Perrq}, the \ac{PDF} of $\deltaVectorAtIteration[\indexCommunicationRound][\indexGradient]$ is needed for given $\numberOFEDsForOptionOne$ and $\numberOFEDsForOptionSecond$, which can be calculated via the properties of exponential random variables under Assumption~\ref{assump:exp} as
\def\varRandom{x}
\begin{align}
	f(\varRandom) = \begin{cases} 
		\frac{\constante^{-\frac{\varRandom}{\meanOptionTwo}}}{\meanOptionOne+\meanOptionTwo}, & \varRandom\le 0 \\
		\frac{\constante^{-\frac{\varRandom}{\meanOptionOne}}}{\meanOptionOne+\meanOptionTwo}, & \varRandom>0 
	\end{cases}~.
	\label{eq:lappro}
\end{align}
By using the \ac{PDF} in \eqref{eq:lappro}, it can be shown that $\probability[\varRandom<0]={\meanOptionTwo}/({\meanOptionOne+\meanOptionTwo})$.	Therefore, by using  Lemma~\ref{lemma:exp} and considering $\gradientPowerCoeffPositive=\gradientPowerCoeffNegative=1$ for \ac{HPA}, we can express the posterior probability as
\if \IEEEsubmission 0
\begin{align}
	\probability[{\signNormal[{\deltaVectorAtIteration[\indexCommunicationRound][\indexGradient]}]\neq 1}|\numberOfEDsWithCorrectChoice=\numberOFEDsForOptionOne,\numberOfEDsWithInCorrectChoice=\numberOFEDsForOptionSecond] &= \frac{\meanOptionTwo}{\meanOptionOne+\meanOptionTwo}
	\nonumber\\&\hspace{-10mm}=\frac{\numberOFEDsForOptionSecond+1/\effectiveSNR}{\numberOFEDsForOptionOne+\numberOFEDsForOptionSecond+2/\effectiveSNR}~,
	\label{eq:probLapResult}
\end{align}
\else
\begin{align}
	\probability[{\signNormal[{\deltaVectorAtIteration[\indexCommunicationRound][\indexGradient]}]\neq 1}|\numberOfEDsWithCorrectChoice=\numberOFEDsForOptionOne,\numberOfEDsWithInCorrectChoice=\numberOFEDsForOptionSecond] &= \frac{\meanOptionTwo}{\meanOptionOne+\meanOptionTwo}
	=\frac{\numberOFEDsForOptionSecond+1/\effectiveSNR}{\numberOFEDsForOptionOne+\numberOFEDsForOptionSecond+2/\effectiveSNR}~,
	\label{eq:probLapResult}
\end{align}
\fi
for $\effectiveSNR\triangleq{\numberOfElementsInAPulse\largeScaleImpactOnLearning\symbolEnergy}/({(\numberOfElementsInAPulse + \uniformGap)\noiseVariance})={2\largeScaleImpactOnLearning}/{\noiseVariance}$. Let $\indexSumKpKm$ denote $ \numberOFEDsForOptionOne+\numberOFEDsForOptionSecond$ for simplifying the notation. We evaluate the sum in \eqref{eq:Perrq} under Assumption~\ref{assump:zeroBound}, by plugging \eqref{eq:binProb} and \eqref{eq:probLapResult} into \eqref{eq:Perrq}, as
\if\IEEEsubmission0
\begin{align}
	\probabilityIncorrect[\indexGradient]=&
	\sum_{\numberOFEDsForOptionOne=0}^{\numberOfEdgeDevices}
	\sum_{\numberOFEDsForOptionSecond=0}^{\numberOfEdgeDevices-\numberOFEDsForOptionOne}\frac{\numberOFEDsForOptionSecond+1/\effectiveSNR}{\numberOFEDsForOptionOne+\numberOFEDsForOptionSecond+2/\effectiveSNR}\probability[\numberOfEDsWithCorrectChoice=\numberOFEDsForOptionOne,\numberOfEDsWithInCorrectChoice=\numberOFEDsForOptionSecond]\nonumber\\
	=&	\sum_{\indexSumKpKm=0}^{\numberOfEdgeDevices}
	\sum_{\numberOFEDsForOptionSecond=0}^{\indexSumKpKm}	\frac{\numberOFEDsForOptionSecond+1/\effectiveSNR}{\indexSumKpKm+2/\effectiveSNR}\binom{\numberOfEdgeDevices}{\indexSumKpKm}\binom{\indexSumKpKm}{\numberOFEDsForOptionSecond}\correctDecision[\indexGradient]^{\indexSumKpKm-\numberOFEDsForOptionSecond}\incorrectDecision[\indexGradient]^{\numberOFEDsForOptionSecond}{\zeroDecision[\indexGradient]}^{\numberOfEdgeDevices-\indexSumKpKm}\nonumber\\
	=&	\sum_{\indexSumKpKm=0}^{\numberOfEdgeDevices}\binom{\numberOfEdgeDevices}{\indexSumKpKm}{\zeroDecision[\indexGradient]}^{\numberOfEdgeDevices-\indexSumKpKm}	\underbrace{\sum_{\numberOFEDsForOptionSecond=0}^{\indexSumKpKm}\frac{\numberOFEDsForOptionSecond+1/\effectiveSNR}{\indexSumKpKm+2/\effectiveSNR}\binom{\indexSumKpKm}{\numberOFEDsForOptionSecond}\correctDecision[\indexGradient]^{\indexSumKpKm-\numberOFEDsForOptionSecond}\incorrectDecision[\indexGradient]^{\numberOFEDsForOptionSecond}}_{\frac{\indexSumKpKm\frac{\incorrectDecision[\indexGradient]}{1-\zeroDecision[\indexGradient]}+\frac{\noiseVariance}{2\largeScaleImpactOnLearning}}{\indexSumKpKm+\frac{\noiseVariance}{\largeScaleImpactOnLearning}}(1-\zeroDecision[\indexGradient])^{\indexSumKpKm}}\nonumber\\
	=&
	\frac{1}{2}\coeffErr[{\zeroDecision[\indexGradient]}]+\frac{\incorrectDecision[\indexGradient]}{1-\zeroDecision[\indexGradient]}(1-\coeffErr[{\zeroDecision[\indexGradient]}])~,\label{eq:properr}
\end{align}
\else
\begin{align}
	\probabilityIncorrect[\indexGradient]=&
	\sum_{\numberOFEDsForOptionOne=0}^{\numberOfEdgeDevices}
	\sum_{\numberOFEDsForOptionSecond=0}^{\numberOfEdgeDevices-\numberOFEDsForOptionOne}\frac{\numberOFEDsForOptionSecond+1/\effectiveSNR}{\numberOFEDsForOptionOne+\numberOFEDsForOptionSecond+2/\effectiveSNR}\probability[\numberOfEDsWithCorrectChoice=\numberOFEDsForOptionOne,\numberOfEDsWithInCorrectChoice=\numberOFEDsForOptionSecond]=	\sum_{\indexSumKpKm=0}^{\numberOfEdgeDevices}
	\sum_{\numberOFEDsForOptionSecond=0}^{\indexSumKpKm}	\frac{\numberOFEDsForOptionSecond+1/\effectiveSNR}{\indexSumKpKm+2/\effectiveSNR}\binom{\numberOfEdgeDevices}{\indexSumKpKm}\binom{\indexSumKpKm}{\numberOFEDsForOptionSecond}\correctDecision[\indexGradient]^{\indexSumKpKm-\numberOFEDsForOptionSecond}\incorrectDecision[\indexGradient]^{\numberOFEDsForOptionSecond}{\zeroDecision[\indexGradient]}^{\numberOfEdgeDevices-\indexSumKpKm}\nonumber\\
	=&	\sum_{\indexSumKpKm=0}^{\numberOfEdgeDevices}\binom{\numberOfEdgeDevices}{\indexSumKpKm}{\zeroDecision[\indexGradient]}^{\numberOfEdgeDevices-\indexSumKpKm}	\underbrace{\sum_{\numberOFEDsForOptionSecond=0}^{\indexSumKpKm}\frac{\numberOFEDsForOptionSecond+1/\effectiveSNR}{\indexSumKpKm+2/\effectiveSNR}\binom{\indexSumKpKm}{\numberOFEDsForOptionSecond}\correctDecision[\indexGradient]^{\indexSumKpKm-\numberOFEDsForOptionSecond}\incorrectDecision[\indexGradient]^{\numberOFEDsForOptionSecond}}_{\frac{\indexSumKpKm\frac{\incorrectDecision[\indexGradient]}{1-\zeroDecision[\indexGradient]}+\frac{1}{\effectiveSNR}}{\indexSumKpKm+\frac{2}{\effectiveSNR}}(1-\zeroDecision[\indexGradient])^{\indexSumKpKm}}=
	\frac{1}{2}\coeffErr[{\zeroDecision[\indexGradient]}]+\frac{\incorrectDecision[\indexGradient]}{1-\zeroDecision[\indexGradient]}(1-\coeffErr[{\zeroDecision[\indexGradient]}])~,\label{eq:properr}
\end{align}
\fi
where $\coeffErr[{\zeroDecision[]}]\in[0,1]$ is defined in \eqref{eq:coeffErr}. 
 To obtain \eqref{eq:properr}, we use the identity given by
\if\IEEEsubmission 0
\begin{align}
	\sum_{\indexSumKpKm=0}^{\numberOfEdgeDevices}&\frac{a\indexSumKpKm +b}{\indexSumKpKm+c}\binom{\numberOfEdgeDevices}{\indexSumKpKm}\correctDecision[]^{\indexSumKpKm}({1-\correctDecision[]})^{\numberOfEdgeDevices-\indexSumKpKm} \nonumber\\&= a+\left(\frac{b}{c}-a\right)(1-\correctDecision[])^{\numberOfEdgeDevices}\hyperGeometricFcn[c][{-\numberOfEdgeDevices}][{c+1}][{\frac{\correctDecision[]}{\correctDecision[]-1}}]~,\nonumber
\end{align}
\else
\begin{align}
	\sum_{\indexSumKpKm=0}^{\numberOfEdgeDevices}&\frac{a\indexSumKpKm +b}{\indexSumKpKm+c}\binom{\numberOfEdgeDevices}{\indexSumKpKm}\correctDecision[]^{\indexSumKpKm}({1-\correctDecision[]})^{\numberOfEdgeDevices-\indexSumKpKm} = a+\left(\frac{b}{c}-a\right)(1-\correctDecision[])^{\numberOfEdgeDevices}\hyperGeometricFcn[c][{-\numberOfEdgeDevices}][{c+1}][{\frac{\correctDecision[]}{\correctDecision[]-1}}]~,\nonumber
\end{align}
\fi
for non-negative $\numberOfEdgeDevices$, $a$, $b$, $c$, and $\correctDecision[]$ \cite{Mathematica}.

\section{Proof of Theorem~\ref{th:convergence}}
\label{app:th:conv}
\begin{IEEEproof}
	By using Assumption~\ref{assump:smoothness} and \eqref{eq:MVsignSGDsche}, we can express the improvement in the loss  as
	\if\IEEEsubmission0
	\begin{align}
		\lossFunctionGlobal[{\modelParametersAtIteration[\indexCommunicationRound+1]}]& - \lossFunctionGlobal[{\modelParametersAtIteration[\indexCommunicationRound]}]\le -\learningRate{\globalGradient[\indexCommunicationRound]}^{\rm T}\majorityVoteScheme[\indexCommunicationRound] + \frac{\learningRate^2}{2}\norm{\nonnegativeConstants}_1\nonumber\\
		=&-\learningRate\norm{\globalGradient[\indexCommunicationRound]}_1+\frac{\learningRate^2}{2}\norm{\nonnegativeConstants}_1\nonumber\\&+2\learningRate\sum_{\indexGradient=1}^{\numberOfModelParameters}|\globalGradientElement[\indexCommunicationRound][\indexGradient]| \indicatorFunction[{\signNormal[{\deltaVectorAtIteration[\indexCommunicationRound][\indexGradient]}]\neq \signNormal[{\globalGradientElement[\indexCommunicationRound][\indexGradient]}]}]\nonumber~.
	\end{align}	
	\else
		\begin{align}
		\lossFunctionGlobal[{\modelParametersAtIteration[\indexCommunicationRound+1]}] - \lossFunctionGlobal[{\modelParametersAtIteration[\indexCommunicationRound]}]&\le -\learningRate{\globalGradient[\indexCommunicationRound]}^{\rm T}\majorityVoteScheme[\indexCommunicationRound] + \frac{\learningRate^2}{2}\norm{\nonnegativeConstants}_1\nonumber\\&=-\learningRate\norm{\globalGradient[\indexCommunicationRound]}_1+\frac{\learningRate^2}{2}\norm{\nonnegativeConstants}_1+2\learningRate\sum_{\indexGradient=1}^{\numberOfModelParameters}|\globalGradientElement[\indexCommunicationRound][\indexGradient]| \indicatorFunction[{\signNormal[{\deltaVectorAtIteration[\indexCommunicationRound][\indexGradient]}]\neq \signNormal[{\globalGradientElement[\indexCommunicationRound][\indexGradient]}]}]\nonumber~.
		\end{align}	
	\fi
	Therefore,
			{
	\begin{align}
&\expectationOperator[{	\lossFunctionGlobal[{\modelParametersAtIteration[\indexCommunicationRound+1]}] - \lossFunctionGlobal[{\modelParametersAtIteration[\indexCommunicationRound]}]|\modelParametersAtIteration[\indexCommunicationRound]}][{\allBatches}] \le  -\learningRate\norm{\globalGradient[\indexCommunicationRound]}_1+\frac{\learningRate^2}{2}\norm{\nonnegativeConstants}_1\nonumber\\&~~~~~~+2\learningRate\underbrace{\sum_{\indexGradient=1}^{\numberOfModelParameters}|\globalGradientElement[\indexCommunicationRound][\indexGradient]| \underbrace{\probability[{\signNormal[{\deltaVectorAtIteration[\indexCommunicationRound][\indexGradient]}]\neq \signNormal[{\globalGradientElement[\indexCommunicationRound][\indexGradient]}]}]}_{\triangleq\probabilityIncorrect[\indexGradient]}}_{\text{Error}}~.\label{eq:errorTerm}
	\end{align}}
	The main challenge is to obtain an upper bound on the error term in \eqref{eq:errorTerm} that is a function of the stochasticity of the local gradients and the detection performance of the proposed scheme. 
	
Let us use definitions in \eqref{eq:prP}, \eqref{eq:prZ}, and \eqref{eq:prZ} for $\correctDecision[\indexGradient]$, $\zeroDecision[\indexGradient]$, and $\incorrectDecision[\indexGradient]$, respectively.
Based on Lemma~\ref{lemma:errorPrHPA} and by Assumption~\ref{assump:zeroBound}, we can obtain a bound on $\probabilityIncorrect[\indexGradient]$
as
\if\IEEEsubmission0
\begin{align}
	\probabilityIncorrect[\indexGradient]
	=&
	\frac{1}{2}\coeffErr[{\zeroDecision[\indexGradient]}]+\frac{\incorrectDecision[\indexGradient]}{1-\zeroDecision[\indexGradient]}(1-\coeffErr[{\zeroDecision[\indexGradient]}])\nonumber
	\\
	\le&
	\frac{1}{2}\coeffErr[{\zeroDecisionMax}]+\frac{\incorrectDecision[\indexGradient]}{1-\zeroDecisionMax}(1-\coeffErr[{\zeroDecisionMax}])
	~.\label{eq:properrpre}
\end{align}
\else
\begin{align}
	\probabilityIncorrect[\indexGradient]
	=&
	\frac{1}{2}\coeffErr[{\zeroDecision[\indexGradient]}]+\frac{\incorrectDecision[\indexGradient]}{1-\zeroDecision[\indexGradient]}(1-\coeffErr[{\zeroDecision[\indexGradient]}])
	\le
	\frac{1}{2}\coeffErr[{\zeroDecisionMax}]+\frac{\incorrectDecision[\indexGradient]}{1-\zeroDecisionMax}(1-\coeffErr[{\zeroDecisionMax}])
	~.\label{eq:properrpre}
\end{align}
\fi

We now need to relate the bound in \eqref{eq:properrpre} to the learning parameters and the channel conditions. To this end, without loss of generality assume that $\globalGradientElement[\indexCommunicationRound][\indexGradient]$ is negative. By Assumption \ref{assump:varBound}, Assumption \ref{assump:mmodal} and utilizing the Gauss inequality, we can obtain a bound on $\incorrectDecision[\indexGradient]$ as 
\if\IEEEsubmission 0
\begin{align}
	\incorrectDecision[\indexGradient]&=\probability[{\signNormal[{\localGradientElement[\indexED,\indexGradient][\indexCommunicationRound]}]=\signNormal[{\globalGradientElement[\indexCommunicationRound][\indexGradient]}]|\gradientWeight[{\localGradientElement[\indexED,\indexGradient][\indexCommunicationRound]}]\neq0}]\nonumber
	\\
	&=\probability[{\localGradientElement[\indexED,\indexGradient][\indexCommunicationRound]>\thresholdForZero(1-\steepnessFactor)+\globalGradientElement[\indexCommunicationRound][\indexGradient]+|\globalGradientElement[\indexCommunicationRound][\indexGradient]|}]~\nonumber
	\\
	&=\probability[{\localGradientElement[\indexED,\indexGradient][\indexCommunicationRound]-\globalGradientElement[\indexCommunicationRound][\indexGradient]>\thresholdForZero(1-\steepnessFactor)+|\globalGradientElement[\indexCommunicationRound][\indexGradient]|}]~\nonumber
	\\
	&=\frac{1}{2}\probability[{|\localGradientElement[\indexED,\indexGradient][\indexCommunicationRound]-\globalGradientElement[\indexCommunicationRound][\indexGradient]|>\thresholdForZero(1-\steepnessFactor)+|\globalGradientElement[\indexCommunicationRound][\indexGradient]||\globalGradientElement[\indexCommunicationRound][\indexGradient]<0}]~\nonumber
	\\
	&\le\frac{1}{2}\begin{cases}
		\frac{4}{9}\frac{\varianceBoundEle[\indexGradient]^2/\batchSize}{(\thresholdForZero(1-\steepnessFactor)+|\globalGradientElement[\indexCommunicationRound][\indexGradient]|)^2}~,&\frac{\thresholdForZero(1-\steepnessFactor)+|\globalGradientElement[\indexCommunicationRound][\indexGradient]|}{\varianceBoundEle[\indexGradient]/\sqrt{\batchSize}}>\frac{2}{\sqrt{3}}~,\\
		1-\frac{\thresholdForZero(1-\steepnessFactor)+|\globalGradientElement[\indexCommunicationRound][\indexGradient]|}{\sqrt{3}\varianceBoundEle[\indexGradient]/\sqrt{\batchSize}}~,&\text{otherwise}~,\\
	\end{cases}~\nonumber
	\\
	&\le
	\frac{1/2}{\frac{1}{\sqrt{3}} \frac{\thresholdForZero(1-\steepnessFactor)+|\globalGradientElement[\indexCommunicationRound][\indexGradient]|}{\varianceBoundEle[\indexGradient]/\sqrt{\batchSize}} +1}
	~.\label{eq:boundGauss}
\end{align}
\else
\begin{align}
	\incorrectDecision[\indexGradient]&=\probability[{\signNormal[{\localGradientElement[\indexED,\indexGradient][\indexCommunicationRound]}]=\signNormal[{\globalGradientElement[\indexCommunicationRound][\indexGradient]}]|\gradientWeight[{\localGradientElement[\indexED,\indexGradient][\indexCommunicationRound]}]\neq0}]=\probability[{\localGradientElement[\indexED,\indexGradient][\indexCommunicationRound]>\thresholdForZero(1-\steepnessFactor)+\globalGradientElement[\indexCommunicationRound][\indexGradient]+|\globalGradientElement[\indexCommunicationRound][\indexGradient]|}]~\nonumber
	\\
	&=\probability[{\localGradientElement[\indexED,\indexGradient][\indexCommunicationRound]-\globalGradientElement[\indexCommunicationRound][\indexGradient]>\thresholdForZero(1-\steepnessFactor)+|\globalGradientElement[\indexCommunicationRound][\indexGradient]|}]=\frac{1}{2}\probability[{|\localGradientElement[\indexED,\indexGradient][\indexCommunicationRound]-\globalGradientElement[\indexCommunicationRound][\indexGradient]|>\thresholdForZero(1-\steepnessFactor)+|\globalGradientElement[\indexCommunicationRound][\indexGradient]||\globalGradientElement[\indexCommunicationRound][\indexGradient]<0}]~\nonumber
	\\
	&\le\frac{1}{2}\begin{cases}
		\frac{4}{9}\frac{\varianceBoundEle[\indexGradient]^2/\batchSize}{(\thresholdForZero(1-\steepnessFactor)+|\globalGradientElement[\indexCommunicationRound][\indexGradient]|)^2}~,&\frac{\thresholdForZero(1-\steepnessFactor)+|\globalGradientElement[\indexCommunicationRound][\indexGradient]|}{\varianceBoundEle[\indexGradient]/\sqrt{\batchSize}}>\frac{2}{\sqrt{3}}~,\\
		1-\frac{\thresholdForZero(1-\steepnessFactor)+|\globalGradientElement[\indexCommunicationRound][\indexGradient]|}{\sqrt{3}\varianceBoundEle[\indexGradient]/\sqrt{\batchSize}}~,&\text{otherwise}~,\\
	\end{cases}\le
	\frac{1/2}{\frac{1}{\sqrt{3}} \frac{\thresholdForZero(1-\steepnessFactor)+|\globalGradientElement[\indexCommunicationRound][\indexGradient]|}{\varianceBoundEle[\indexGradient]/\sqrt{\batchSize}} +1}
	~.\label{eq:boundGauss}
\end{align}
\fi
Hence, by using \eqref{eq:boundGauss} in \eqref{eq:properrpre}, an upper bound on the error term in \eqref{eq:errorTerm}  can be obtained as
\begin{align}
	&\sum_{\indexGradient=1}^{\numberOfModelParameters}|\globalGradientElement[\indexCommunicationRound][\indexGradient]|\probabilityIncorrect[\indexGradient]=\sum_{\indexGradient=1}^{\numberOfModelParameters}|\globalGradientElement[\indexCommunicationRound][\indexGradient]|\left(\frac{1}{2}\coeffErr[{\zeroDecisionMax}]+\frac{\incorrectDecision[\indexGradient]}{1-\zeroDecisionMax}(1-\coeffErr[{\zeroDecisionMax}])\right)\nonumber
	\\
	&\le\sum_{\indexGradient=1}^{\numberOfModelParameters}|\globalGradientElement[\indexCommunicationRound][\indexGradient]|\left(\frac{1}{2}\coeffErr[{\zeroDecisionMax}]+\frac{\frac{1/2}{\frac{1}{\sqrt{3}} \frac{\thresholdForZero(1-\steepnessFactor)+|\globalGradientElement[\indexCommunicationRound][\indexGradient]|}{\varianceBoundEle[\indexGradient]/\sqrt{\batchSize}} +1}}{1-\zeroDecisionMax}(1-\coeffErr[{\zeroDecisionMax}])\right)\nonumber
	\\
&= \frac{1}{2} \left( \coeffErr[{\zeroDecisionMax}]\norm{\globalGradient[\indexCommunicationRound]}_1 +\frac{1-\coeffErr[{\zeroDecisionMax}]}{1-\zeroDecisionMax}\sum_{\indexGradient=1}^{\numberOfModelParameters}\frac{\frac{\sqrt{3}\varianceBoundEle[\indexGradient]}{\sqrt{\batchSize}}|\globalGradientElement[\indexCommunicationRound][\indexGradient]|}{ \thresholdForZero(1-\steepnessFactor)+|\globalGradientElement[\indexCommunicationRound][\indexGradient]| +\frac{\sqrt{3}\varianceBoundEle[\indexGradient]}{\sqrt{\batchSize}}}\right) \nonumber
	\\
&\le \frac{1}{2} \left( \coeffErr[{\zeroDecisionMax}]\norm{\globalGradient[\indexCommunicationRound]}_1 +\frac{\sqrt{3}}{\sqrt{\batchSize}}\frac{1-\coeffErr[{\zeroDecisionMax}]}{1-\zeroDecisionMax}\frac{\maxGradient}{ \thresholdForZero(1-\steepnessFactor)+\maxGradient } \norm{\varianceBound}_1\right)~. \nonumber	
\end{align}
Based on Assumption~\ref{assump:boundedLoss},  we perform a telescoping sum over the iterations and
calculate the expectation over the randomness in the trajectory
as
\if \IEEEsubmission 0
	\begin{align}
	\lossFunctionGlobalMinimum-\lossFunctionGlobal[{\modelParametersAtIteration[0]}]&\le \expectationOperator[{\lossFunctionGlobal[{\modelParametersAtIteration[\communicationRounds]}]}][]-\lossFunctionGlobal[{\modelParametersAtIteration[0]}]\nonumber\\&=\expectationOperator[{\sum_{\indexCommunicationRound=0}^{\communicationRounds-1}\lossFunctionGlobal[{\modelParametersAtIteration[\indexCommunicationRound+1]}]-\lossFunctionGlobal[{\modelParametersAtIteration[\indexCommunicationRound]}] }][]\nonumber\\
	&\le
	\expectationOperator[{	\sum_{\indexCommunicationRound=0}^{\communicationRounds-1}-\learningRate(1-\coeffErr[{\zeroDecisionMax}])\norm{\globalGradient[\indexCommunicationRound]}_1+\frac{\learningRate^2}{2}\norm{\nonnegativeConstants}_1 \nonumber \right.\\&\left.~~~+\frac{\sqrt{3}\learningRate}{\sqrt{\batchSize}}\frac{1-\coeffErr[{\zeroDecisionMax}]}{1-\zeroDecisionMax}\frac{\maxGradient}{ \thresholdForZero(1-\steepnessFactor)+\maxGradient }\norm{\varianceBound}_1  }][]~.
	\label{eq:finaleq}
\end{align}
\else
	\begin{align}
	&\lossFunctionGlobalMinimum-\lossFunctionGlobal[{\modelParametersAtIteration[0]}]\le \expectationOperator[{\lossFunctionGlobal[{\modelParametersAtIteration[\communicationRounds]}]}][]-\lossFunctionGlobal[{\modelParametersAtIteration[0]}]=\expectationOperator[{\sum_{\indexCommunicationRound=0}^{\communicationRounds-1}\lossFunctionGlobal[{\modelParametersAtIteration[\indexCommunicationRound+1]}]-\lossFunctionGlobal[{\modelParametersAtIteration[\indexCommunicationRound]}] }][]\nonumber\\
	&\le
	\expectationOperator[{	\sum_{\indexCommunicationRound=0}^{\communicationRounds-1}-\learningRate(1-\coeffErr[{\zeroDecisionMax}])\norm{\globalGradient[\indexCommunicationRound]}_1+\frac{\learningRate^2}{2}\norm{\nonnegativeConstants}_1 +\frac{\sqrt{3}\learningRate}{\sqrt{\batchSize}}\frac{1-\coeffErr[{\zeroDecisionMax}]}{1-\zeroDecisionMax}\frac{\maxGradient}{ \thresholdForZero(1-\steepnessFactor)+\maxGradient }\norm{\varianceBound}_1  }][]~.
	\label{eq:finaleq}
\end{align}
\fi
By rearranging the terms in \eqref{eq:finaleq}, we obtain \eqref{eq:convergence}.
\end{IEEEproof}

\acresetall

\bibliographystyle{IEEEtran}
\bibliography{references}

\end{document}